\documentclass[oldversion]{aa}

\usepackage{natbib}
\usepackage{graphicx}

\makeatletter
\def\gsim{\compoundrel>\over\sim}

\def\compoundrel#1\over#2%
  {\mathpalette\compoundreL{{#1}\over{#2}}}
\def\compoundreL#1#2{\compoundREL#1#2}
\def\compoundREL#1#2\over#3%
  {\mathrel{\vcenter{\hbox{$\m@th \buildrel{#1#2}\over{#1#3}$}}}}
\makeatother

\begin{document}

\title{Axisymmetric simulations of magnetorotational core collapse:
       Approximate inclusion of general relativistic effects}
\titlerunning{Magnetorotational core collapse}

\author{M. Obergaulinger\inst{1} \and M.A. Aloy\inst{1,2} \and
H. Dimmelmeier\inst{1} \and E. M\"uller\inst{1}} 
\institute{Max-Planck-Institut f\"ur Astrophysik, 
           Karl-Schwarzschild-Str. 1, 
           D-85741 Garching, Germany
\and
           Departamento de Astronom\'{\i}a y Astrof\'{\i}sica,
           Universidad de Valencia,
           46100 Burjassot, Spain}
 \offprints{M. Obergaulinger}

\date{Received / Accepted }


\abstract{
  We continue our investigations of the magnetorotational collapse of
  stellar cores by discussing simulations performed with a modified
  Newtonian gravitational potential that mimics general relativistic
  effects. The approximate TOV gravitational potential used in our
  simulations captures several basic features of fully relativistic
  simulations quite well.  In particular, it is able to correctly
  reproduce the behavior of models that show a qualitative change both
  of the dynamics and the gravitational wave signal when switching from
  Newtonian to fully relativistic simulations.  For models where the
  dynamics and gravitational wave signals are already captured
  qualitatively correctly by a Newtonian potential, the results of the
  Newtonian and the approximate TOV models differ quantitatively.  The
  collapse proceeds to higher densities with the approximate TOV
  potential, allowing for a more efficient amplification of the magnetic
  field by differential rotation.  The strength of the saturation fields
  ($\sim 10^{15} \ \mathrm{G}$ at the surface of the inner core) is a
  factor of two to three higher than in Newtonian gravity.
  Due to the more efficient field amplification, the influence of
  magnetic fields is considerably more pronounced than in the Newtonian
  case for some of the models.  As in the Newtonian case, sufficiently
  strong magnetic fields slow down the core's rotation and trigger a
  secular contraction phase to higher densities.  More clearly than in
  Newtonian models, the collapsed cores of these models exhibit two
  different kinds of shock generation.  Due to magnetic braking, a first
  shock wave created during the initial centrifugal bounce at subnuclear
  densities does not suffice for ejecting any mass, and the temporarily
  stabilized core continues to collapse to supranuclear densities.
  Another stronger shock wave is generated during the second bounce as
  the core exceeds nuclear matter density.  The gravitational wave
  signal of these models does not fit into the standard classification.
  Therefore, in the first paper of this series we introduced a new type
  of gravitational wave signal, which we call type\,IV or ``magnetic
  type''.  This signal type is more frequent for the approximate
  relativistic potential than for the Newtonian one.  Most of our
  weak-field models are marginally detectable with the current LIGO
  interferometer for a source located at a distance of 10 kpc. Strongly
  magnetized models emit a substantial fraction of their GW power at
  very low frequencies. A flat spectrum between 10 Hz and $\la 100$ kHz
  denotes the generation of a jet-like hydromagnetic outflow.
}



\maketitle


\section{Introduction}
\label{Sek:Intro}

In a core collapse supernova, the iron core of an evolved massive star
with mass $M \ga 8 - 10 M_{\odot}$ collapses to a neutron star, thereby
releasing a large amount of gravitational binding energy.  Although this
basic picture is commonly accepted, state-of-the-art supernova
calculations still do not yield explosions that match the observations
\citep{Buras_etal_03, ThJ_Ringberg2004}.  These calculations incorporate
a detailed and thus computationally very expensive treatment of the
microphysics of core matter (equation of state, radiation transport,
neutrino physics, etc.).  Additionally, they have to be performed in at
least two, or even better three, spatial dimensions in order to be able
to follow the development of genuine non-spherical effects, such as
convection or rotation, and to explain observed explosion asymmetries
and neutron star kicks.  Due to their inherent complexity, these
simulations are still subject to some limitations.  In particular, most
of them neglect the possible influence of magnetic fields, and they
usually treat gravity in the Newtonian limit.  Thus, it is desirable
that these calculations are complemented by investigations that focus
particularly on some selected aspects of the full scenario not studied
yet in great detail, but that avoid some of the computationally most
expensive and physically crucial aspects.

Since the end product of gravitational core collapse is a compact object
with a radius not much larger than its Schwarzschild radius, general
relativity (GR) rather than Newtonian gravity is the appropriate theory
for describing the gravitational field of a supernova core. This issue
has been addressed recently by several studies that are concerned with
the GR collapse and the subsequent evolution of stellar cores (see,
e.g.\,\citealp[hereafter DFM]{DFM1, DFM2},
\citealp{2005PhRvD..71b4014S}, and references therein). In addition, the
near success of detailed supernova simulations in producing an explosion
(see, e.g.\,\citealp{Buras_etal_03, Janka_etal05, Mezzacappa05})
suggests that not only the microphysics of core matter but also other
ingredients of the complex problem, such as GR and magnetohydrodynamic
(MHD) effects, should be treated as accurately as possible.

Extending a comprehensive parameter study of the gravitational collapse
of rotating cores in Newtonian gravity \citep[ henceforth ZM]{ZM97} to
the conformal-flatness approximation of full GR, DFM showed that, in
principle, the same types of dynamic behavior and gravitational
radiation result with Newtonian, as well as with GR, gravity.  In both
cases, the collapse of a core can be stopped by the stiffening of the
nuclear equation of state at supra-nuclear densities (standard-type
regular bounce), or -- for sufficiently fast initial rotation -- by
centrifugal forces (multiple bounce).  But DFM found both quantitative
and qualitative differences between the evolution of the Newtonian and
GR variants of the same initial configuration calculated with the same
equation of state. They observed a shift of the borderline separating
regions of parameter space with models of different dynamic behavior
resulting in different types of gravitational wave signals. In GR the
collapse is generically deeper, i.e.\,higher maximum densities are
reached, and some models suffering a centrifugal bounce in Newtonian
gravity collapse to supra-nuclear densities and experience a pressure
bounce.  However, the study of DFM (as that of ZM) was based on rotating
polytropes in hydrostatic equilibrium as initial configurations,
involved only models with simplified microphysics, and
completely neglected transport physics.  On the other hand,
multi-dimensional full GR simulations with detailed microphysics are not
yet available.

Since the compactness of a neutron star is still moderate, one may ask
oneself whether it is indeed necessary to perform full-scale GR
simulations, or whether it is possible to capture the essentials of the
GR effects by using some approximative treatment, such as relativistic
corrections to the Newtonian gravitational potential.  To explore this
possibility, we applied the effective \emph{Tolman--Oppenheimer--Volkoff
  (TOV) potential} proposed by \citet{RamJan02} and \citet{TOVPot1D} in
the simulations described in this publication.

In addition to GR effects, magnetic fields are often neglected in
simulations of supernova core collapse, which may not be justified. In
the past only a few authors \citep[]{LBW, Bisnovatyi, MEAS76, Ewald79,
  Ohnishi, Symba} have considered MHD effects, but during the past few
years magnetorotational core collapse has become an active research
field \citep[]{WMW02, Aki1, KotakeSN, KotakeMGW, TakiwakiMHD, YamSaw,
  Ardeljan05, KotakeMHDnuan, SawKotYam}.

We joined this effort very recently and performed a parameter study of
the magnetorotational collapse of stellar cores in Newtonian gravity
\citep[ hereafter Paper\,I]{OAM1} by considering the evolution of a set
of initial models with different rotation rates and rotation profiles,
and with different initial magnetic fields (field strength $|\vec b|
\sim 10^{10} - 10^{13}\ \mathrm{G}$) that are purely poloidal.  The
properties of the non-magnetized initial configurations and the
microphysics included in the simulations are the same as those used in
the studies of ZM and DFM, i.e.\,we neglected radiation transport and
nuclear reactions, and used a simplified analytic equation of state
allowing for different values of the sub-nuclear adiabatic index.  The
gravitational wave (GW) signal was calculated using the standard
quadrupole formula, as implemented by \citet{MoMey}, and extended to MHD
by \citet{KotakeMGW} and \citet{YamSaw}.  The main findings of Paper\,I
are:
\begin{itemize}
\item The initial magnetic field is amplified by the differential
  rotation of the core to magnetic energies that are $\sim 10\%$ of the
  rotational energy, the initially poloidal field being wound up in a
  dominant toroidal component.
\item 
According to \citet{Aki1}, the magnetorotational instability (MRI)
\citep{BalHaw91, Balbus95} may play an important role during core
collapse leading to MHD turbulence, very efficient field amplification,
and angular momentum transport.  Our simulations indeed showed the
growth of MRI-like modes in a number of models with intermediate
initial field strengths.  
\item If the initial field becomes sufficiently strong after core
  bounce, it extracts rotational energy from the core by such a large
  amount that it loses centrifugal support and begins to contract,
  evolving from its post-bounce rotational equilibrium state towards
  another more compact equilibrium state.  The magnetic field can thus
  transform a centrifugally supported core into one that is supported
  against gravity (mainly) by pressure forces.
\item Cores with very strong initial magnetic fields ($|\vec b| >
  10^{12}\ \mathrm{G}$) develop collimated bipolar outflows along the
  rotation axis.
\item The gravitational wave signals of weakly magnetized cores do not
  differ from those of the corresponding non-magnetized cores studied by
  ZM and DFM. However, the peak signal amplitudes for strong magnetic
  fields differ by several percent at bounce. In strongly magnetized
  models evolving from a centrifugally to a pressure-supported
  configuration, the wave signal changes from type\,II to type\,I.
\item The presence of a collimated outflow causes a positive GW
  amplitude after bounce, which can become comparable to the amplitude at
  bounce.
\end{itemize}

In the following we present a continuation of our investigation of
magnetorotational core collapse by extending our previous, purely
Newtonian treatment of gravity to an approximately relativistic one. To
this end, we re-calculated a subset of the models discussed in Paper\,I,
substituting the Newtonian potential by an effective TOV potential
\citep{RamJan02, TOVPot1D}, which approximates the effects of GR gravity
on the dynamics of the core.

The paper is organized as follows.  We describe the physics underlying
our models in Sect.\,\ref{Sek:Modelle}, including the approximate
treatment of relativistic gravity. In Sect.\,\ref{Sek:Ergebnissse} we
present the results of our simulations and discuss how these results
obtained with the effective TOV potential differ from those of the
previous Newtonian simulations (Paper\,I).  Our findings are
summarized in Sect.\,\ref{Sek:Schluss}, which also gives some
conclusions.  A compilation of some data of our 12 models is provided
in tabular form in Appendix \ref{Sek:Syn}.  For more detailed
information about our numerical method and physical model the reader
is referred to Paper\,I.


\section{Physics of our models}
\label{Sek:Modelle}

\subsection{Magnetohydrodynamic evolution}
\label{Suk:Modelle:Glgn}

We evolve the density $\rho$, the velocity $\vec v$, the total energy
density $e_{\star} = e + e_{\mathrm{kin}} + e_{\mathrm{mag}}$ ($e$,
$e_{\mathrm{kin}} = \rho \vec v^2 / 2$, and $e_{\mathrm{mag}} = \vec
B^2 / 8\pi$ are internal, kinetic, and magnetic energy density,
respectively), and the magnetic field $\vec B$ of our models according
to the equations of Newtonian ideal magnetohydrodynamics (MHD):
\begin{eqnarray}
  \label{Gl:Konti}
  \partial_t \rho + \nabla_m (\rho v^m ) & = & 0, \\
  \partial_t 
  \left(
    \rho v_n\right) + \nabla_m \left(\rho v_n v^m 
    + P_{\star} - b_n b^m
  \right) & =  & f_n \label{Gl:Euler}, 
  \\
  \partial_t e_{\star} + \nabla_m 
  \left(\left(e_{\star}+P_{\star}\right) v^m 
    - b^m b_n v^n\right) & = &  q.
  \label{Gl:Energie}
\end{eqnarray}
Here, Latin indices run from $1$ to $3$ and Einstein's summation
convention applies.  In the following, we use natural units where
$G=c=1$. The total pressure $P_{\star} = P_{\mathrm{gas}} + \vec b^2 / 2$
is the sum of the gas pressure $P_{\mathrm{gas}}$ and the isotropic
magnetic pressure $P_{\mathrm{mag}} = \vec b^2 / 2$, with $\vec b = \vec
B / \sqrt{4\pi}$.  We integrate the MHD equations in spherical
coordinates $(r, \theta, \phi)$ assuming axisymmetry and equatorial
symmetry.
 
Neutrino transport is not included in the code. We use a simple hybrid
ideal gas equation of state that consists of a polytropic contribution
describing the degenerate electron pressure and (at supra-nuclear
densities) the pressure due to repulsive nuclear forces, and a thermal
contribution that accounts for the heating of the matter by shocks:
\begin{equation}
  P = P_\mathrm{p} + P_\mathrm{th},
  \label{eq:hybrid_eos}
\end{equation}
where
\begin{equation}
  P_\mathrm{p} = K \rho^{\gamma},
  \qquad
  P_\mathrm{th} = \rho \epsilon_\mathrm{th} (\gamma_\mathrm{th} - 1),
  \label{eq:hybrid_eos_terms}
\end{equation}
and $ \epsilon_\mathrm{th} = \epsilon - \epsilon_\mathrm{p} $. The
polytropic specific internal energy $ \epsilon_\mathrm{p} $ is
determined from $ P_\mathrm{p} $ by the ideal gas relation in
combination with continuity conditions in the case of a discontinuous
$ \gamma $. In that case, the polytropic constant $ K $ also has to be
adjusted \citep[for more details, see][]{DFM1, JZM93}.

The initial models are rotating polytropes in equilibrium, which mimic
an iron core supported by electron degeneracy pressure, with a central
density $ \rho_{\rm c\,i} = 10^{10} {\rm\ g\ cm}^{-3} $ and equation of
state parameters $ \gamma_\mathrm{i} = 4 / 3 $ and $ K = 4.897 \times
10^{14} $ (in cgs units). To initiate the collapse, the initial
adiabatic index is reduced to $ \gamma_1 < \gamma_\mathrm{i} $.  At
densities above nuclear-matter density, $ \rho > \rho_{\rm nuc}$, the
adiabatic index is increased to $ \gamma_2 \gsim 2.5 $ to model the
abrupt stiffening that a realistic equation of state exhibits at the
phase transition to nuclear matter. The density at which this transition
occurs depends on the details of the equation of state. With variations
of at most a few $10\%$, $\rho_{\rm nuc}\equiv 2.0 \times
10^{14}\,$g\,cm$^{-3}$ can be considered a representative value.
Furthermore, as this transition density value has been used in all
previous studies employing a similar simplified equation of state, we
adopt this value too, in order to allow for a comparison with those
studies.

The initial models are characterized by their rotational energy
parameter $\beta_{\mathrm{rot}} = {E_{\mathrm{rot}}} /
{|E_{\mathrm{grav}}|}$, where $E_{\mathrm{rot}}$ and $E_{\mathrm{grav}}$
denote rotational and gravitational energy, respectively, and their
degree of differential rotation, respectively (for details see
Paper\,I).  The angular velocity profile $\Omega(\varpi)$ is given by
the so-called $j$-constant law,
\begin{equation}
  \label{Gl:jconst}
  \Omega(\varpi) = \frac{\Omega_0}{1+\left(\frac{\varpi}{A}\right)^2},
\end{equation}
where $\Omega_0$, $\varpi$, and $A$ are the angular velocity at the
center, the distance from the rotational axis, and a characteristic
length scale, respectively.

The initial models are obtained with the method and code of
\citet{Komatsu_etal_89}, which allows for both Newtonian and GR gravity.
The initial magnetic field is purely poloidal. It is generated by a
current loop of a given radius $r_{\mathrm{mag}}$ and has a prescribed
field strength $|\vec b_{0}| = 10^{10} - 10^{13} \ \mathrm{G}$ in the
center of the core. We choose $r_{\mathrm{mag}}= 400\,$km in most of our
models.

In the subsequent discussion, we follow the same naming convention as in
Paper\,I. The allocation of model names to physical model parameters is
explained in Table \ref{Tab:InitPars} for the hydrodynamic initial data and in
Table \ref{Tab:MinitPars} for the initial magnetic field configuration,
respectively.  The model names defined in this way are extended further
by the letters ``N'' (for Newtonian) or ``T'' (for TOV), respectively.

\begin{table}[ht]
  \caption[Initial Models] 
{ Initial models and their parameterization: $A$ and
  $\beta_{\mathrm{rot}}$ are the rotation law parameter
  (Eq.\,\ref{Gl:jconst}) and the ratio of rotational energy to
  gravitational energy, respectively. Higher values of $A$ correspond
  to more rigidly rotating cores, and $\Gamma_1$ is the sub-nuclear
  adiabatic index of our hybrid equation of state.  }
  \label{Tab:InitPars}
  \centering
  \begin{tabular}{cl|cl|cl}
    \hline \hline
    Model & $A [\mathrm{cm}]$ & Model & $\beta_{\mathrm{rot}} [\%]$ 
                              & Model & $\Gamma_1$ \\
    \hline$\rule{0 em}{1.3 em}$
    A1 &  $5\cdot 10^{9}$  & B1 & $\approx 0.25$ & G1 & $1.325$\\
    A2 &  $1\cdot 10^{8}$  & B2 & $\approx 0.45$ & G2 & $1.32$ \\
    A3 &  $5\cdot 10^{7}$  & B3 & $\approx 0.9$  & G3 & $1.31$ \\
    A4 &  $1\cdot 10^{7}$  & B4 & $\approx 1.8$  & G4 & $1.30$ \\
       &               & B5 & $\approx 4.0$  & G5 & $1.28$ \\
    \hline \hline
  \end{tabular}
\end{table}

\begin{table}[ht]
  \caption[Initial magnetic fields] 
{ Parameterization of the initial magnetic field configuration for the
  models of series AaBbGg-DdMm by the radius of the field-generating
  current loop centered at $r_{\mathrm{mag}}$ (parameterized by $d = 1,
  2, 3, 4, 0$) and the field strength in the core's center
  $B_0=\sqrt{4\pi}\,b_0$ (parameterized by $ m = 10, 11, 12, 13$).  For
  models AaBbGg-D0Mm, the field-generating current loop is located at
  infinity, yielding a uniform magnetic field throughout the entire
  core. Unlike Paper\,I, we discuss here only models with
  $r_{\mathrm{mag}} = 400\,$km (i.e., $d=3$).  }
  \label{Tab:MinitPars}
  \centering
  \begin{tabular}{cc|cl}
    \hline\hline
    Model & $r_{\mathrm{mag}} \ [\mathrm{km}]$ & Model 
          & $b_0  \ [\textrm{G}]$ \\
    \hline$\rule{0 em}{1.3 em}$
    D1 & $100$   & M10 & $10^{10}$ \\
    D2 & $200$   & M11 & $10^{11}$ \\
    D3 & $400$   & M12 & $10^{12}$ \\
    D4 & $800$   & M13 & $10^{13}$ \\
    D0 & $\infty$  & & \\
    \hline\hline
  \end{tabular}
\end{table}

The simulations described in this paper were performed using a
second-order conservative Eulerian code based on the relaxing TVD scheme
\citep{JX} for the solution of the fluid equations and the constraint
transport method \citep{EvHaw} to ensure the solenoidal character of the
magnetic field. The same numerical code was used to compute the
simulations presented in Paper\,I.

For the calculation of the GW amplitude, we employ the quadrupole
formula as numerically implemented by \citet{MoMey}.  Because of the
assumption of axisymmetry, the GW signal is determined completely by the
quadrupole amplitude $A^{\mathrm{E}2}_{20}$, which is a function of
density, velocity, gravitational potential, and magnetic field strength
(for details see Paper\,I).  The dimensionless GW strain measured by an
observer located in the equatorial plane at a distance $R$ from the core
is given by
\begin{equation}
  \label{Gl:hTT-AE220}
  h =   
  \frac{1}{8} 
  \sqrt {\frac{15}{\pi}} \frac{A^{\mathrm{E}2}_{20}}{R}
  =
  8.8524\times 10^{-21} 
  \!\!
  \left(
    \frac{A^{\mathrm{E}2}_{20}}{10^3\,\mathrm{cm}}
  \right)
  \!\!
  \left(
    \frac{10\,\mathrm{kpc}}{R}
  \right)
  \!\!
  .
\end{equation}

\subsection{Gravity}
\label{Suk:Modelle:Grav}

The inclusion of the effects of gravitational forces into the MHD
equations introduces sources of momentum and energy in the
conservation laws (\ref{Gl:Euler}, \ref{Gl:Energie}):
\begin{eqnarray}
  \vec f_{\mathrm{grav}} & = & - \rho \vec \nabla \Phi,
  \label{Gl:Grimp} \\
  q_{\mathrm{grav}} & = & - \rho \vec v \cdot \vec \nabla \Phi.
  \label{Gl:Grerg}
\end{eqnarray}
The Newtonian gravitational potential $\Phi_{\mathrm{N}}$ obeys the
Poisson field equation
\begin{equation}
  \bigtriangleup \Phi_{\mathrm{N}} = 4 \pi \rho,
  \label{Gl:Poisson}
\end{equation}
where $G$ is the gravitational constant.  The gravitational potential
is determined from the density distribution using the computationally
efficient Poisson solver of \citet{MueSte}, which is based on the
integral form of Poisson's equation and (in axisymmetry) on an
expansion of the density distribution into Legendre polynomials (up to
order 12 in our simulations).

In order to take into account the effects of general relativity into
account in an approximate way, we follow the approach proposed by
\citet{RamJan02}, which was extended and further investigated by
\citet{TOVPot1D}.  In this approach the {\em 1D} spherical
Newtonian potential $\Phi^{1\mathrm{d}}_{\mathrm{N}}$ is replaced by an
effective GR potential $\Phi^{1\mathrm{d}}_{\mathrm{TOV}}$, which is
constructed using the TOV equation (see, e.g.\,\citealp{ShapTeuk}) of
hydrostatic equilibrium in GR:
\begin{equation}
  \label{Gl:TOV}
  \frac{\mathrm{d} P}{\mathrm{d} r} = - \frac{m}{r^2} \left(1+\frac{P}{\rho}\right)
                          \left(1+\frac{4\pi r^3 P}{m}\right)
                          \left(1-\frac{2m}{r} \right)^{-1}
                          \!\!\!\!\!\!\!.
\end{equation}
Here, $r, \rho$, and $P$ are the radial coordinate, the density, and
the pressure, respectively.  The gravitational mass
\begin{equation}
  m(r) = \int_{0}^{r} (\rho + e)\, 4\pi r'^2 \mathrm{d} r'
\end{equation}
includes contributions from the mass density $\rho$ and the internal
energy density $e$.  Comparing the TOV equation with its Newtonian limit
($c \to \infty$), corrections to the potential can be defined to take
into account that in GR every form of energy, including pressure, acts
as a source of gravity.  Additionally, the radial dependence of the
potential is corrected for the Schwarzschild radius $R_\mathrm{S}(r) = 2
m(r)$ of the gravitational mass inside a radius $r$, and a term
depending on the radial motion of the fluid is included, yielding the
effective relativistic potential \citep{TOVPot1D}
\begin{eqnarray}
  \Phi^{\mathrm{1d}}_{\mathrm{TOV}} (r) 
  & = & G \int_{\infty}^{r}  
             \left( m(r') + 4\pi r'^3 P \right) 
             \left( \frac{\rho + e + P}{\rho} \right)
  \nonumber \\
  & \phantom{=} & \phantom{G \int_{\infty}^r} \times  
             \left( 1 + v_r^2 - \frac{2 m(r')}{r' }
             \right)^{-1} \frac{\mathrm{d} r'}{r'^2} \, .
\label{Gl:TOV1D}
\end{eqnarray}

This spherically symmetric effective relativistic potential is also
applied in our 2D axisymmetric simulations,
where we first compute angular averages of the relevant hydrodynamic
variables.  These are then used to calculate the spherical Newtonian
potential $\Phi_{\mathrm{N}}^{\mathrm{1d}}(r)$ and the spherical TOV
potential $\Phi_{\mathrm{TOV}}^{\mathrm{1d}}(r)$.  Finally, we modify
the 2D Newtonian potential $\Phi_{\mathrm{N}}^{\mathrm{2d}} (r,\theta)$
to obtain the two-dimensional TOV potential
\begin{equation}
  \Phi_{\mathrm{TOV}}^{\mathrm{2d}}(r,\theta) = 
  \Phi_{\mathrm{N}}^{\mathrm{2d}}(r,\theta) + 
        \left( \Phi_{\mathrm{TOV}}^{\mathrm{1d}}(r) - 
               \Phi_{\mathrm{N}  }^{\mathrm{1d}}(r)   \right).
\label{Gl:TOV2D}
\end{equation}


\section{Results}
\label{Sek:Ergebnissse}

\subsection{Hydrodynamic simulations}
\label{Suk:Hydro}

Recently, \citet{TOVPot1D} presented a comprehensive investigation of
different approximative treatments of relativistic gravity within
Newtonian hydrodynamics codes for supernova simulations.  They find that
the effective relativistic potential produces excellent agreement with a
fully relativistic solution in spherical symmetry and that it
approximates relativistic solutions for rotational core collapse
qualitatively well.  A few years earlier, \citet{DFM1, DFM2} compared
the results of 1D and 2D supernova core collapse calculations obtained
with their approximate (exact for spherically symmetric models) GR code
based on the conformal flatness condition (CFC) with those of Newtonian
simulations.

To calibrate our implementation of the effective TOV potential,
Eqs.\,(\ref{Gl:TOV1D}, \ref{Gl:TOV2D}), and to compare the results of
our MHD code with those of \citet{TOVPot1D} and \citet{DFM2}, we perform
several (purely) hydrodynamic core collapse calculations both in
spherical symmetry and in axisymmetry.  The initial equilibrium models
are constructed using either Newtonian or full GR gravity (with the same
numerical codes as in DFM).

Using a sub-nuclear adiabatic index $\gamma_1 = 1.31$ (see Sect.\,%
\ref{Suk:Modelle:Glgn}) and assuming spherical symmetry, the TOV
potential yields results that agree very well with the fully
relativistic ones, the error in the bounce density $\rho_{\mathrm{b}}$
being about 7\,\% compared to about 28\,\% for a Newtonian potential
(Table\,\ref{Tab:1d-Dichten}). Note that the choice of initial data
(Newtonian or GR) has little effect ($<0.5\,\%$). Concerning global core
quantities, like e.g.\,the radius where $\rho(r) = 10^{12}\,%
\mathrm{g\,cm^{-3}}$ and the mass inside this radius, we find very good
agreement, with the errors less than 2\,\%.

\begin{table}[ht]
  \centering
  \caption{ Summary of the simulations of the spherically symmetric
            collapse of a core.  The table gives the values of the
            time of bounce $t_{\mathrm{b}}$ and the density at bounce
            $\rho_{\mathrm{b}}$ for the evolution of Newtonian (N) and
            GR (G) initial models using Newtonian, TOV, or GR gravity
            (assuming CFC). The last data are from DFM. The last two
            columns give the radius where $\rho(r) = 10^{12}\,%
            \mathrm{g\,cm^{-3}}$, and the mass inside this radius at
            $t\approx 55\,$ms, respectively.}
  \label{Tab:1d-Dichten}
  \begin{tabular}{cccccc}
    \hline \hline
    gravity & initial data & $t_{\mathrm{b}}$ & $\rho_{\mathrm{b}}$ 
            & $R_{>12}$ & $M_{>12}$ \\
    & & $[\mathrm{ms]}$  & $[10^{14} \mathrm{g cm^{-3}}]$ 
    & [km] & $M_{\sun}$ \\
    \hline
    N   & N  & 47.9 & 3.97 & 24.5 & 0.58 \\
    TOV & N  & 46.7 & 4.73 & 25.0 & 0.55 \\
    TOV & G & 47.8 & 4.75 & 25.1 & 0.55 \\
    GR  & G & 48.0 & 5.10 & 24.8 & 0.54 \\
    \hline \hline
  \end{tabular}
\end{table}

\begin{table}[ht]
  \centering
  \caption{Comparison of the rotating models A1B3G3 and A3B3G3
           calculated using the Newtonian potential (N), the effective
           TOV potential (T), and GR (G) gravity (assuming CFC). The
           last data are from DFM.}
  \label{Tab:2d-Hydro}
  \begin{tabular}{cccc}
    \hline \hline
    model & $t_{\mathrm{b}}$ $^{\mathrm{a}}$ & $\rho_{\mathrm{b}}$ $^{\mathrm{b}}$ & 
            $\beta_{\mathrm{rot}}^{\mathrm{max}}$ $^{\mathrm{c}}$ \\
            & $[\mathrm{ms]}$ & $[10^{14} \mathrm{g \ cm^{-3}}]$ & $[\%]$ \\
    \hline
    A1B3G3-N & $48.6$ & $3.40$ & $ 8.1$ \\
    A1B3G3-T & $48.5$ & $4.14$ & $ 7.3$ \\
    A1B3G3-G & $48.6$ & $4.23$ & $10.6$ \\
    A3B3G3-N & $49.7$ & $2.41$ & $15.8$ \\
    A3B3G3-T & $49.5$ & $3.19$ & $17.1$ \\
    A3B3G3-G & $49.7$ & $3.35$ & $20.3$ \\
    \hline \hline
  \end{tabular}
  \begin{list}{}{}
  \item[$^{\mathrm{a}}$] Time of bounce.
  \item[$^{\mathrm{b}}$] Maximum density at bounce.
  \item[$^{\mathrm{c}}$] Maximum ratio of rotational to gravitational
    energy.
  \end{list}
\end{table}

\begin{figure*}[htbp]
  \centering
  \includegraphics[width=6.7cm]{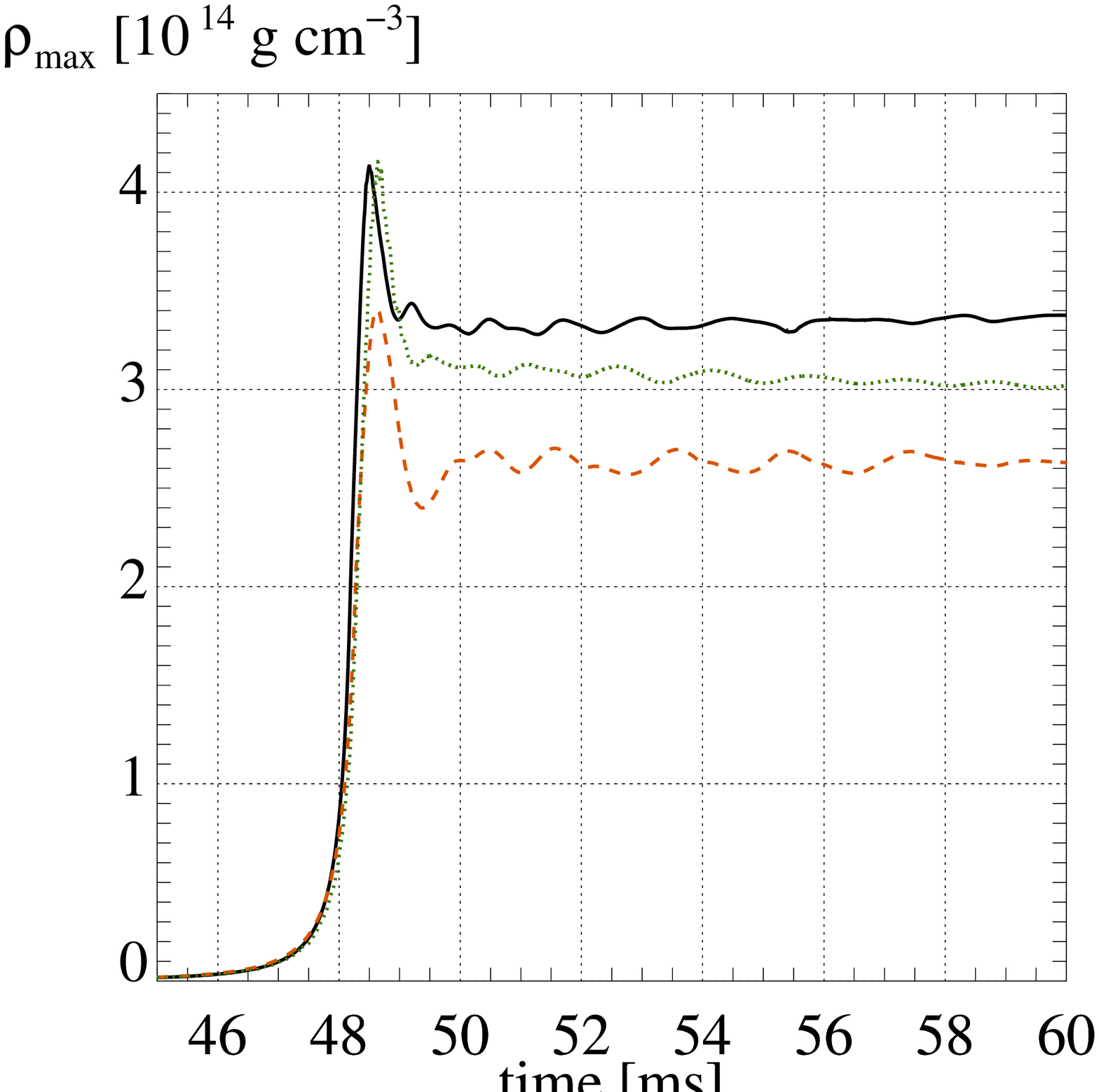}
  \includegraphics[width=6.7cm]{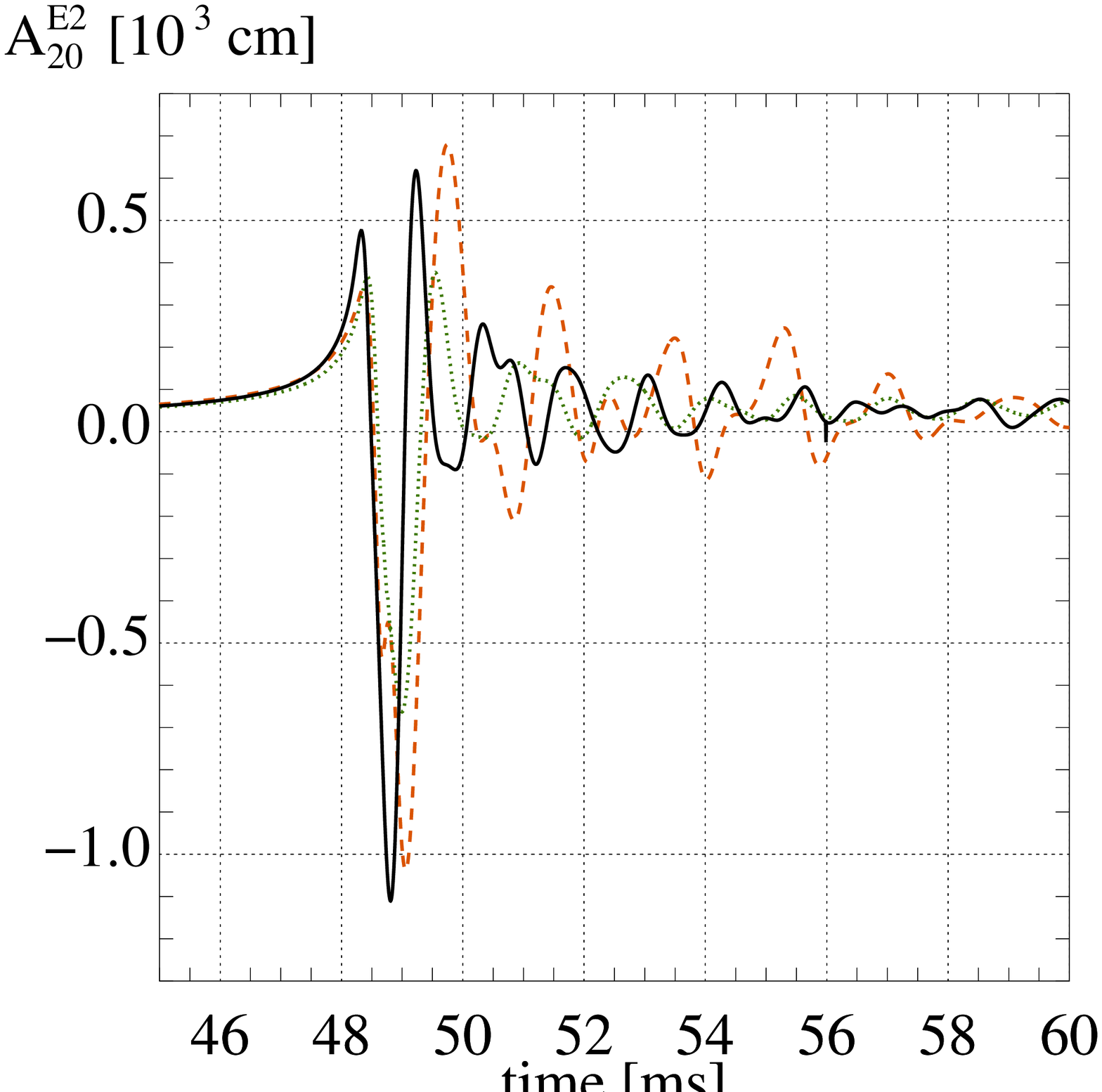}
  \includegraphics[width=6.7cm]{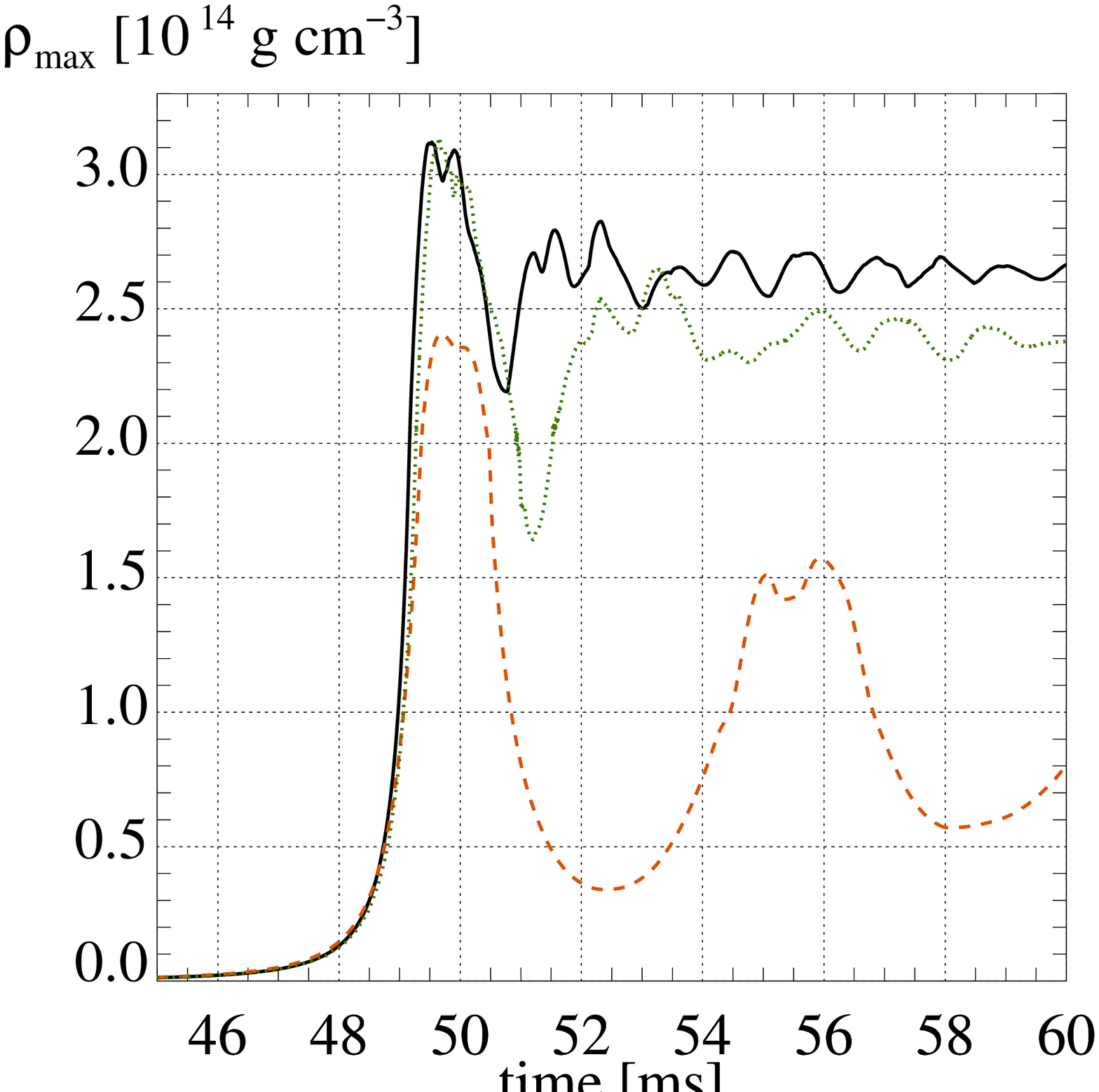}
  \includegraphics[width=6.7cm]{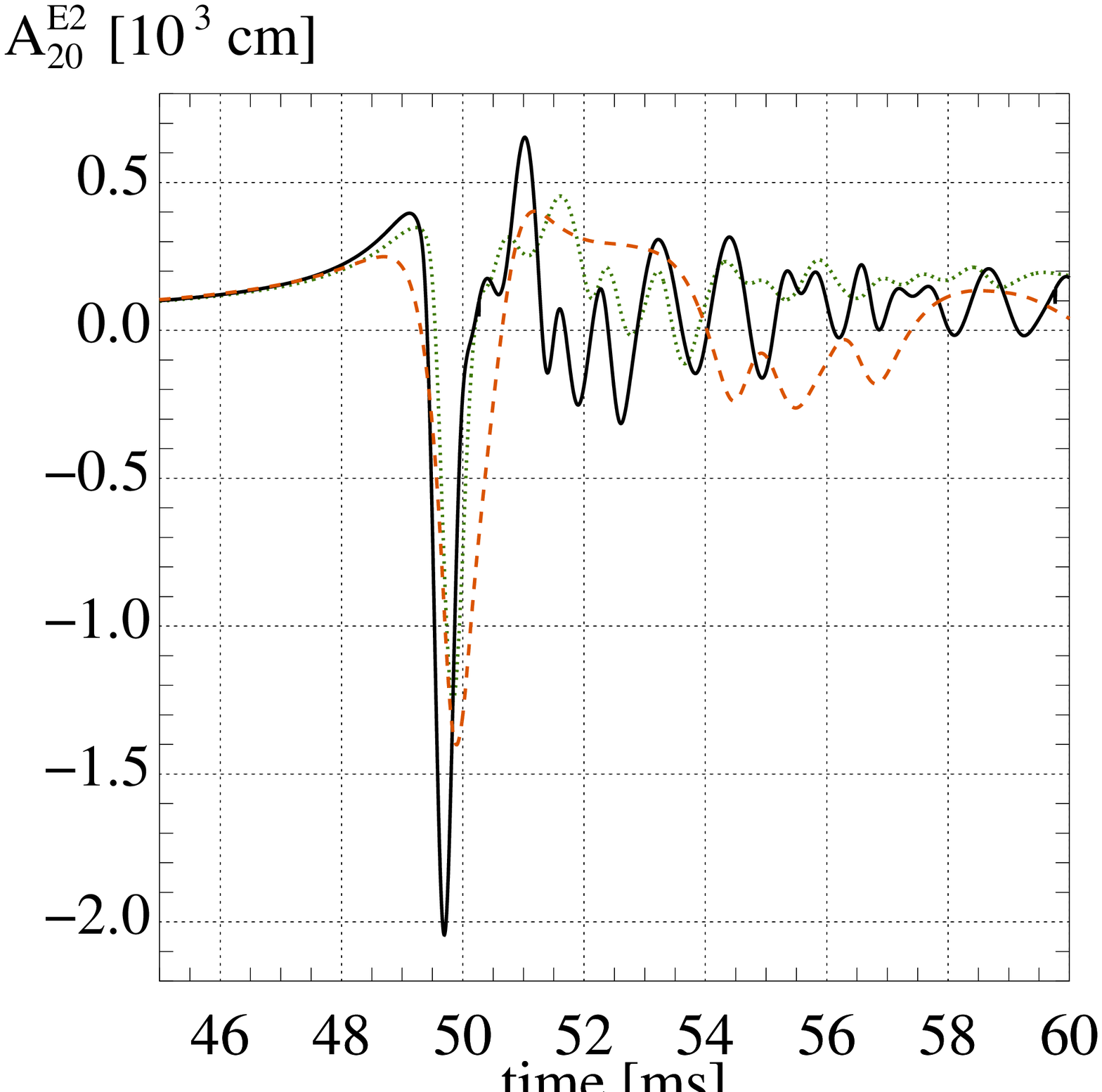}
  \caption{Evolution of the maximum density $\rho_{\mathrm{max}}$
           (left panels) and the GW amplitude $A^{E2}_{20}$ (right
           panels) of models A1B3G3 (upper panels) and A3B3G3 (lower
           panels) computed with the Newtonian (dashed red lines) and
           the effective TOV (solid black lines) potential, and in
           full GR (dotted green lines) assuming CFC.
           A color version of the figure can be found in the on-line
           edition of the journal.
         }
  \label{Fig:rho_GW}
\end{figure*}

We also test the performance of the TOV potential in axisymmetric
simulations of non-magnetic, rotating models using two prototypical
models from the model set of Paper\,I (Table\,\ref{Tab:2d-Hydro},
Fig.\,\ref{Fig:rho_GW}).

Model A1B3G3 bounces both in Newtonian and GR gravity due to the
stiffening of the equation of state beyond nuclear density (type\,I
model).  Using the effective TOV potential, the model reaches a maximum
density at bounce, which is about 20\,\% higher than in the Newtonian
case and only 2\,\% lower than in GR.  The GW signal obtained with the
effective TOV potential shows similar qualitative features as the one
calculated in GR, but overestimates the signal amplitudes at bounce by
about 50\,\% (Fig.\,\ref{Fig:rho_GW}, upper left). Using the Newtonian
potential, the deviations are of comparable order, but persist also
during the ring-down phase. The frequencies of the ring-down
oscillations are off by $\sim 30\,\%$ in the Newtonian case, while those
resulting from the effective TOV potential agree very well with the
frequencies of the GR model.

Model A3B3G3 rotates initially quite differentially and very rapidly
($\beta_{\mathrm{rot}}^{\mathrm{ini}} = 1.8\,\%$).  In the Newtonian
case, the core bounces mainly due to centrifugal forces, although it
exceeds nuclear matter density during bounce.  After bounce, the core
expands to sub-nuclear densities and exhibits large-scale pulsations
with little damping, giving rise to a GW signal intermediate to ZM's
type\,I and type\,II signals (Fig.\,\ref{Fig:rho_GW}, lower panels).
In GR the core reaches a $\sim 40\,\%$ higher density during bounce and
settles into a pressure supported equilibrium of supra-nuclear central
density after some ring-down oscillations.  The gravitational wave
signal is of type\,I.  The gross features of the GR density evolution
are reproduced by the effective TOV potential, and the maximum density
at bounce agrees within 5\,\% with that of the GR simulation. However,
there are considerable differences in the GW signal.  Contrary to the
Newtonian potential, the TOV potential gives the correct signal type
(I), but the amplitude of the dominant negative peak at bounce exceeds
that of the Newtonian run by $\sim 30\,\%$, whereas the correct GR
amplitude is $\sim 10\,\%$ smaller than the Newtonian one
(Fig.\,\ref{Fig:rho_GW}, lower right).

The above results show that the TOV potential is able to reproduce the
results of full GR simulations quite well for slowly rotating cores.
For rapidly rotating models the evolution of the maximum density is
reproduced very well, and the GW signal is of the correct type, but
considerable differences are found concerning the amplitude of the GW
signal.  A more comprehensive investigation of the performance of
effective TOV potentials in hydrodynamic simulations of rotational
core collapse will be presented elsewhere (M\"uller et.\,al., in
preparation).

\subsection{Magnetohydrodynamic simulations}
\label{Suk:Hydmag}

We consider four models from Paper\,I to explore the effects of
relativistic gravity on the dynamics and the GW signal of
magnetorotational core collapse.  Three of these models exhibit the same
type of signal as the corresponding Newtonian models of ZM: A1B3G3
(type\,I), A2B4G1 (type\,II), and A3B3G5 (type\,III). In the fourth
model (A3B3G3), the GW signal changes from transition type\,I/II to
type\,I when changing from Newtonian to relativistic gravity.  For each
of these selected models, we substitute the Newtonian potential by the
effective TOV potential Eqs.\,(\ref{Gl:TOV1D}, \ref{Gl:TOV2D}) and
perform three MHD simulations with an initially weak ($10^{10} \ 
\mathrm{G}$), strong ($10^{12} \ \mathrm{G}$), and very strong magnetic
field ($10^{13} \ \mathrm{G}$), respectively (see
Figs.\,\ref{Fig:A1B3G3-rho-GW} to \ref{Fig:A2B4G1-rho-GW}, and the
tables in Appendix\,\ref{Sek:Syn}).

Before we compare the results of the Newtonian versions of these MHD
models (see Paper\,I) with those obtained with the effective TOV
potential, we outline some relevant findings of Paper\,I.  In Newtonian
gravity, type\,I and type\,III models show a secular contraction when a
strong initial magnetic field ($\ga 10^{12}\ \mathrm{G}$) is imposed,
while the structure of the core remains unchanged from the hydrodynamic
case. Type\,II models lose their centrifugal support and begin to
depart from the rotational equilibrium established in the corresponding
non-magnetic model. Eventually, they transform into a pressure supported
configuration.  All initially strongly magnetized models develop
collimated bipolar outflows.

Overall, the MHD TOV models show the same qualitative dynamic behavior
and the same GW signal types as the corresponding Newtonian ones, but
they also exhibit several quantitative differences.  The efficiency of
the field amplification by winding due to differential rotation differs
from the Newtonian case as the collapse proceeds to higher densities in
the TOV models.  This effect slightly shifts the borders between the
various signal types in parameter space.  As a consequence the new
type\,IV GW signal, which was observed with the Newtonian potential for
one single MHD model only (Paper\,I), is encountered more
frequently when using the effective TOV potential.

\begin{figure*}[htbp]
  \centering
  \includegraphics[width=5.6cm]{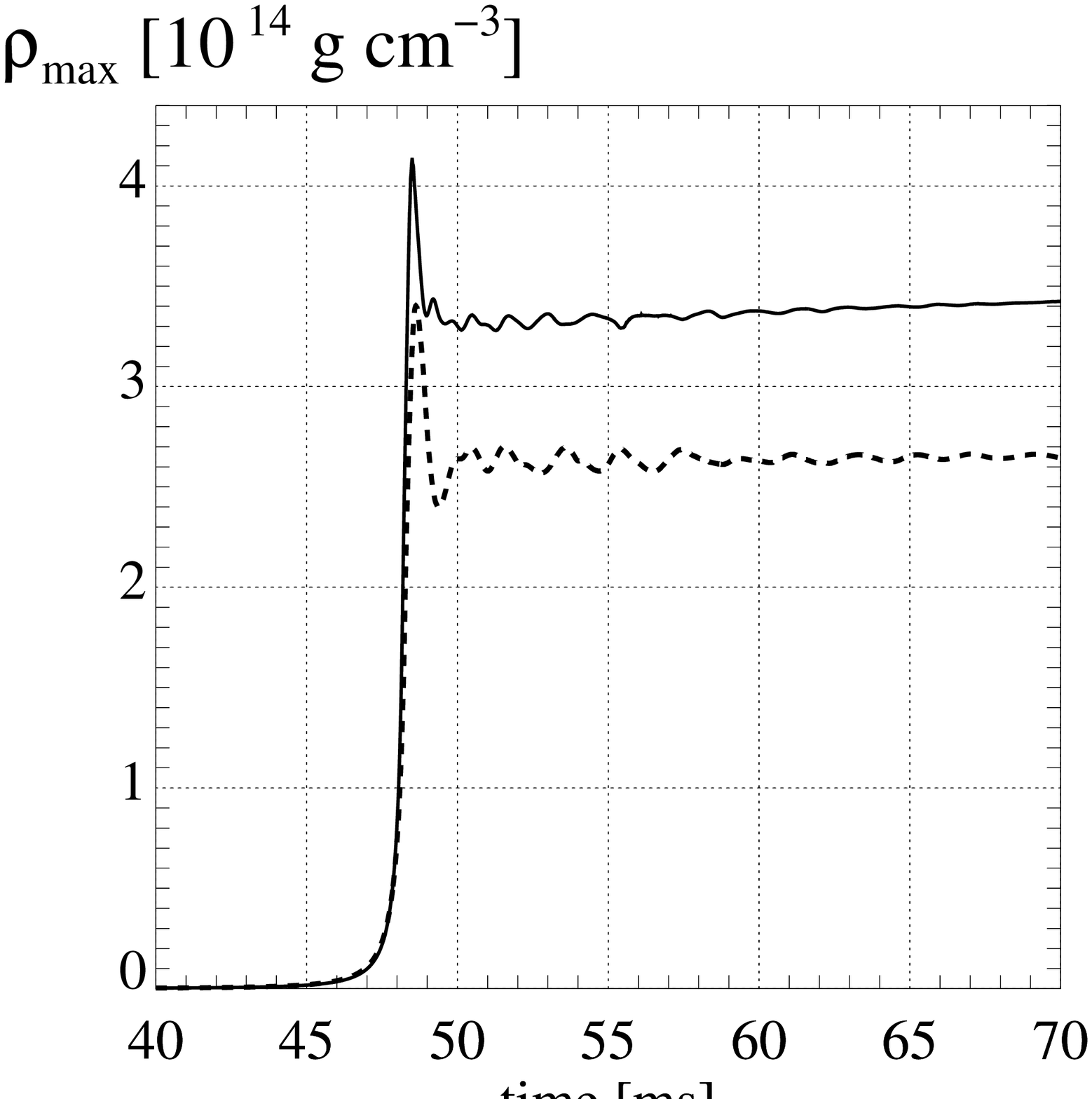}
  \includegraphics[width=5.6cm]{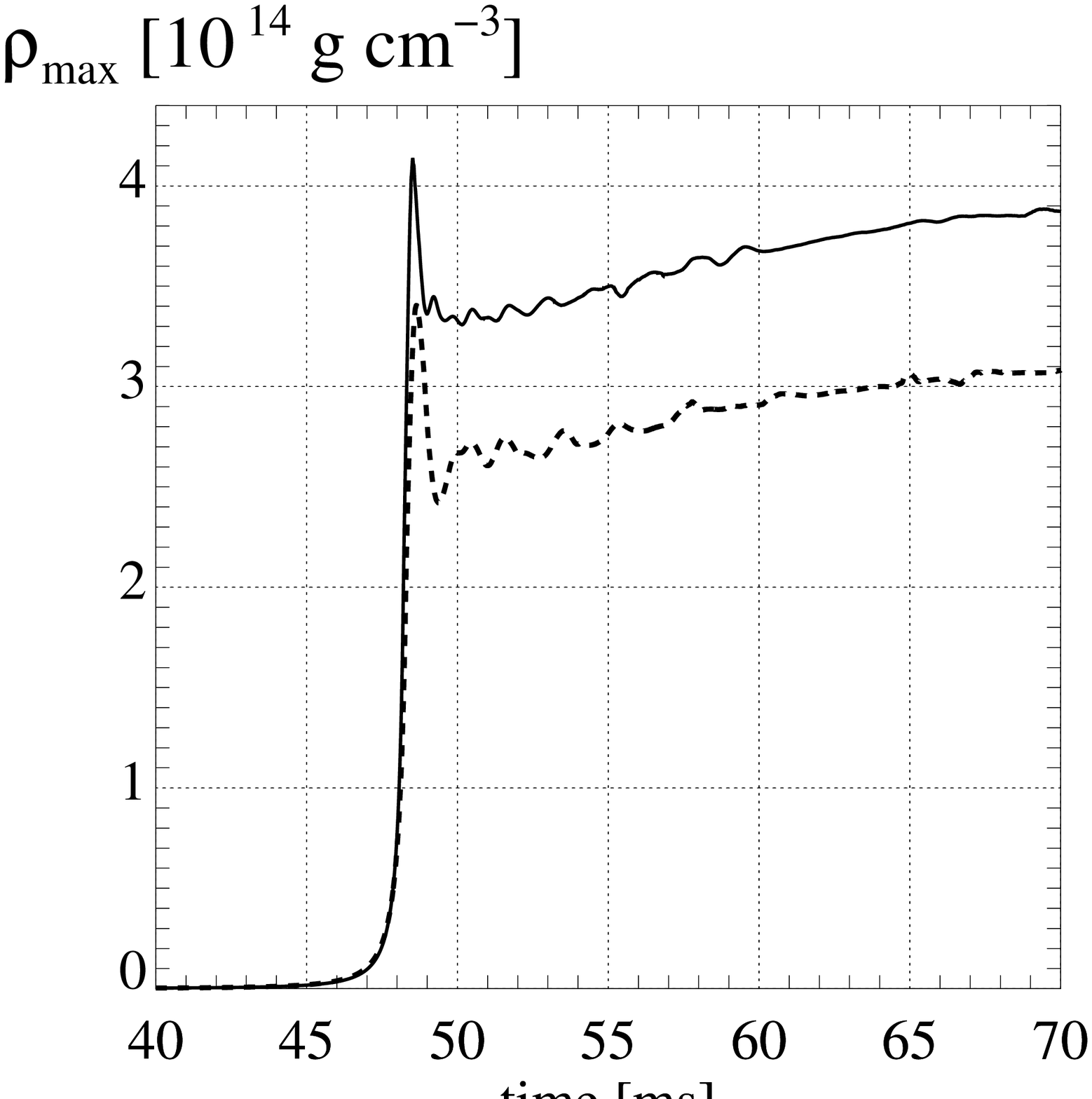}
  \includegraphics[width=5.6cm]{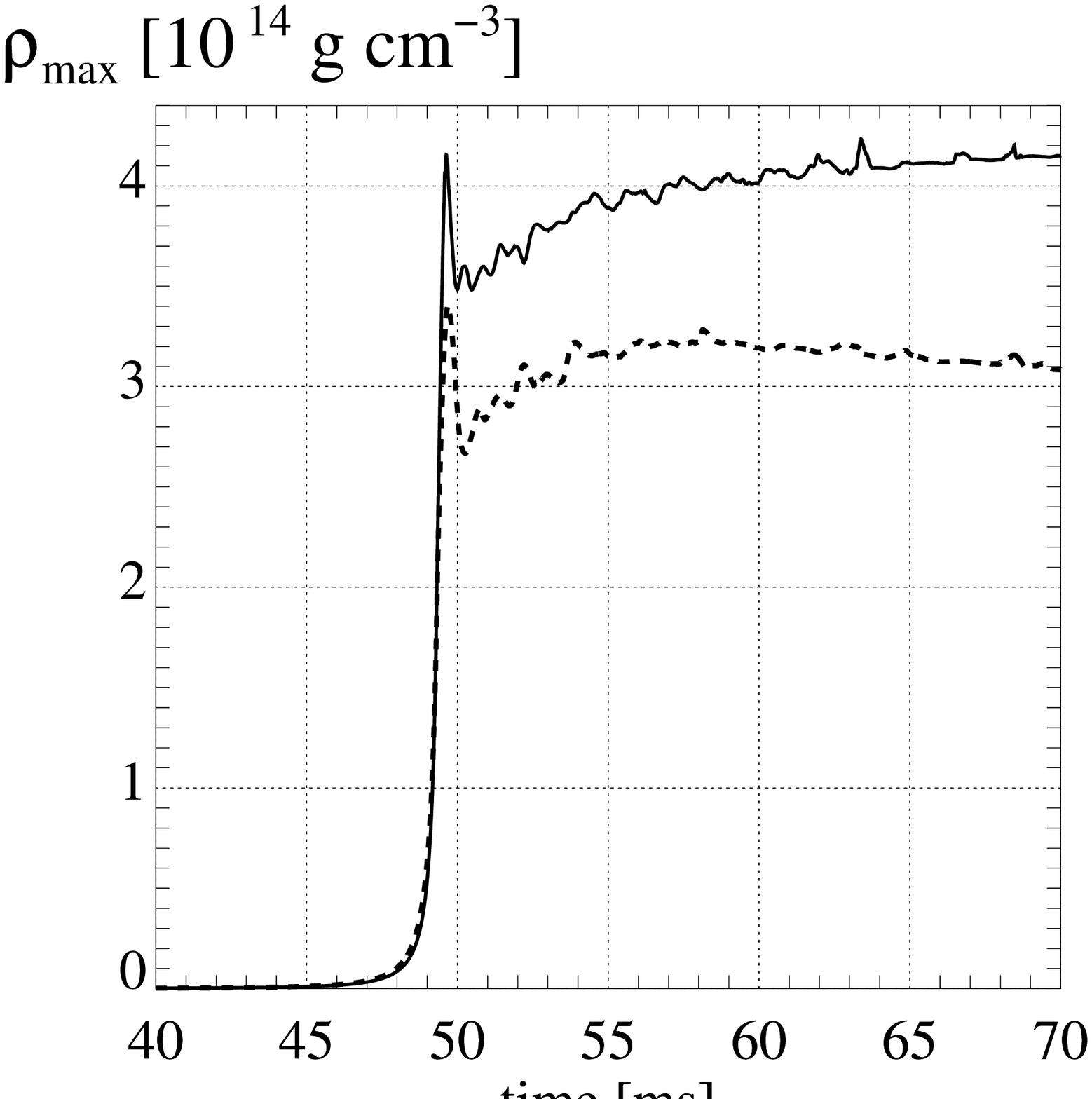}\\
  \includegraphics[width=5.6cm]{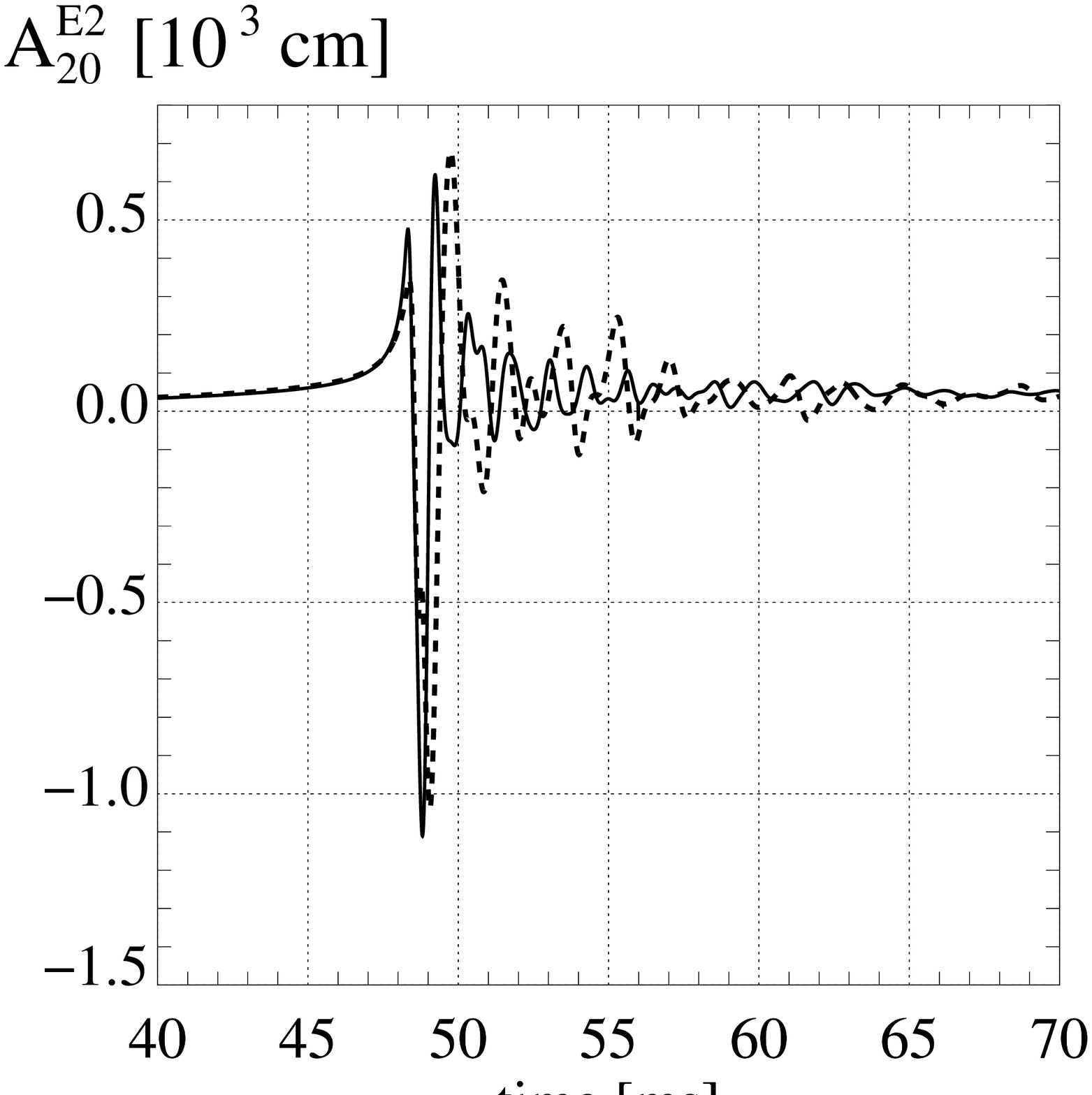}
  \includegraphics[width=5.6cm]{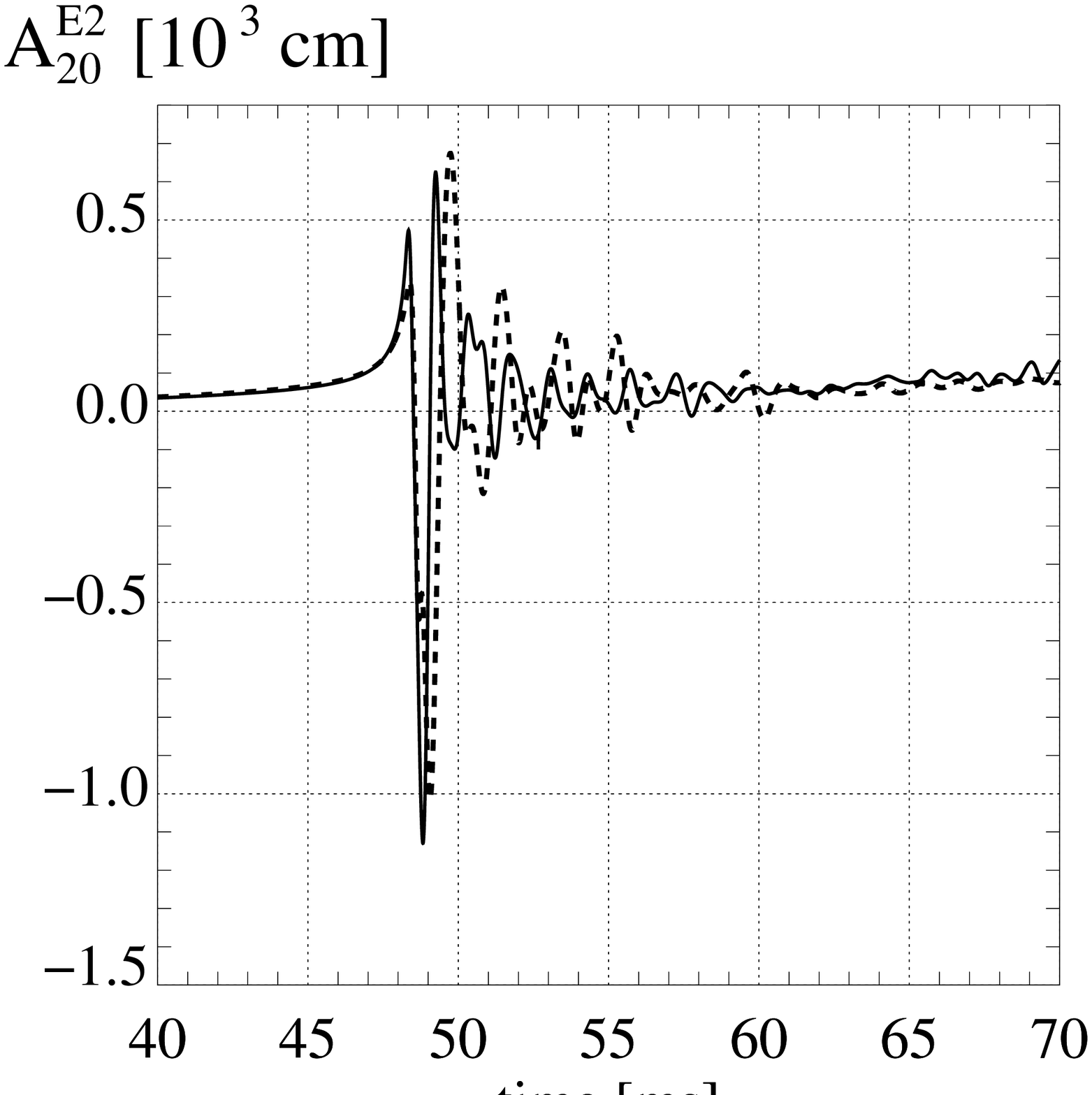}
  \includegraphics[width=5.6cm]{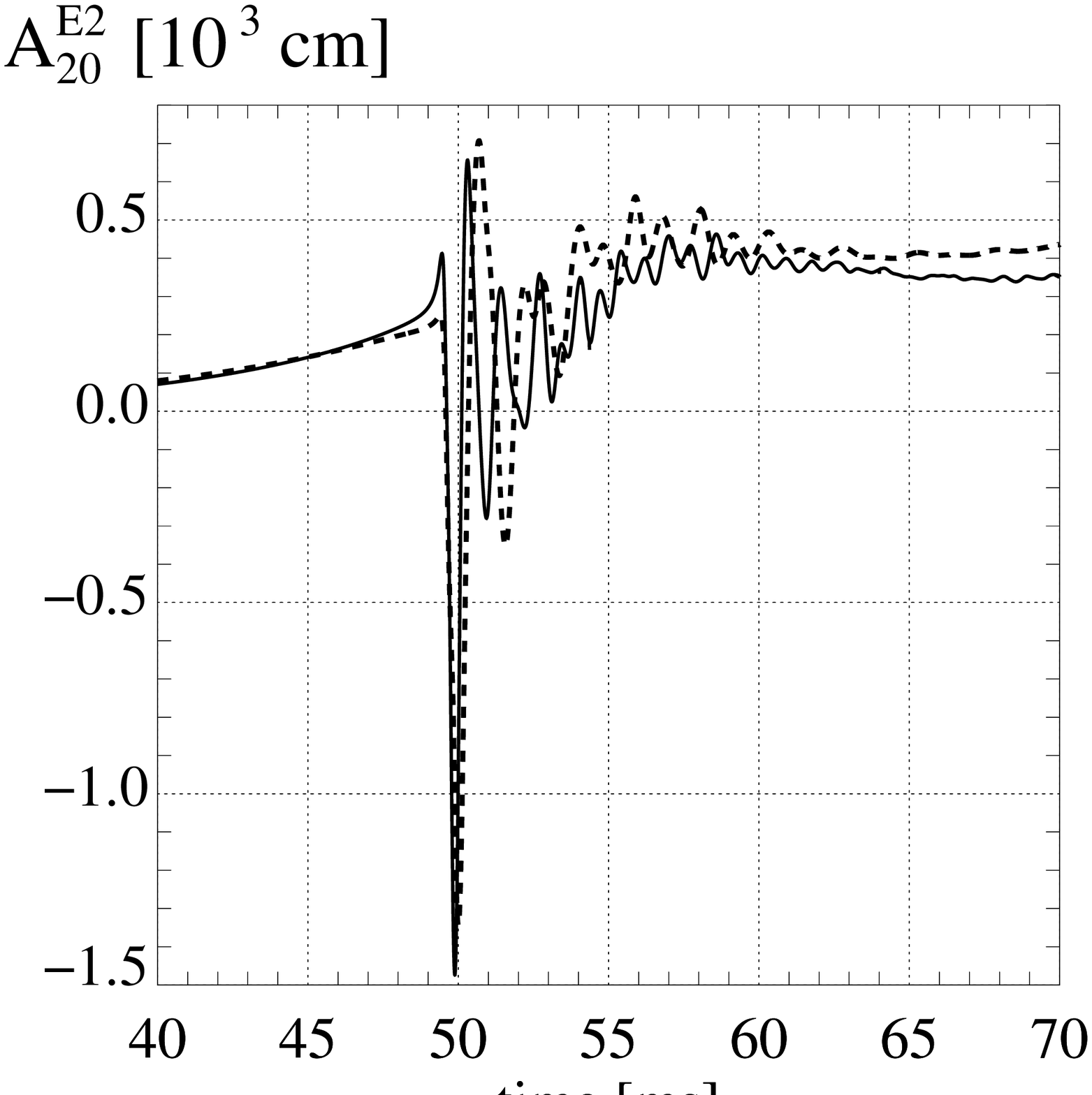}
  \caption{The evolution of the maximum density (upper panels) and GW
           amplitudes (lower panels) of models A1B3G3-D3M10-N/T
           (left), A1B3G3-D3M12-N/T (middle), and A1B3G3-D3M13-N/T
           (right).  Solid and dashed lines show TOV and Newtonian
           models, respectively.  }
  \label{Fig:A1B3G3-rho-GW}
\end{figure*}

Models of series A1B3G3-D3Mm-N/T show a qualitatively very similar
behavior, but several small quantitative differences are observed.  For
the weak-field model A1B3G3-D3M10-T, both the maximum density at core
bounce and in the post-bounce equilibrium state are larger than the
corresponding Newtonian values by $\sim 25\,\%$
(Fig.\,\ref{Fig:A1B3G3-rho-GW}, top left).  The magnetic field is
amplified by the differential rotation of the core in the same way as in
the Newtonian case, and the TOV model also satisfies the MRI condition
\citep{BalHaw91, Balbus95, Aki1} in large regions of the post-bounce
core. The MRI growth times and saturation fields are of a similar order
as in the Newtonian case, i.e.\,a few milliseconds and $\sim 10^{16}\ 
\mathrm{G}$, respectively (see also Paper\,I), but due to the faster
rotation of the collapsed inner core, the growth times are slightly
smaller and the saturation fields are slightly stronger than in the
Newtonian case.  The topologies and the energies of the magnetic field
of the cores are quite similar in the post-bounce quasi-equilibrium
state, the latter differing by only about 30\,\% at $t = 89\,$ms
($E_{\mathrm{mag}}^{\mathrm{T}} = 7.0\times 10^{51}\,$erg compared to
$E_{\mathrm{mag}}^{\mathrm{N}} = 9.8\times10^{51}\,$erg).  The magnetic
field is predominantly toroidal, but also exhibits an additional complex
structure consisting of cylindrical sheets and regions of field lines
wound up like balls of wool.  The GW amplitudes at bounce agree very
well, the differences being smaller than 10\,\%, while the immediate
post-bounce ring-down amplitudes are about 50\,\% smaller in the TOV
model (Fig.\,\ref{Fig:A1B3G3-rho-GW}, lower left).  Concerning these
results we point out that the evolution of the model in full GR is only
approximated by the use of an effective TOV potential, and hence some
additional, but probably small, modifications of the results are expected
when repeating the simulations in full GR.

In the strong-field models A1B3G3-D3M12-T
(Fig.\,\ref{Fig:A1B3G3-rho-GW}, upper middle) and A1B3G3-D3M13-T
(Fig.\,\ref{Fig:A1B3G3-rho-GW}, upper right), considerable amounts of
rotational energy are extracted from the central core by the transport
of angular momentum caused by magnetic field stresses.  Consequently,
the core loses centrifugal support and begins to contract.  This effect
is qualitatively the same in Newtonian and TOV gravity, and with respect
to the time scales, the amount of rotational energy lost, and the
increase in the central density also are quantitatively very similar.
The GW amplitude (Fig.\,\ref{Fig:A1B3G3-rho-GW}, lower right) at bounce
is enhanced by about 40\,\% compared to the non-magnetic or weak-field
case (Fig.\,\ref{Fig:A1B3G3-rho-GW}, lower left).  The post-bounce GW
signal shows the typical type\,I ring-down behavior superimposed on an
initially rising ($t_{\mathrm{b}} \la t \la t_{\mathrm{b}}+ 5\,$ms) and
then roughly constant positive mean GW amplitude.  The later
contribution to the GW signal is due to the emergence of a high speed
($v \sim {c}/{3}$) collimated outflow (jet) along the rotation axis.

As in the Newtonian case, the magnetic field of the initially most
strongly magnetized model A1B3G3-D3M13-T already affects the angular
momentum distribution of the core considerably during core collapse.  At
bounce the rotational energy of this core
($E_{\mathrm{rot}}^{\mathrm{T}} = 4.2\times 10^{51}\,$erg) is lower by
18\,\% compared to that of models A1B3G3-D3M12-T and A1B3G3-D3M10-T
($E_{\mathrm{rot}}^{\mathrm{TOV}} = 5.1\times 10^{51}\,$erg). In model
A1B3G3-D3M12-T, the magnetic field is too weak at bounce to be
dynamically important. However, a few milliseconds after bounce the
poloidal field energy starts to grow exponentially when meridional
circulation flow develops near the surface of the inner core. The flow
winds up the radial magnetic field component, and magnetic stresses
begin to transport angular momentum outwards.  Furthermore, a weak
bipolar outflow develops in this model towards the end of the
simulation, which is driven primarily by magnetic hoop stresses.
Because of this outflow, the GW amplitude rises slowly towards a
positive mean value, which is however smaller than in the case of model
A1B3G3-D3M13-T (Fig.\,\ref{Fig:A1B3G3-rho-GW}, lower middle and right).

\begin{figure*}[htbp]
  \centering
  \includegraphics[width=5.6cm]{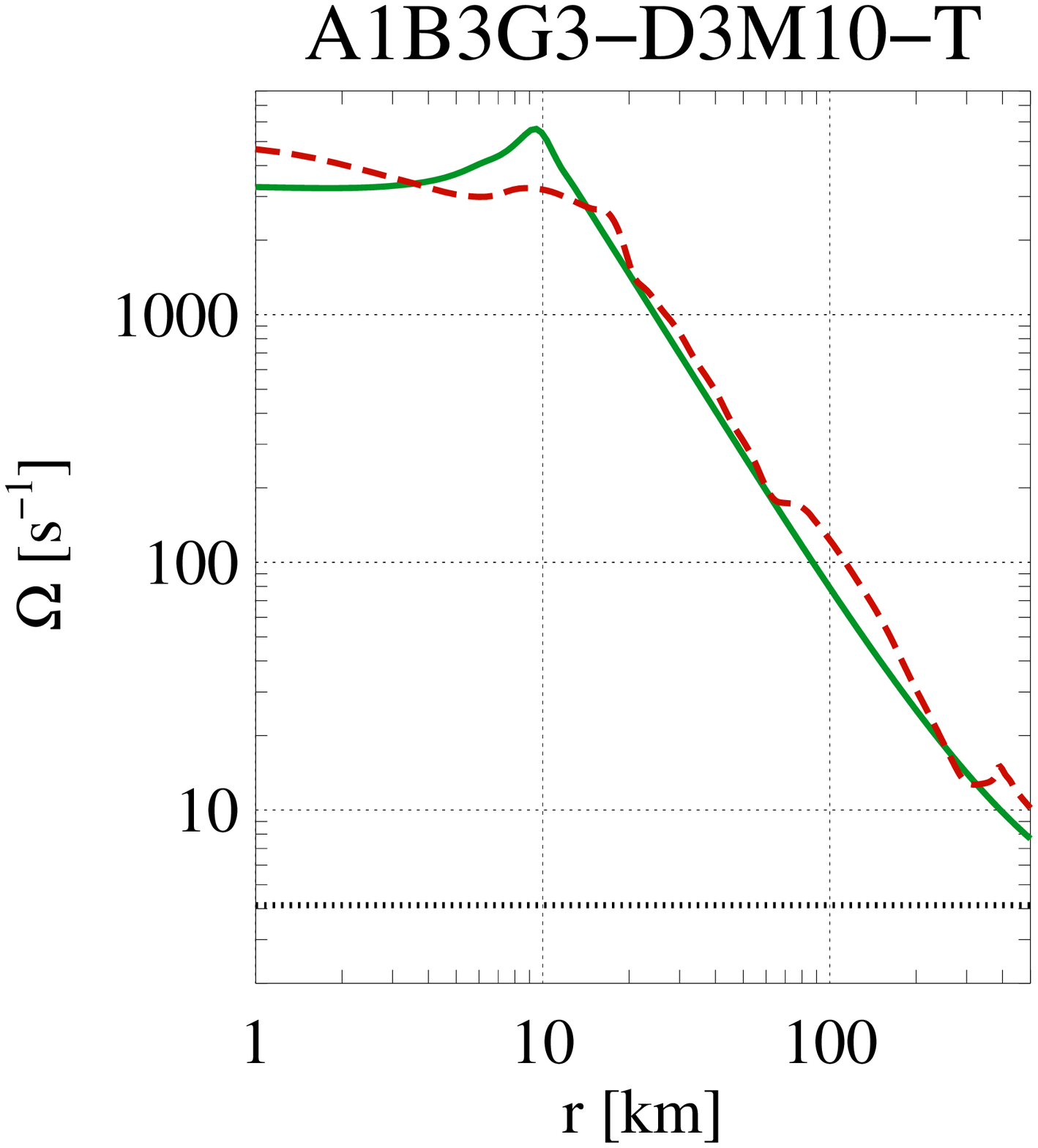}
  \includegraphics[width=5.6cm]{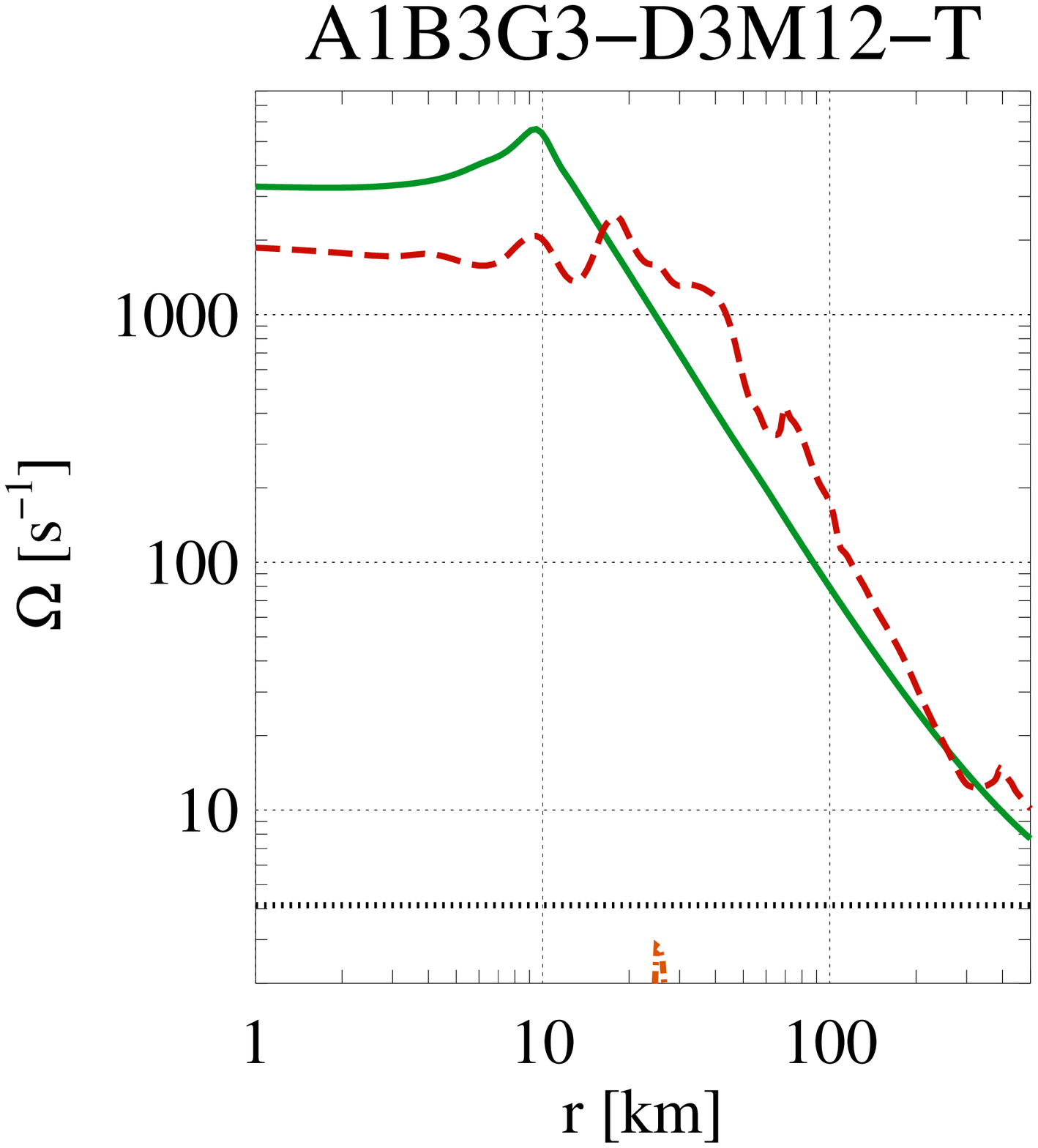}
  \includegraphics[width=5.6cm]{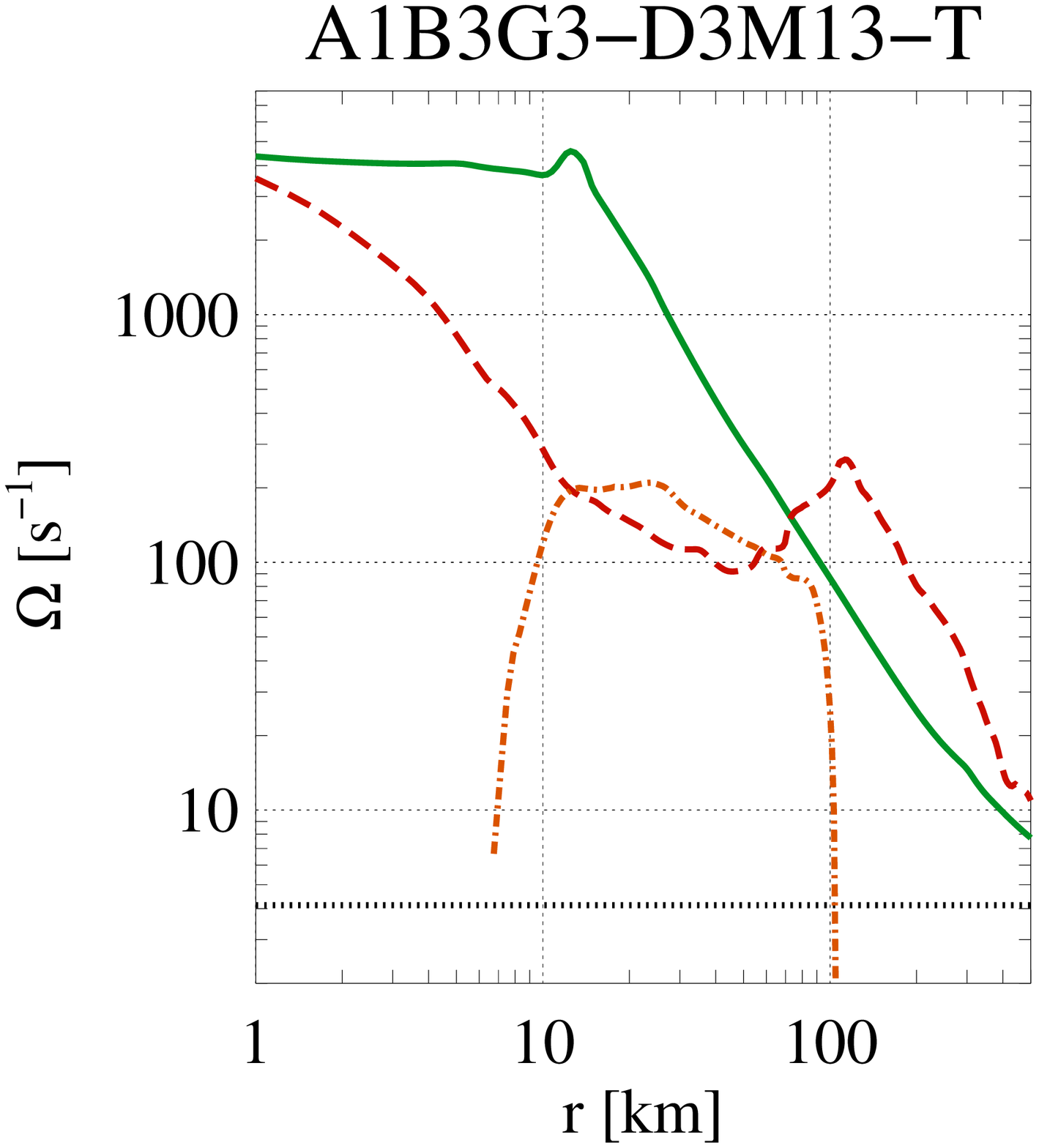}
  \caption{Temporal evolution of the rotation profiles for the models
           of series A1B3G3-D3Mm-T. The panels show the angular
           averaged profiles $\Omega(r)$ of models A1B3G3-D3M10-T
           (left), A1B3G3-D3M12-T (middle), and A1B3G3-D3M13-T
           (right), respectively.  The initial rotation profiles and
           the profiles at core bounce are given by the black dotted
           and green solid lines, respectively. The additional lines
           show the profiles at $t\approx 65\,$ms, the red dashed and
           the orange dash-dotted lines corresponding to the angular
           averaged rotation profiles of the prograde and of the
           retrograde rotating parts of the core, respectively.  While
           there is no region of retrograde rotation present in model
           A1B3G3-D3M10-T (left), and only a very small one in model
           A1B3G3-D3M12-T (middle), large amounts of matter rotate in
           a retrograde way near the equator in model A1B3G3-D3M13-T
           (right). }
  \label{Fig:angvelo}
\end{figure*}

\begin{figure*}[htbp]
  \centering
  \includegraphics[width=8cm]{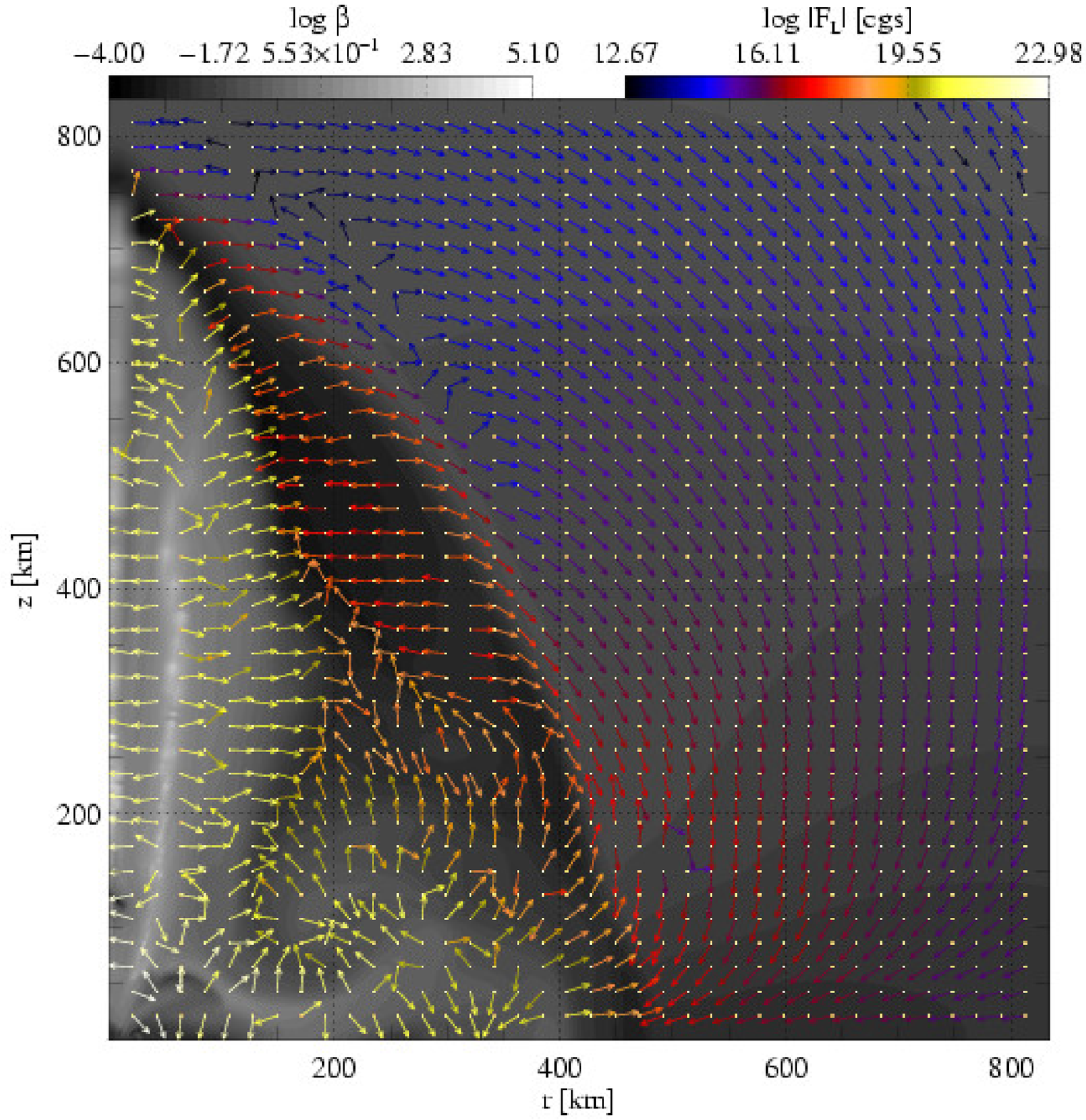}
  \includegraphics[width=8cm]{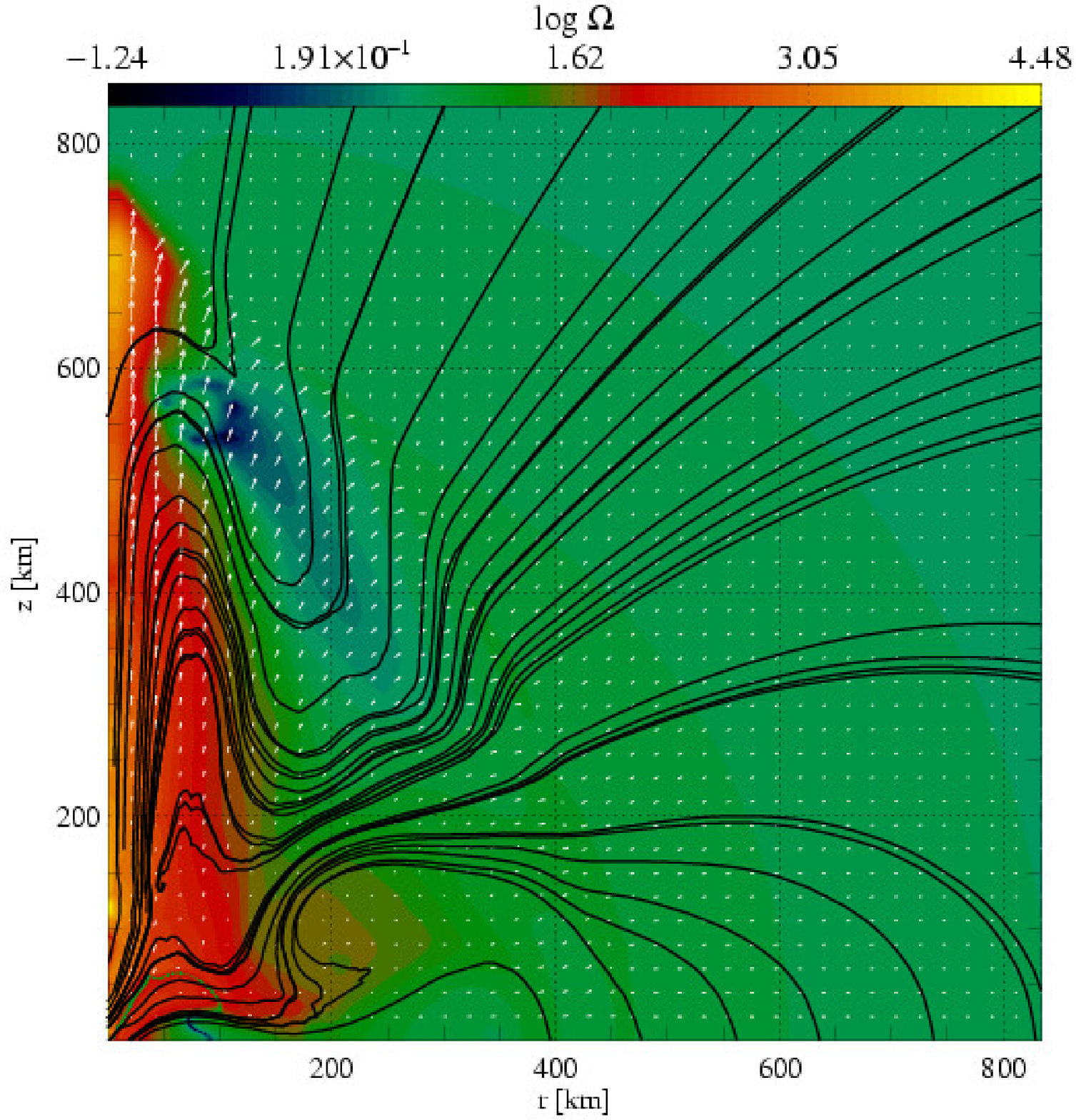}
  \caption{
      The left panel shows the distribution of $\beta=P_{\mathrm{mag}} /
      P_{\mathrm{gas}}$ in the outflow of model A1B3G3-D3M13-T at
      $t\approx65\,$ms (grey scale) and the direction and modulus
      of the Lorentz force (arrows). The arrows are normalized to
      the same length, but show the strength of the force color-coded.
      The right panel displays the angular velocity distribution
      (color scale), the velocity field (arrows), and
      the magnetic field (field lines) at the same time.
    }
  \label{Fig:Lorentz}
\end{figure*}

To illustrate the effects of the magnetic field on the angular momentum
distribution we compare the evolution of the angular averaged rotation
profiles $\Omega(r)$ for three cores from our model series at different
times (Fig.\,\ref{Fig:angvelo}).  Although rotating almost rigidly
initially, all three cores develop a strongly differential rotation
profile at bounce. At this evolutionary stage, there is only very little
difference between the profiles of models A1B3G3-D3M10-T and
A1B3G3-D3M12-T, whereas angular momentum transport by magnetic fields
has already slightly altered the rotation profile of model
A1B3G3-D3M13-T. At $t\approx 65\,$ms, the weak-field core still has
essentially the same rotational profile as it had at bounce, while the
the rotation profile of the strong-field cores has changed quite
considerably.  The rotation rate of the inner core of model
A1B3G3-D3M12-T is down by a factor of $\sim 2$, and angular momentum
transport by magnetic fields has created a fast rotating region outside
of the inner core.

In the case of model A1B3G3-D3M13-T, angular momentum transport is
even more efficient, and at intermediate radii, $7\,\mathrm{km} \la r
\la 100\, \mathrm{km}$, a region of slow retrograde rotation develops
near the equator. At similar radii, the fluid along the polar axis
still rotates in a prograde direction but quite slowly, whereas the
rotation rate of matter outside $\sim 70\,$km is much faster than in
the corresponding models with weaker initial fields. Spatially, the
rapidly rotating matter is concentrated along the axis in the jet-like
outflow driven by the magnetic fields.

The differences between the rotation profiles reflect different modes of
angular momentum redistribution to some extent.  When amplified by
compression during collapse, the initially strong field in model
A1B3G3-D3M13-T manages to launch a prominent outflow, which carries
angular momentum away from the center towards the outflow axis.
In the less strongly magnetized model A1B3G3-D3M12-T, on the other
hand, most transport is due to the magnetic field growing near the
boundary of the inner core ($r \approx 20\,\mathrm{km}$) at all
latitudes by the action of MHD instabilities. A few vortices develop,
where the off-diagonal Maxwell-stress components responsible for
angular momentum transport become large. Consequently, a considerable
loss of rotational energy from the inner core occurs due to these
vortices that extend further outwards with time. Only later in the
evolution the appearance of a weak polar outflow opens up an
additional channel of angular momentum transport similar to the one
discussed above.

The efficient transport of angular momentum along the outflow and the
collimation of the fluid by magnetic stresses together give rise to the
very characteristic structure of the outflow (Fig.\,\ref{Fig:Lorentz}).
The outflow drags along and bends the poloidal field lines that are
initially located near the surface of the core, giving rise to the
formation of a cylindrically shaped magnetic sheet
(Fig.\,\ref{Fig:Lorentz}). This magnetic sheet separates the outflow
into two concentric regions: an interior region resembling the {\it
beam} of a jet and an exterior region resembling the jet {\it
cocoon}.
The fluid is collimated mainly by the magnetic field and
predominantly by its hoop stress. Fig.\,\ref{Fig:Lorentz} (left
panel) shows the ratio $\beta\equiv P_{\mathrm{mag}} /
P_{\mathrm{gas}}$ of the magnetic pressure and the gas pressure, as
well as the direction and magnitude of the Lorentz force exerted by
the magnetic field on the gas. The outflow is magnetically dominated,
$\beta$ being much larger than unity.
Regions of differently oriented Lorentz force can be identified, which
gives rise to the formation of two regions, contractive and expansive,
in the outflow that roughly match the division into beam (contractive)
and cocoon (expansive) as sketched above. This feature appears to be
inherent to the evolution of magnetized jets (see, e.g.,
\citet{Leismann_etal__2005__ApJ__RMHD-Jets}). As the gas is compressed
towards the rotational axis, both the field strength and the angular
velocity increase, i.e, the gas in the jet beam begins to rotate very
rapidly (see Fig.\,\ref{Fig:Lorentz}, right panel).

\begin{figure*}[htbp]
  \centering
  \includegraphics[width=5.6cm]{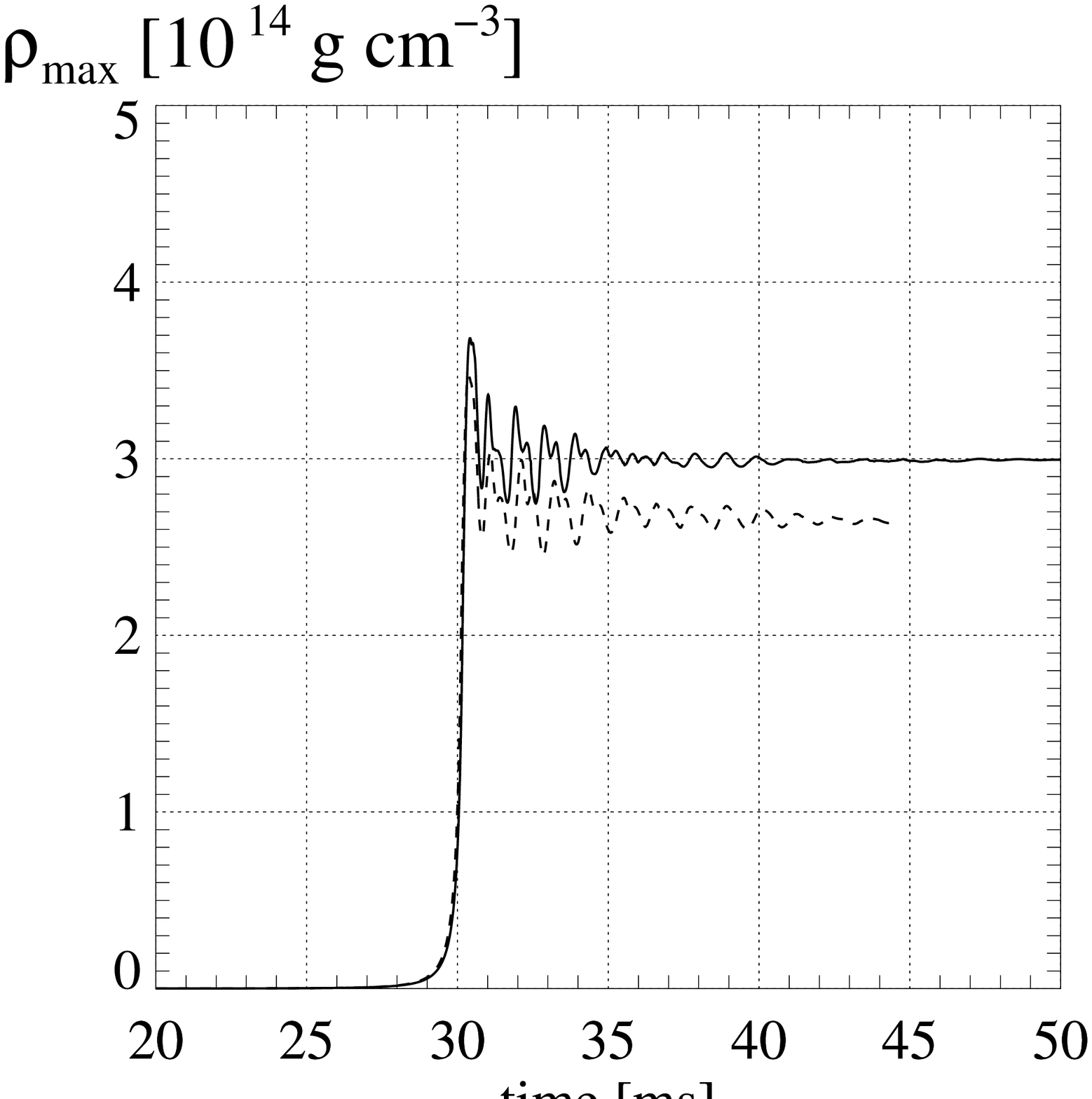}
  \includegraphics[width=5.6cm]{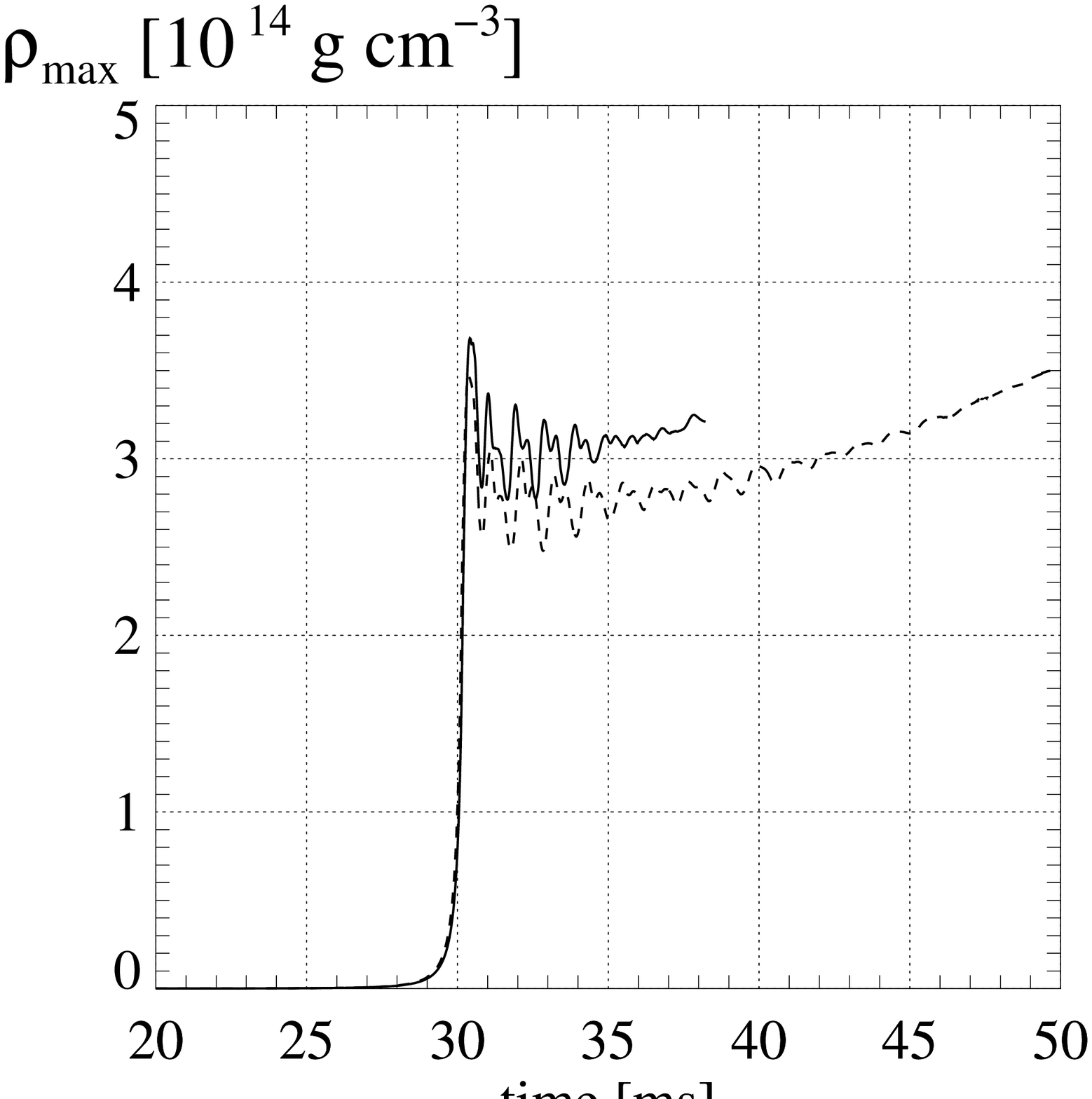}
  \includegraphics[width=5.6cm]{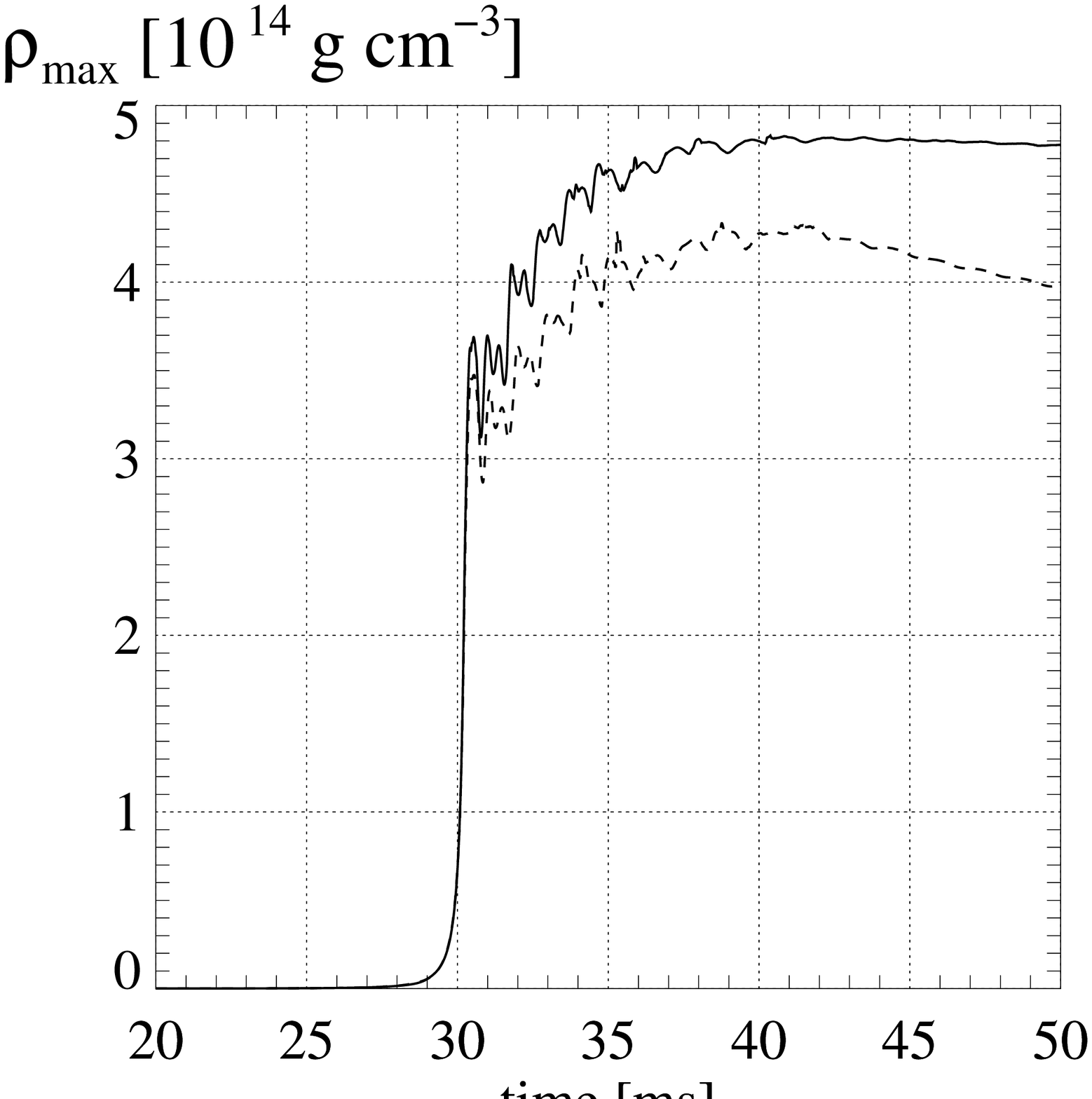}\\
  \includegraphics[width=5.6cm]{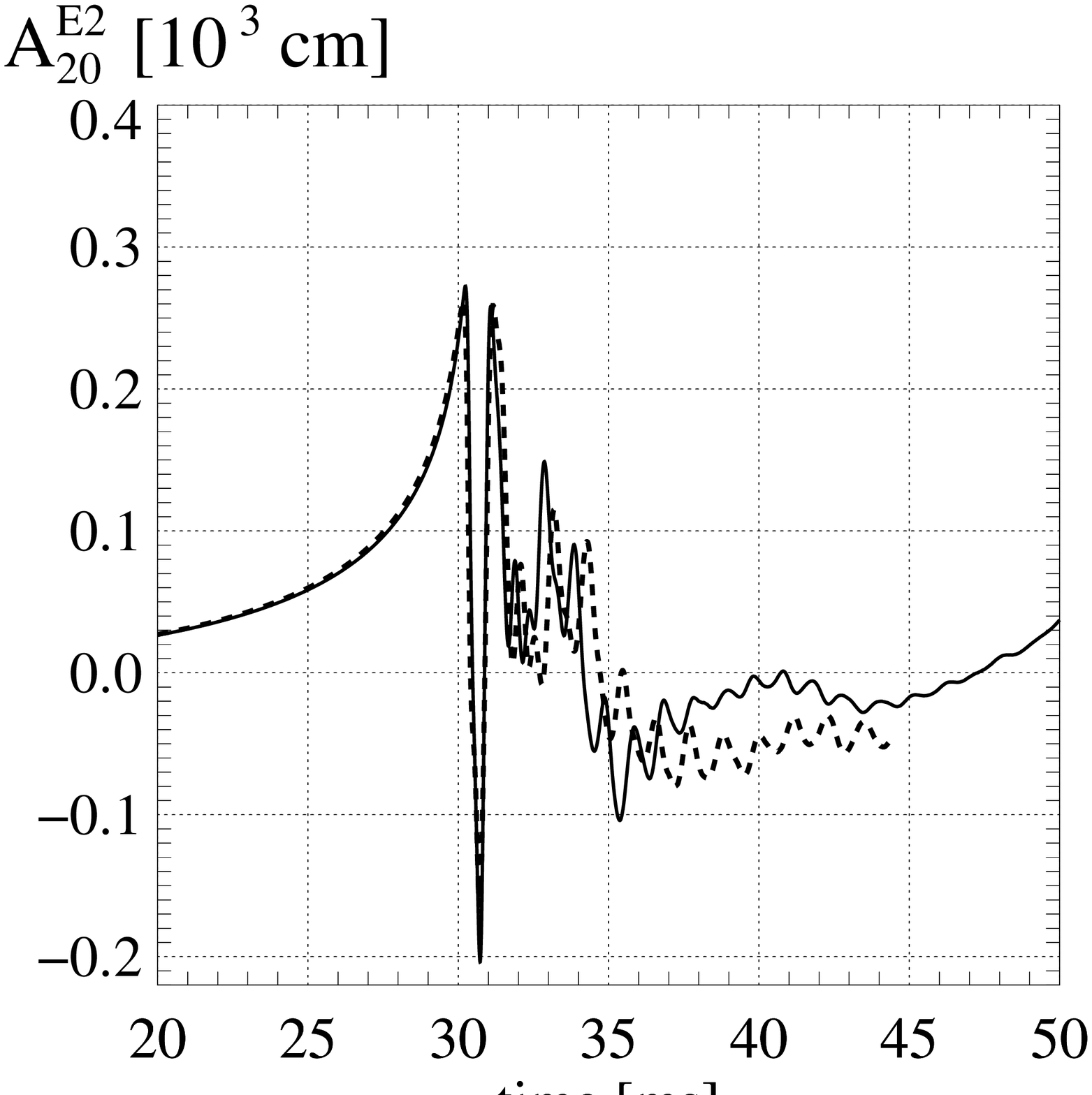}
  \includegraphics[width=5.6cm]{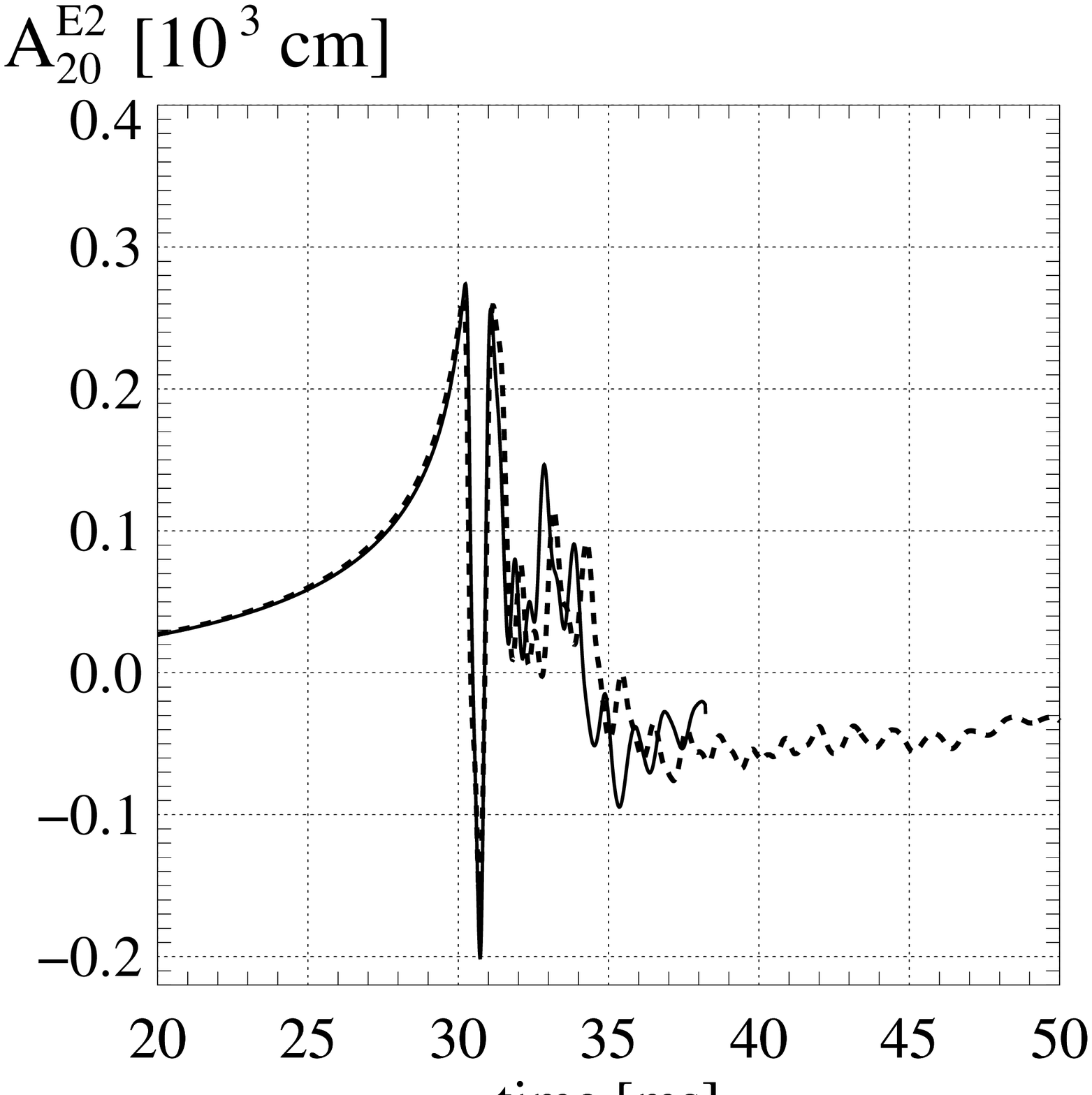}
  \includegraphics[width=5.6cm]{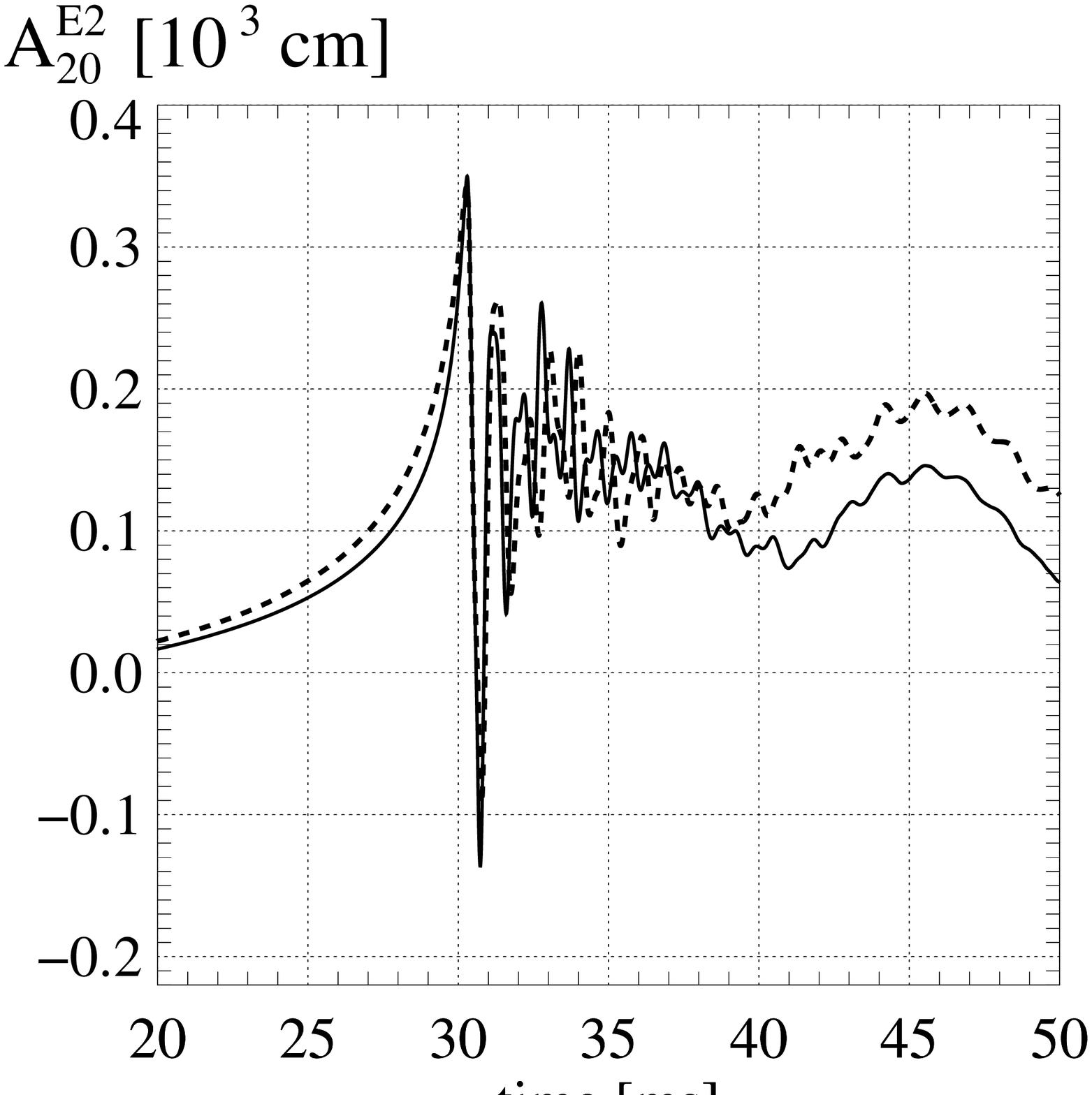}
  \caption{The evolution of the maximum density (upper panels) and GW
    amplitudes (lower panels) of models A3B3G5-D3M10-N/T (left),
    A3B3G5-D3M12-N/T (middle), and A3B3G5-D3M13-N/T (right).  Solid and
    dashed lines show TOV and Newtonian models, respectively. }
  \label{Fig:A3B3G5-rho-GW}
\end{figure*}

\begin{figure*}[htbp]
  \centering
  \includegraphics[width=5.6cm]{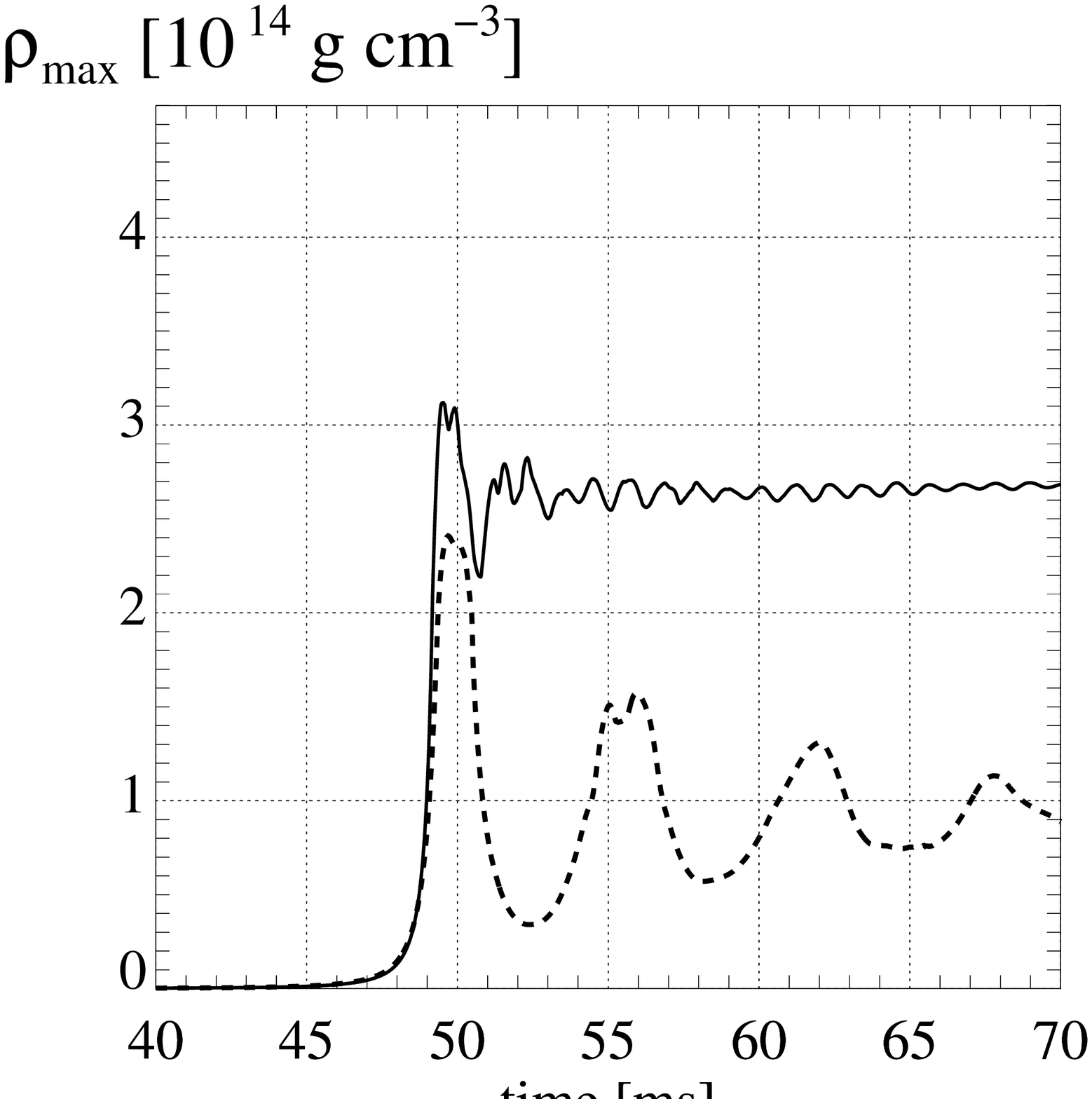}
  \includegraphics[width=5.6cm]{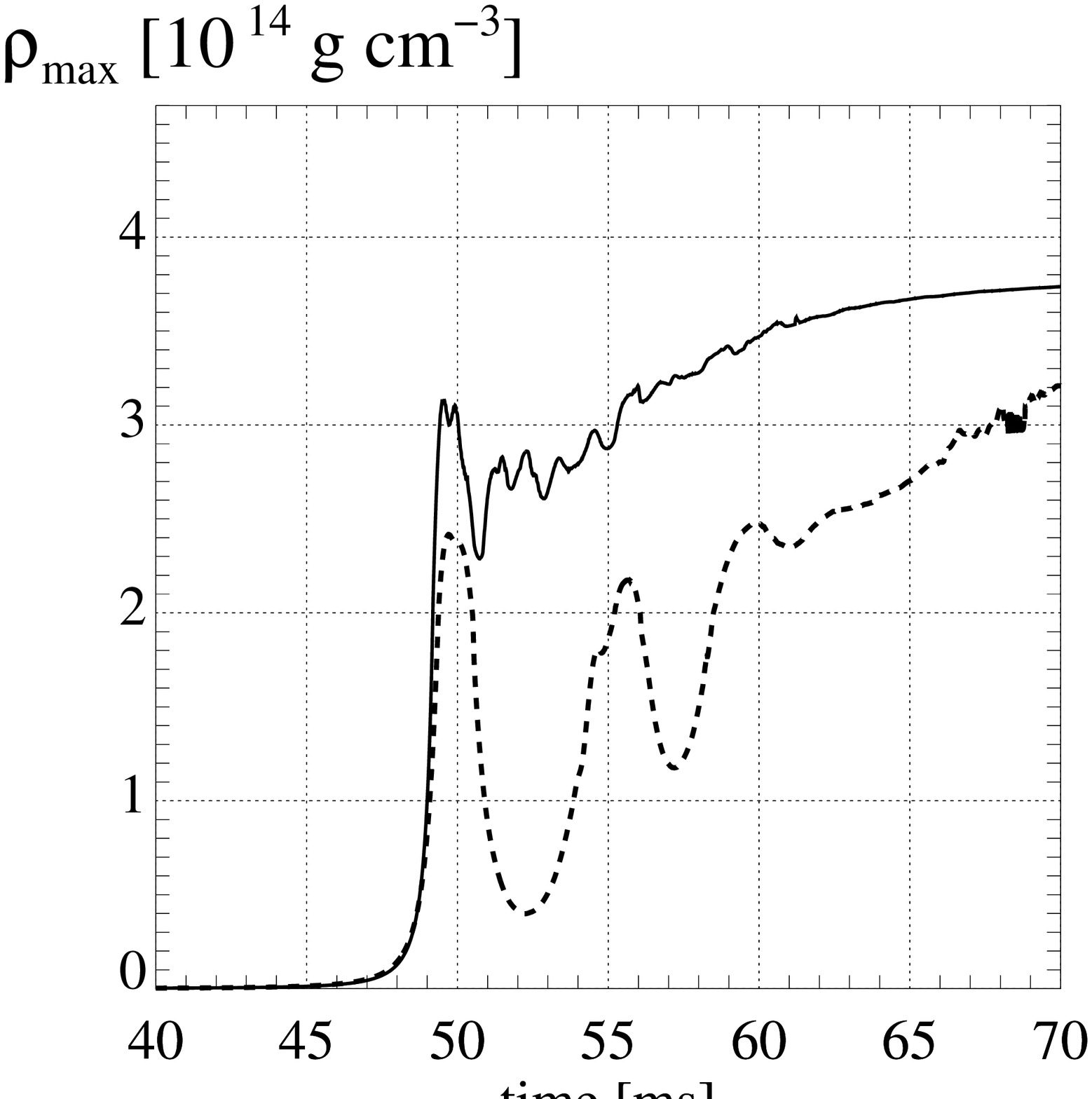}
  \includegraphics[width=5.6cm]{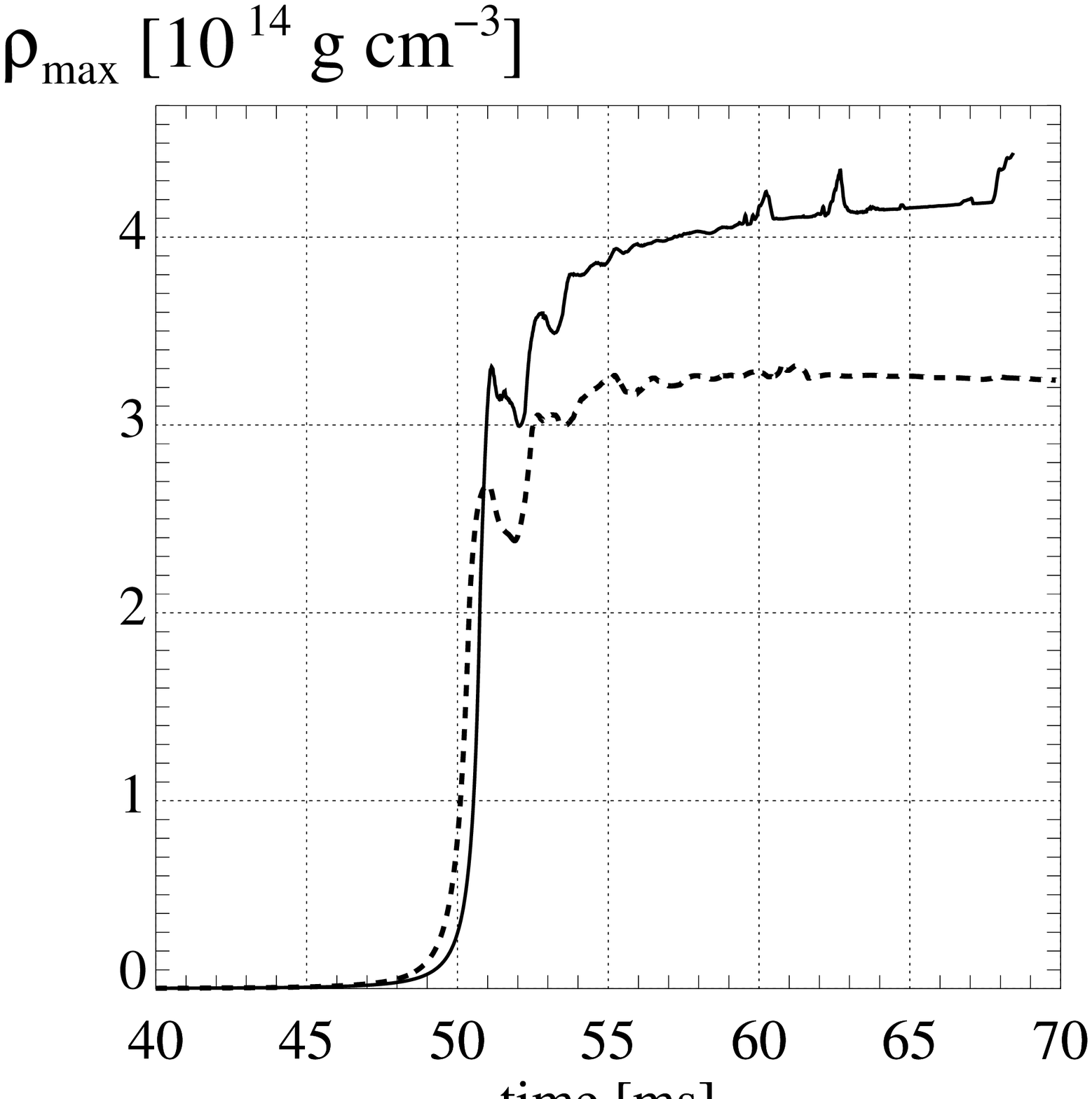}\\
  \includegraphics[width=5.6cm]{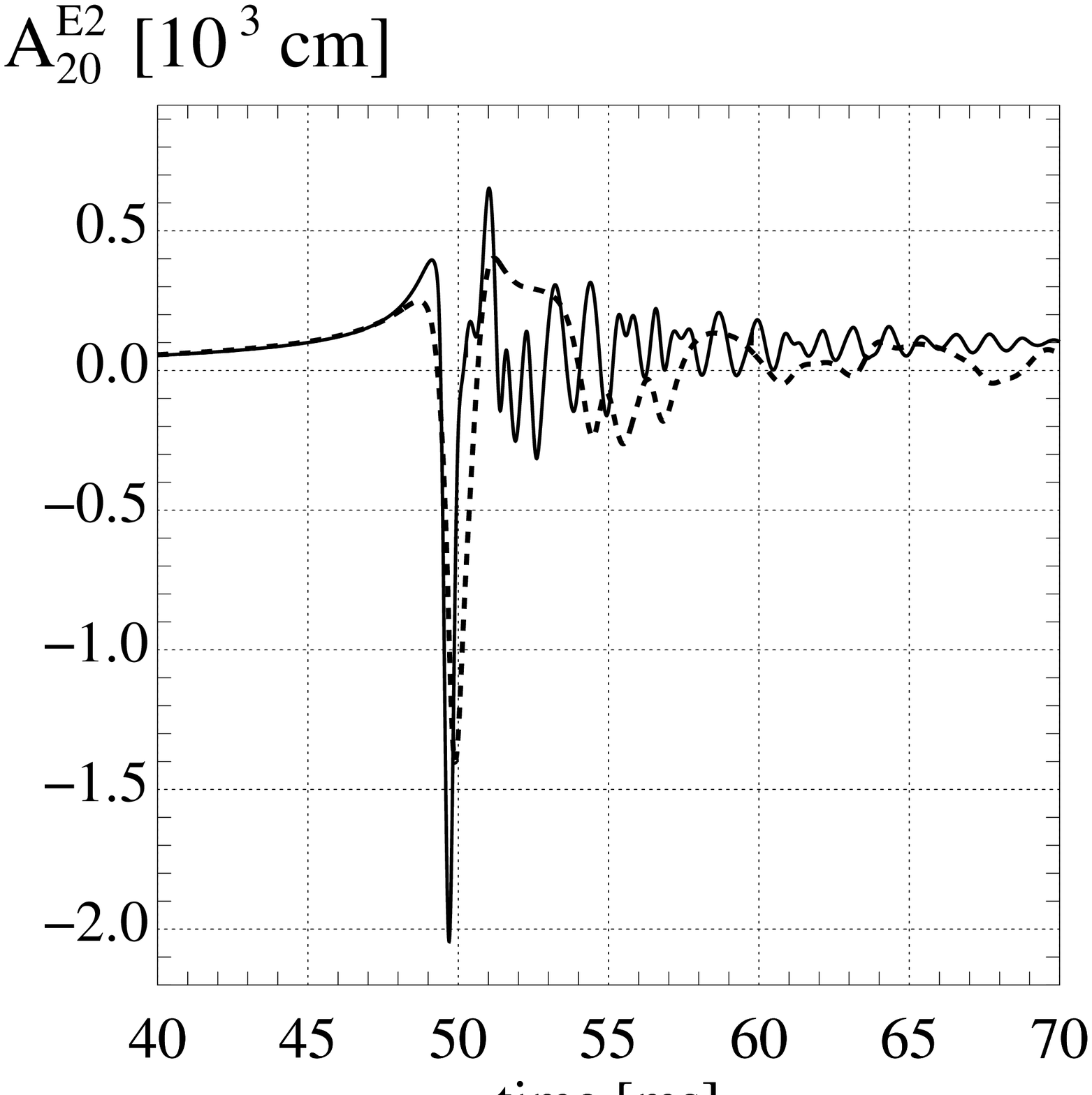}
  \includegraphics[width=5.6cm]{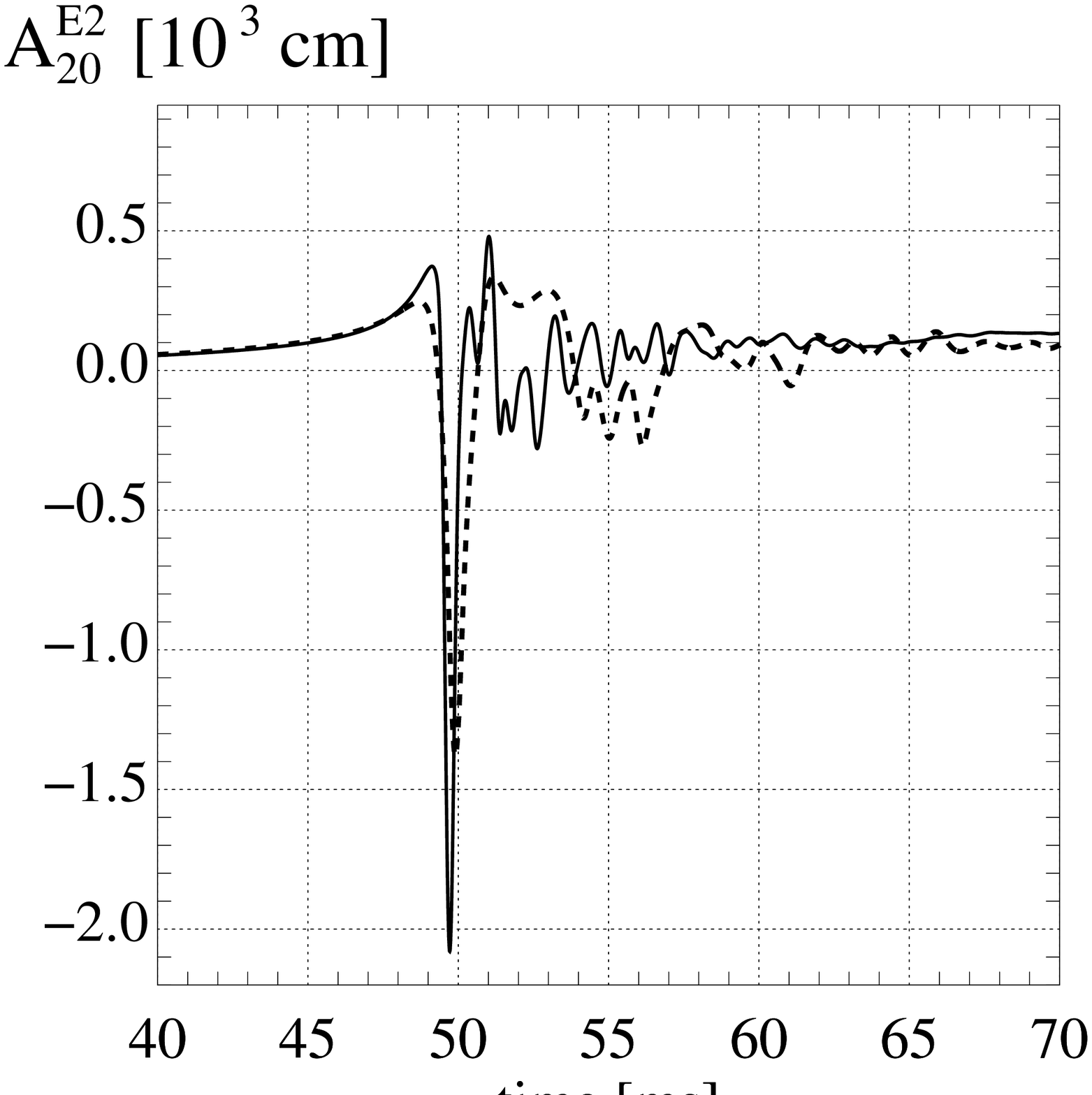}
  \includegraphics[width=5.6cm]{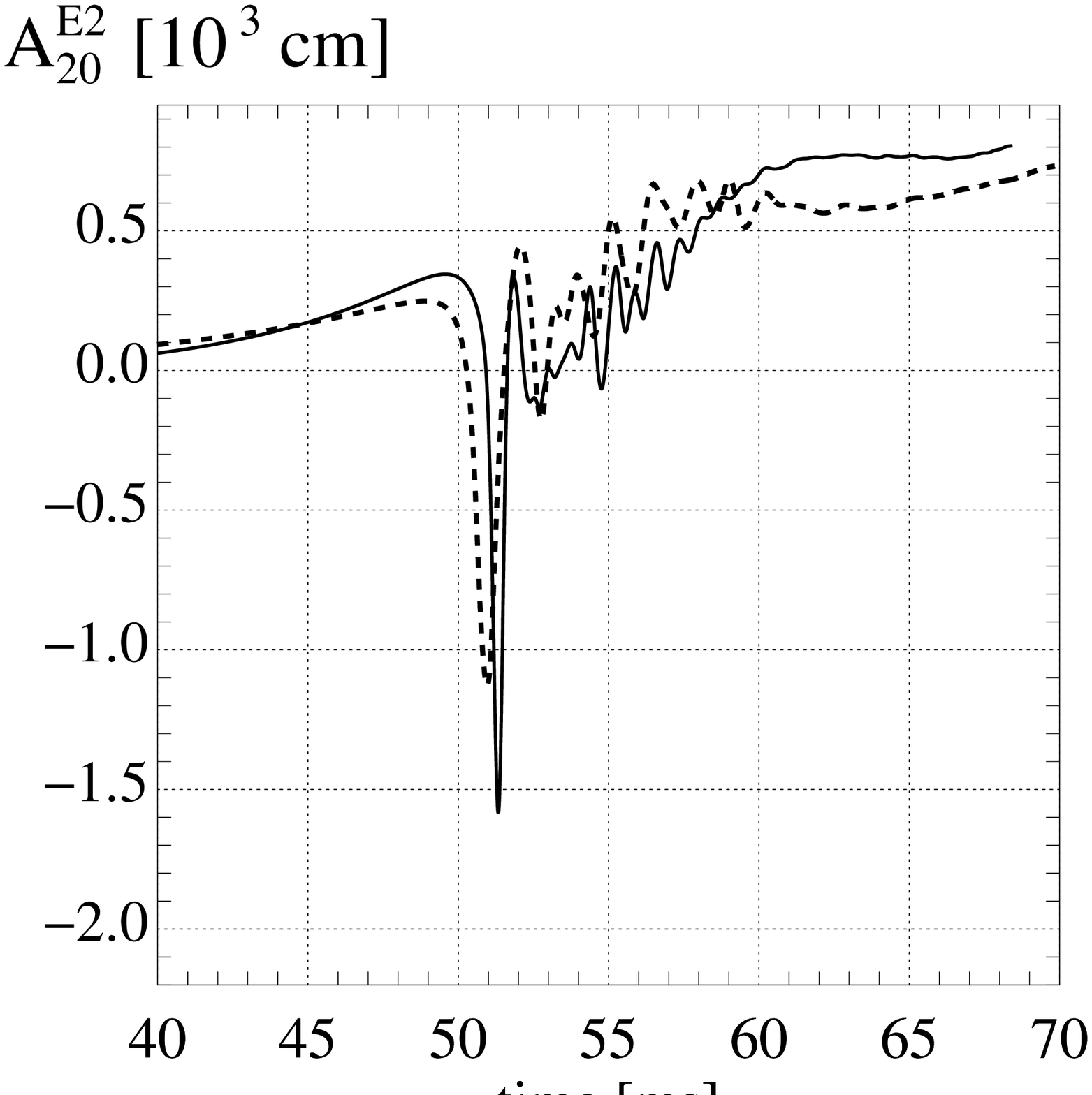}
  \caption{The evolution of the maximum density (upper panels) and GW
           amplitudes (lower panels) of models A3B3G3-D3M10-N/T
           (left), A3B3G3-D3M12-N/T (middle), and A3B3G3-D3M13-N/T
           (right).  Solid and dashed lines show TOV and Newtonian
           models, respectively.  }
  \label{Fig:A3B3G3-rho-GW}
\end{figure*}

\begin{figure*}[htbp]
  \centering
  \includegraphics[width=5.6cm]{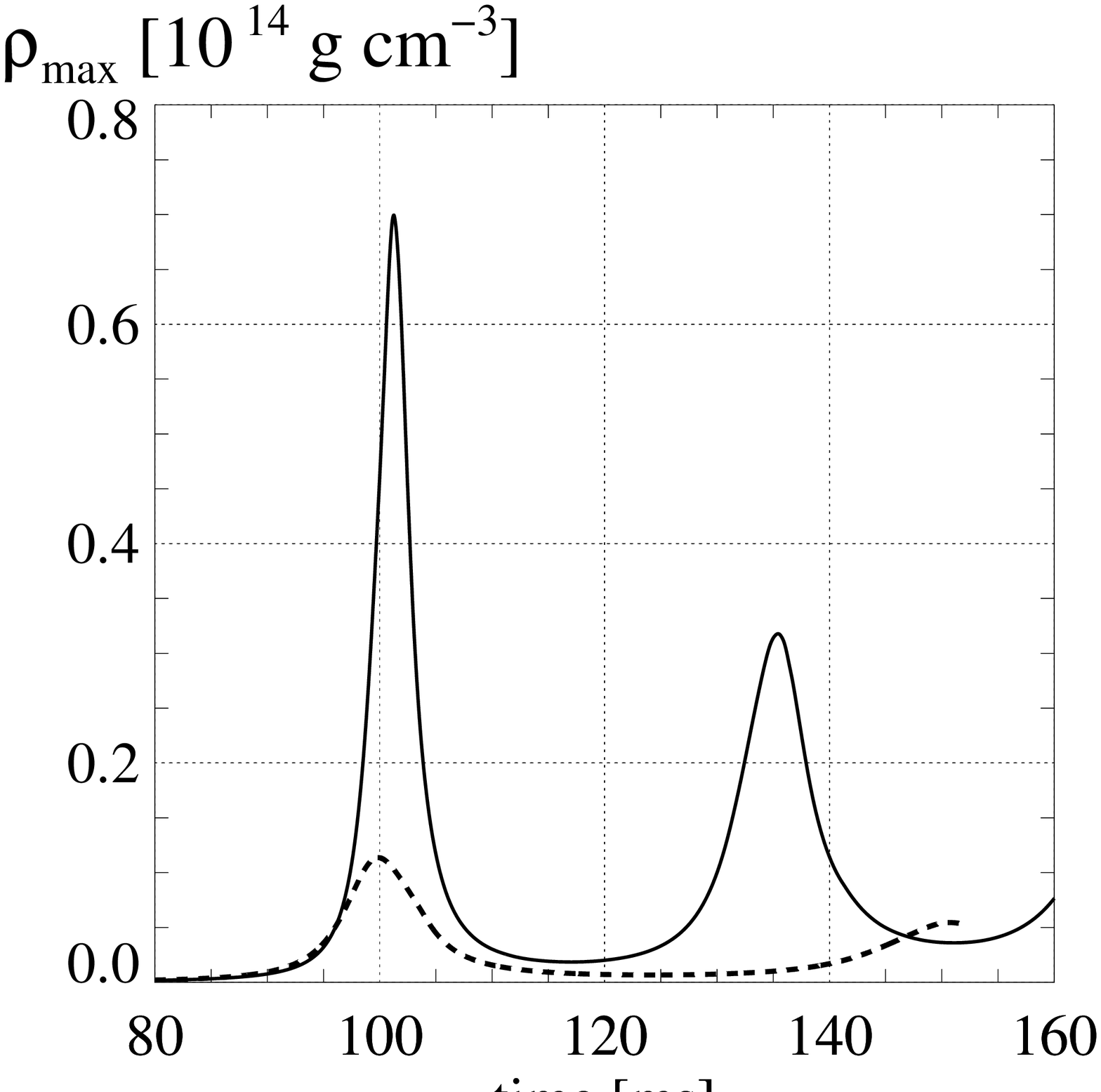}
  \includegraphics[width=5.6cm]{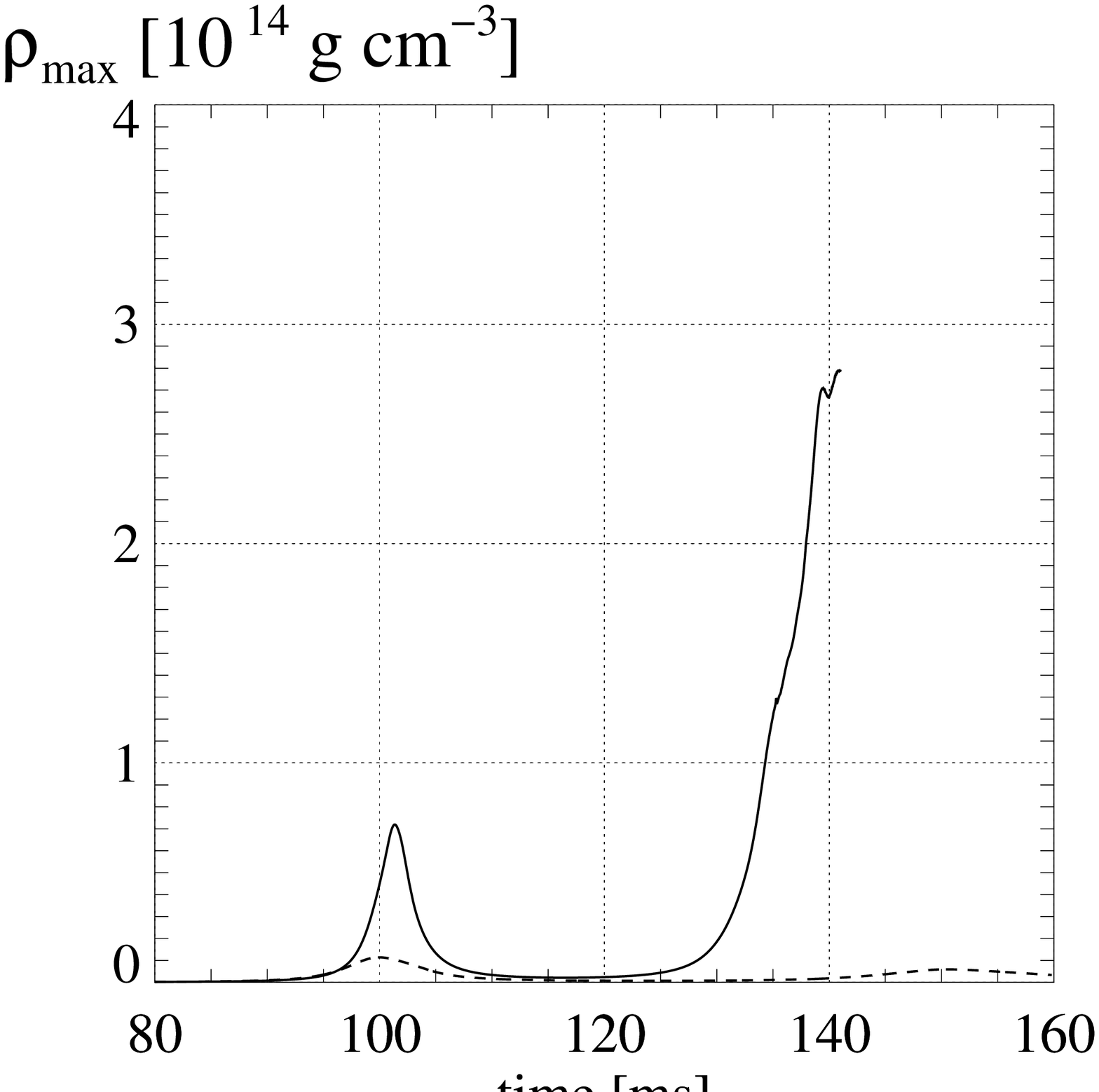}
  \includegraphics[width=5.6cm]{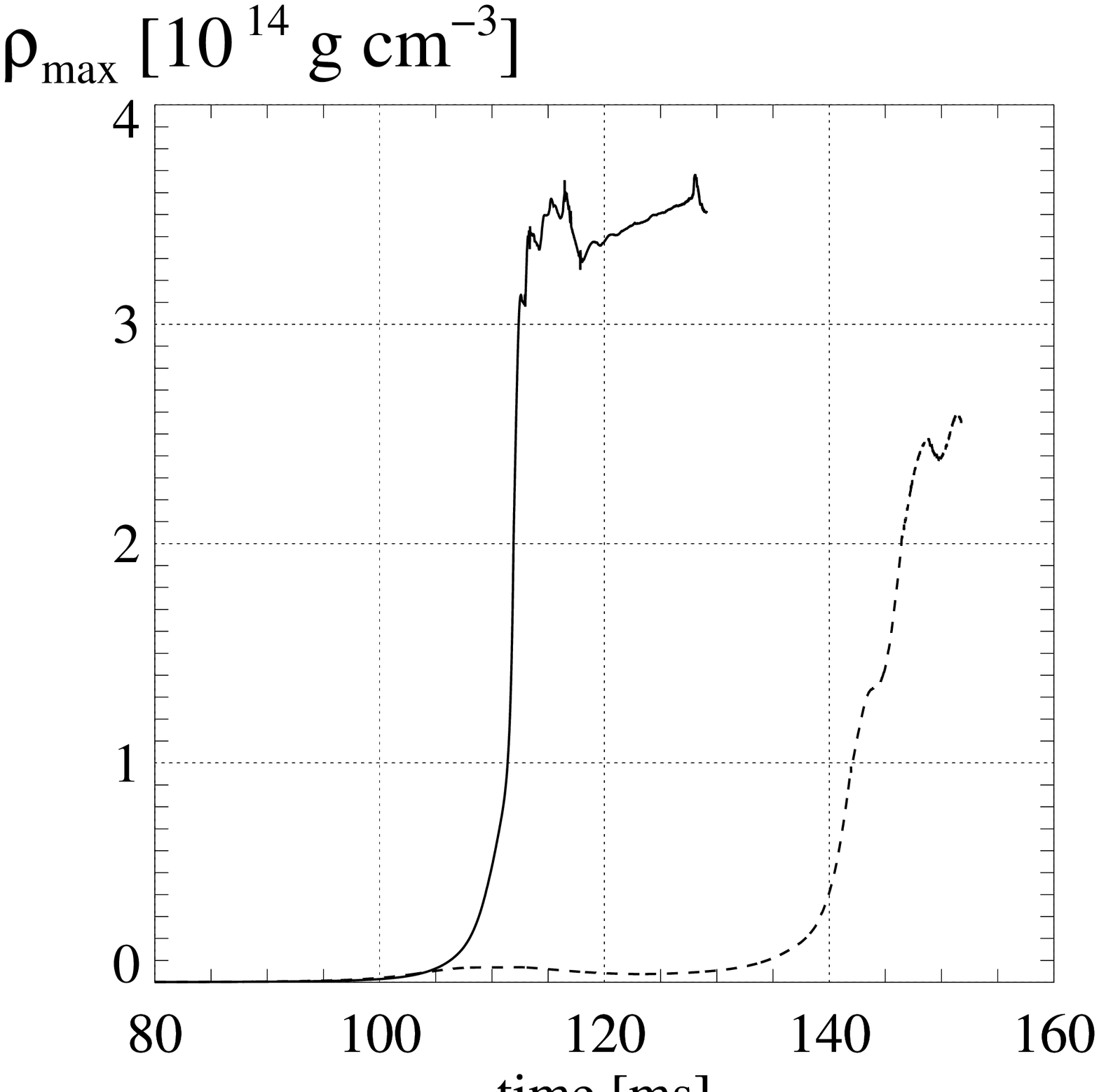}\\
  \includegraphics[width=5.6cm]{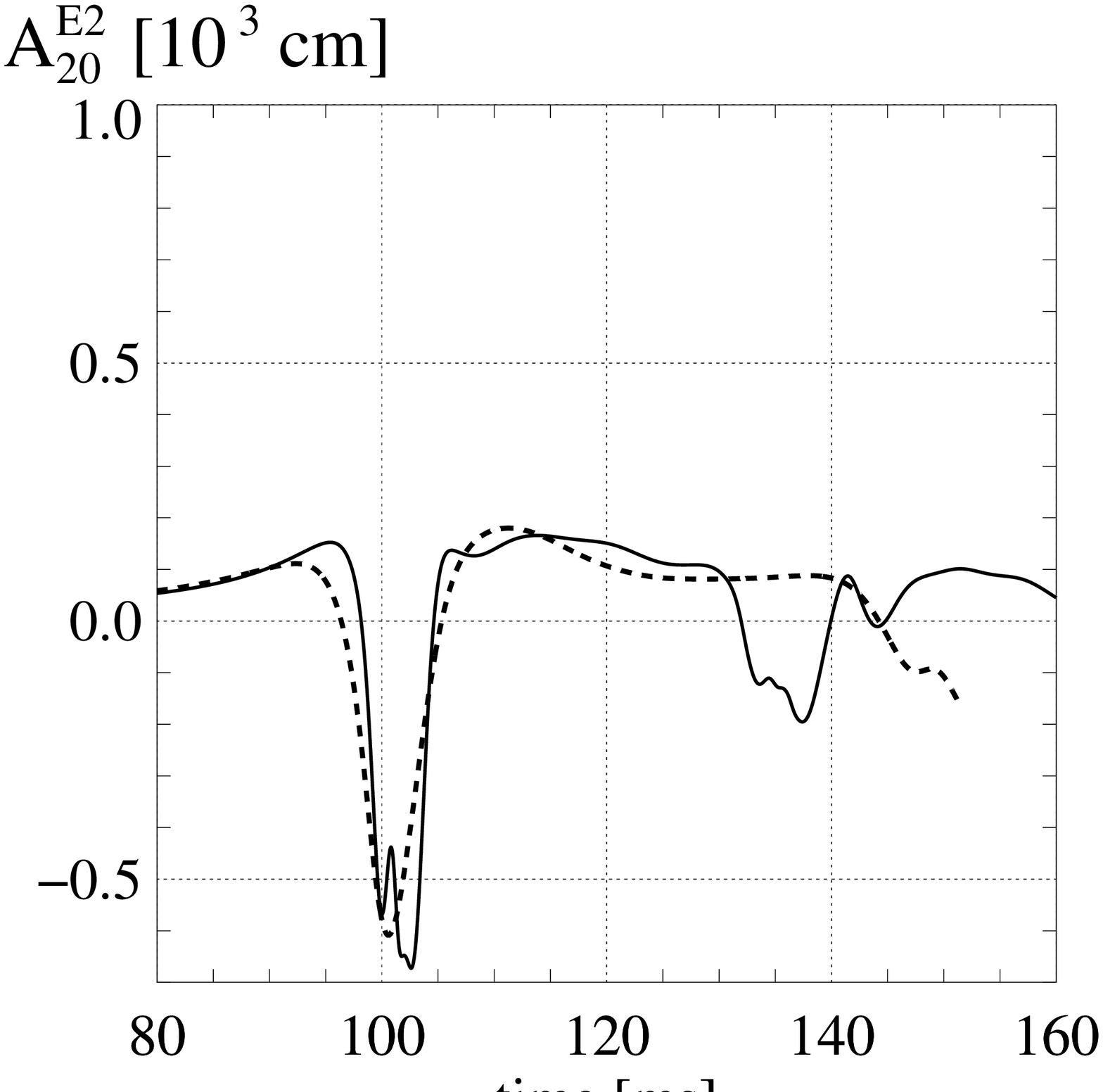}
  \includegraphics[width=5.6cm]{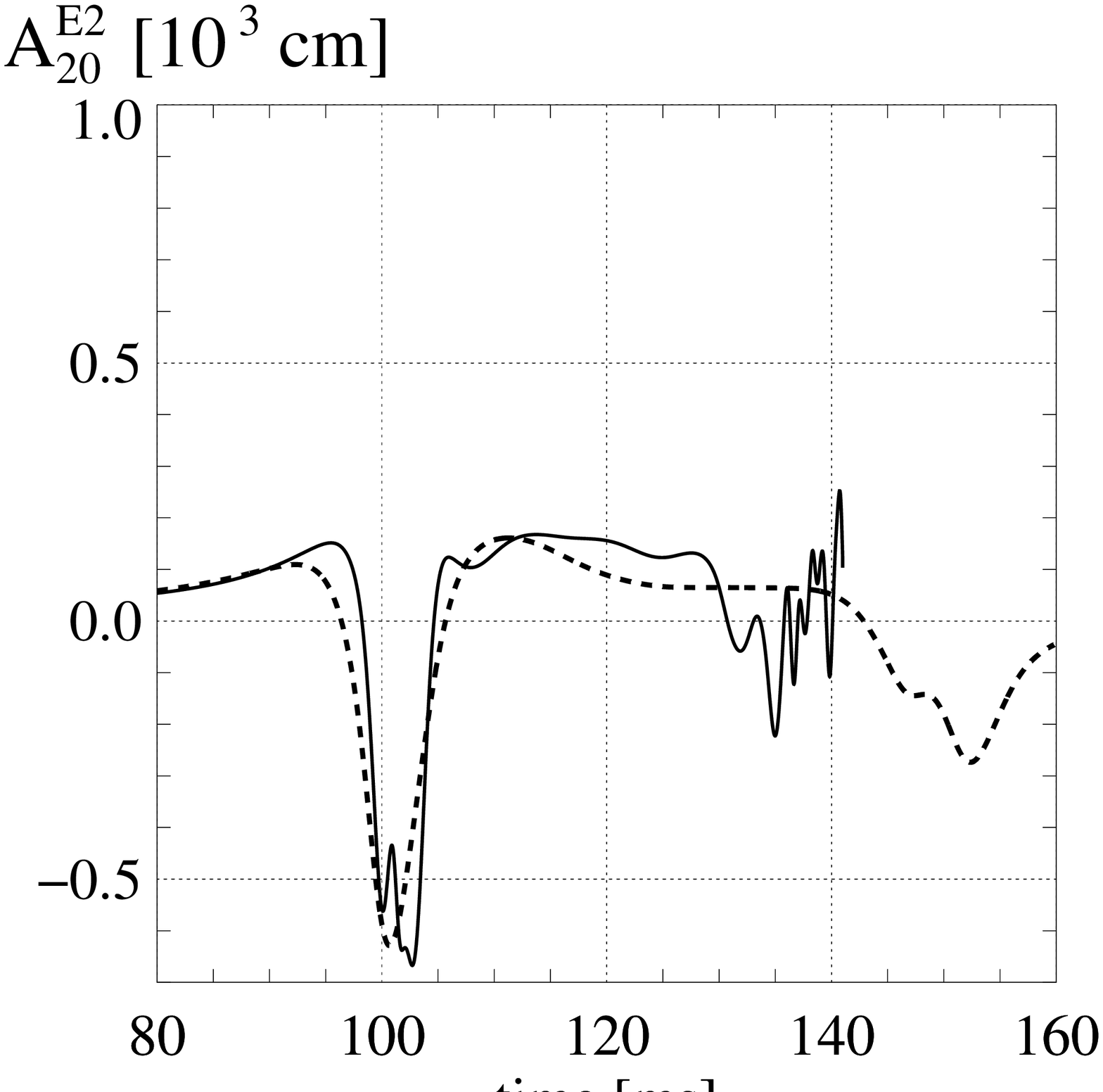}
  \includegraphics[width=5.6cm]{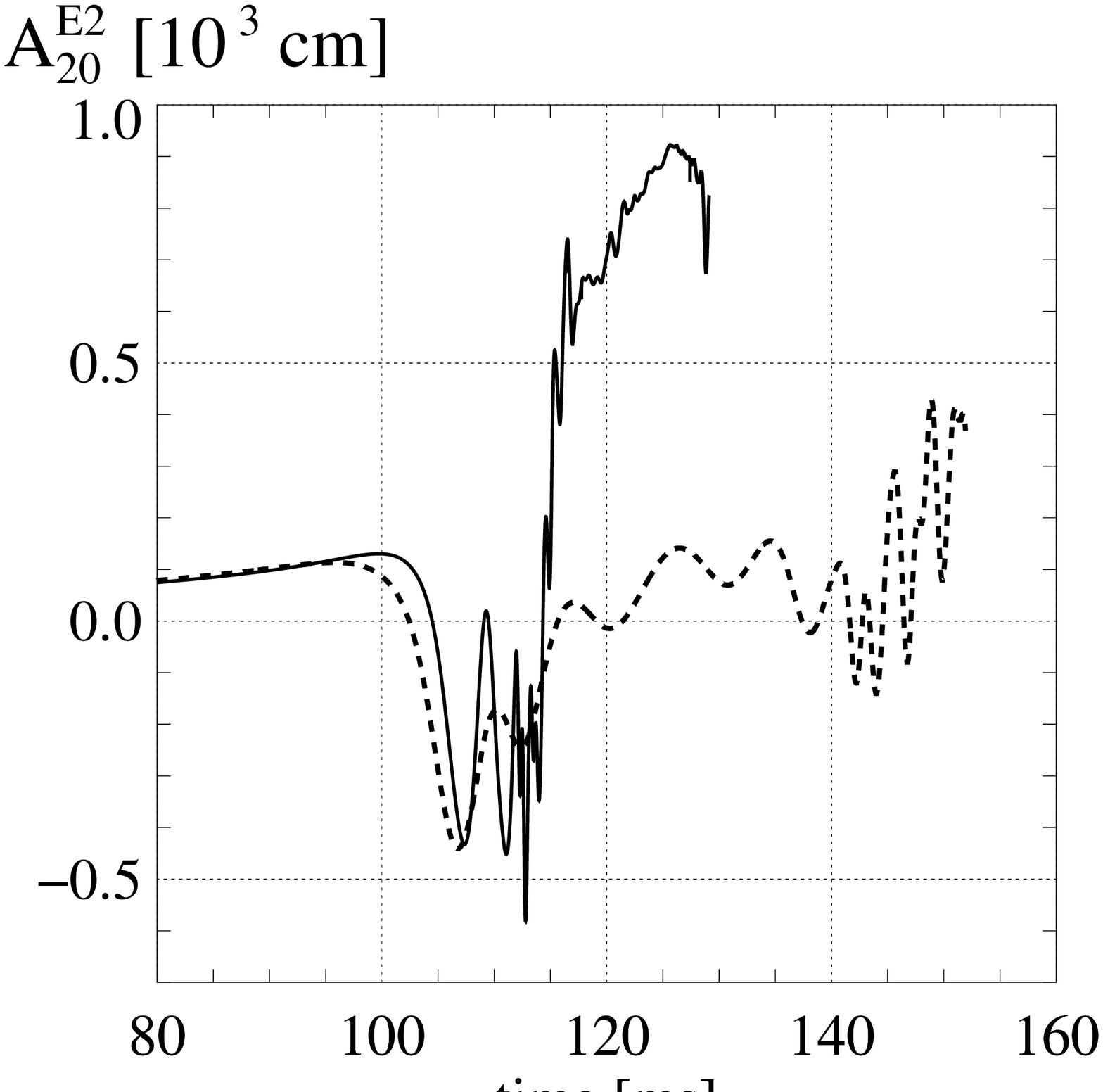}
  \caption{The evolution of the maximum density (upper panels) and GW
           amplitudes (lower panels) of models A2B4G1-D3M10-N/T
           (left), A2B4G1-D3M12-N/T (middle), and A2B4G1-D3M13-N/T
           (right).  Solid and dashed lines show TOV and Newtonian
           models, respectively.  }
  \label{Fig:A2B4G1-rho-GW}
\end{figure*}

The next series of models to be discussed is A3B3G5-D3Mm.  Simulating
these type III models with the effective TOV potential significantly
changes neither the dynamics nor the GW signal compared to runs
performed with the Newtonian potential (Fig.\,\ref{Fig:A3B3G5-rho-GW}).
The TOV models reach a slightly higher rotation rate than the
corresponding Newtonian ones, which leads to a slightly more efficient
amplification of the magnetic field. At 14\,ms post bounce, the total
magnetic energies of the Newtonian and TOV cores are
$E_{\mathrm{mag}}^{\mathrm{N}} = 3.8\times 10^{46}\,$erg, and
$E_{\mathrm{mag}}^{\mathrm{TOV}} = 6.0\times 10^{46}\,$erg,
respectively. For models with a sufficiently strong magnetic field, the
maximum density of the core increases like in the Newtonian case as the
core loses rotational support due to the redistribution of angular
momentum by magnetic field stresses.  In model A3B3G5-D3M13-T the
maximum density reached in the post-bounce quasi-equilibrium
configuration is about 20\,\% higher than in the corresponding Newtonian
model (Fig.\,\ref{Fig:A3B3G5-rho-GW}, upper right). This statement also
roughly holds for the weak-field model A3B3G5-D3M10
(Fig.\,\ref{Fig:A3B3G5-rho-GW}, upper left; but note the different
evolution of the maximum density for models with weak and strong initial
magnetic fields). The type\,III GW signals of all three magnetized
models are quite similar for the Newtonian and the effective TOV
potential (Fig.\,\ref{Fig:A3B3G5-rho-GW}, lower panels).

For the third series of models, A3B3G3-D3Mm, the changes resulting from
the use of the effective TOV potential instead of the Newtonian one
depend on the strength of the initial magnetic field
(Fig.\,\ref{Fig:A3B3G3-rho-GW}).  The weak-field Newtonian model
A3B3G3-D3M10-N, which bounces due to a combination of (mainly)
centrifugal and pressure forces at $\rho_{\mathrm{b}} = 2.4\times
10^{14}\ \mathrm{g\,cm}^{-3}$ just above nuclear matter density, shows
several distinct post-bounce oscillations of the maximum density
(Fig.\,\ref{Fig:A3B3G3-rho-GW} upper left) and emits a GW signal
intermediate between a type\,I and a type\,II signal
(Fig.\,\ref{Fig:A3B3G3-rho-GW}, lower left). With the effective TOV
potential, the model collapses deeper ($\rho_{\mathrm{b}} = 3.1 \times
10^{14}\ \mathrm{g\,cm}^{-3}$), thus spinning faster than in the
Newtonian case. Its GW signal is almost a pure type\,I signal showing
the typical ring-down oscillations instead of coherent large-scale
oscillations (Fig.\,\ref{Fig:A3B3G3-rho-GW}, lower left). The with
stronger initial fields (A3B3G3-D3M12-T and A3B3G3-D3M13-T) collapse to
about 30\,\% higher densities than their Newtonian counterparts. The
initial magnetic fields of these models are sufficiently strong for both
potentials to trigger a secular contraction of the core, due to
angular-momentum redistribution by magnetic field stresses, and to cause
a collimated outflow.

The cores of the fourth series of models (A2B4G1-D3Mm) considered in
our study bounce due to centrifugal forces as do their Newtonian
counterparts (and the GR-CFC models, see DFM) exhibiting multiple
bounces and large-scale pulsations (Fig.\,\ref{Fig:A2B4G1-rho-GW}).
In the weak-field case (A2B4G1-D3M10-T), the maximum density never
exceeds nuclear matter density, and the magnetic field is amplified
less efficiently than in the models of series A1B3G3-D3Mm-T due to the
longer rotation period of the less compact core of the models of
series A2B4G1-D3Mm-T.  Compared to the corresponding Newtonian model,
we find a much higher field amplification rate. Both
$E_{\mathrm{rot}}$ and $\beta_{\mathrm{mag}}$ have about the same
magnitude in the TOV model at bounce as the corresponding quantities
in the Newtonian model about 50\,ms past bounce (during the second
pulsation of the core centered at $t \approx 150\,$ms), i.e.\,after a
significantly longer period of amplification.  This is a consequence
of the deeper collapse of the TOV model, whose maximum density exceeds
that of the Newtonian model A2B4G1-D3M10-N by a factor of about seven
(Fig.\,\ref{Fig:A2B4G1-rho-GW}, upper left).  This, in turn, leads to
a more compact core with a shorter rotation period favoring a more
efficient field amplification.  The core emits a type\,II GW signal
like the Newtonian counterpart (Fig.\,\ref{Fig:A2B4G1-rho-GW}, lower
left).

The effects of a strong magnetic field on models of series A2B4G1-D3Mm-T
are even more pronounced than in the Newtonian case due to the deeper
relativistic potential.  For an initial field of $\sim 10^{13}\ 
\mathrm{G}$ the Newtonian model exhibits one small amplitude pulsation
(centered at $\sim 110\,$ms) before angular-momentum transport induced
by the magnetic field triggers a rapid contraction at $\sim 140\,$ms
(Fig.\,\ref{Fig:A2B4G1-rho-GW}, upper right).  The dynamic impact of a
field of $\sim 10^{12}\ \mathrm{G}$ on the core of the TOV model is very
similar to that of a field of $\sim 10^{13}\ \mathrm{G}$ in the
Newtonian case.  Both cores (A2B4G1-D3M12-T and A2B4G1-D3M13-N) undergo
one single post-bounce pulsation, where the amplitude is more pronounced
in the TOV model (Fig.\,\ref{Fig:A2B4G1-rho-GW}, upper middle and
right), and then rapidly contract to densities slightly above nuclear
matter density ($\rho_{\mathrm{max}} \approx 2.7\times 10^{14} \ 
\mathrm{g \ cm^{-3}}$).

During the immediate post-bounce evolution, the GW signals emitted by
the strong field models are very similar to those of the weak-field
model A2B4G1-D3M10-T. However, later in the evolution, the GW signals
are radically different from those of the corresponding non-magnetic and
initially weakly magnetized cores (Fig.\,\ref{Fig:A2B4G1-rho-GW}, lower
panels).  For $t \ga 130\,$ms, the GW amplitude of model A2B4G1-D3M12-T
exhibits rapid oscillations with periods in the range of milliseconds
(Fig.\,\ref{Fig:A2B4G1-rho-GW}, lower middle), while it rises to high
positive values in the case of model A2B4G1-D3M13-T at $t \approx
115\,$ms (Fig.\,\ref{Fig:A2B4G1-rho-GW}, lower right) also showing
superimposed oscillations. The frequency of the oscillations increases
as the density of the core grows. When it reaches supra-nuclear
densities, we observe oscillation periods in the sub-millisecond range
and oscillation amplitudes on the order of 100\,cm. The large positive
amplitude ($\sim 900\,$cm) of model A2B4G1-D3M13-T at late times is due
to the very prolate shape of its shock wave (jet).

As the GW signals of models A2B4G1-D3M12-T and A2B4G1-D3M13-T do not
belong to any of the familiar types\,I, II, or III, we classify them as
belonging to a new (magnetic) type\,IV GW signal, introduced in Paper\,I.

Finally, we discuss the process of shock formation in model
A2B4G1-D3M13-T in some more detail, which is also relevant to models
A2B4G1-D3M12-T and A2B4G1-D3M13-N.  In a type\,I model, the shock forms
by the steepening of pressure waves created as successive shells of core
matter feel the stiffening of the equation of state during bounce.  The
pressure waves move outwards and evolve into a shock as they accumulate
near the sonic point.  The typical time scale for this process is on the
order of the sound-crossing time of the inner core (that part of the
core that is located inside the sonic point), which is approximately
1\,ms.  In contrast, the shock wave is launched in a type\,II model at
the edge of the inner core, when it bounces due to the effect of the
centrifugal force and expands into the still infalling matter of the
outer core with supersonic speed.  The strength of this shock may vary
strongly with polar angle and may even develop for different angles at
different times.

Both modes of shock generation are at work in model A2B4G1-D3M13-T.
During its collapse the rotational energy rises considerably, but less
than in the non-magnetic or weak-field case. The rotation rate
$\beta_{\mathrm{rot}}$ already reaches a maximum {\em during} collapse
when $\rho_{\mathrm{max}} = 8.4\times 10^{13} \, \mathrm{g\,cm}^{-3}$,
i.e.\,well below the bounce density $\rho_{\mathrm{b}} = 3.5\times
10^{14} \, \mathrm{g\,cm}^{-3}$, as very efficient angular momentum
transport by the strong magnetic fields extracts rotational energy from
the core. This effect creates a rotation profile in the core where
matter with the same density as in the corresponding non-magnetic or
weakly magnetized models rotates faster near the pole.  The extraction
of rotational energy from the inner core also causes the post-shock gas
to continue to fall towards the center, and further shock waves are
created by the pressure-bounce mechanism.

\begin{figure*}[htbp]
  \centering
  \includegraphics[width=5.6cm]{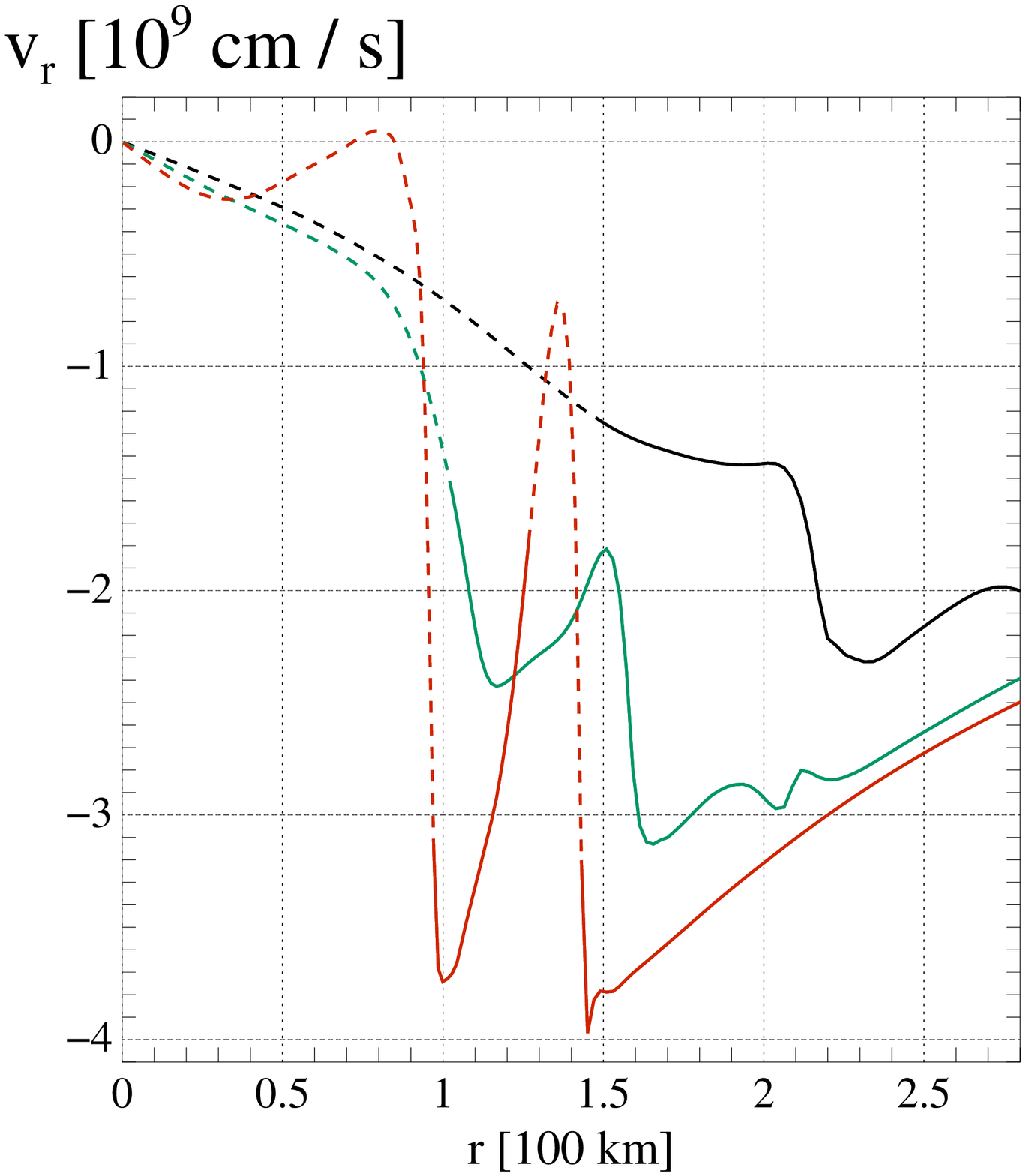}
  \includegraphics[width=5.6cm]{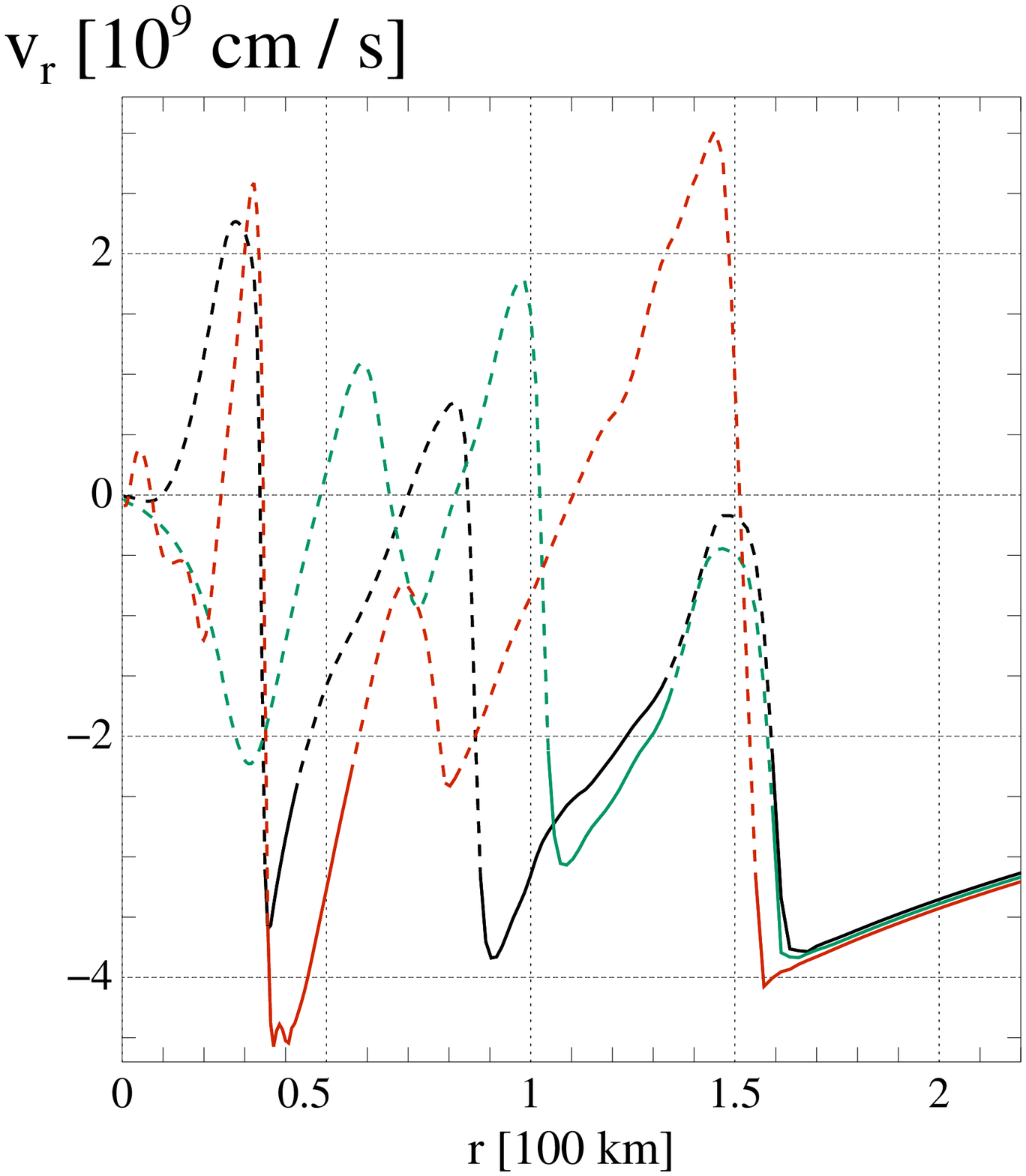}
  \includegraphics[width=5.6cm]{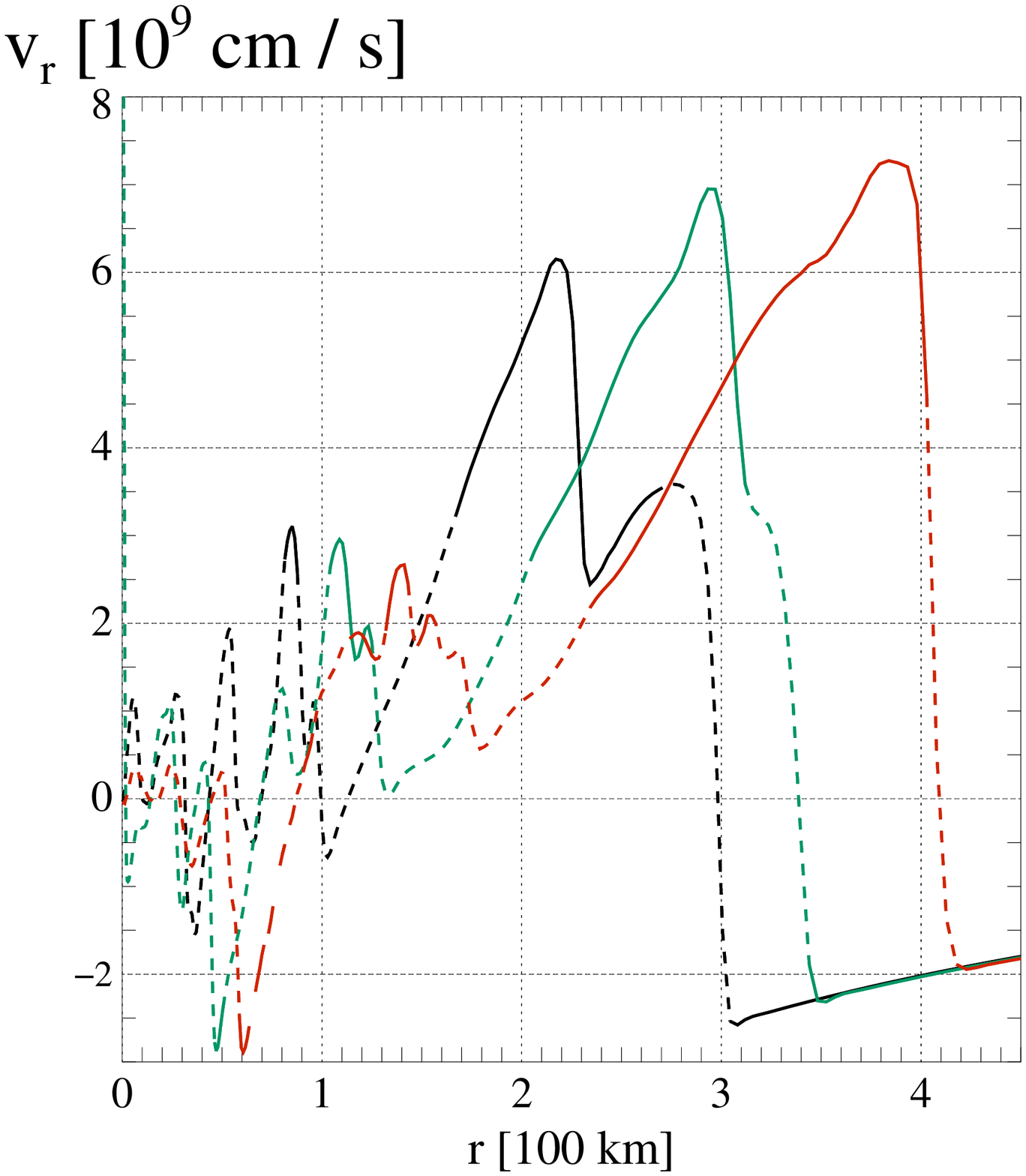}\\
  \includegraphics[width=5.6cm]{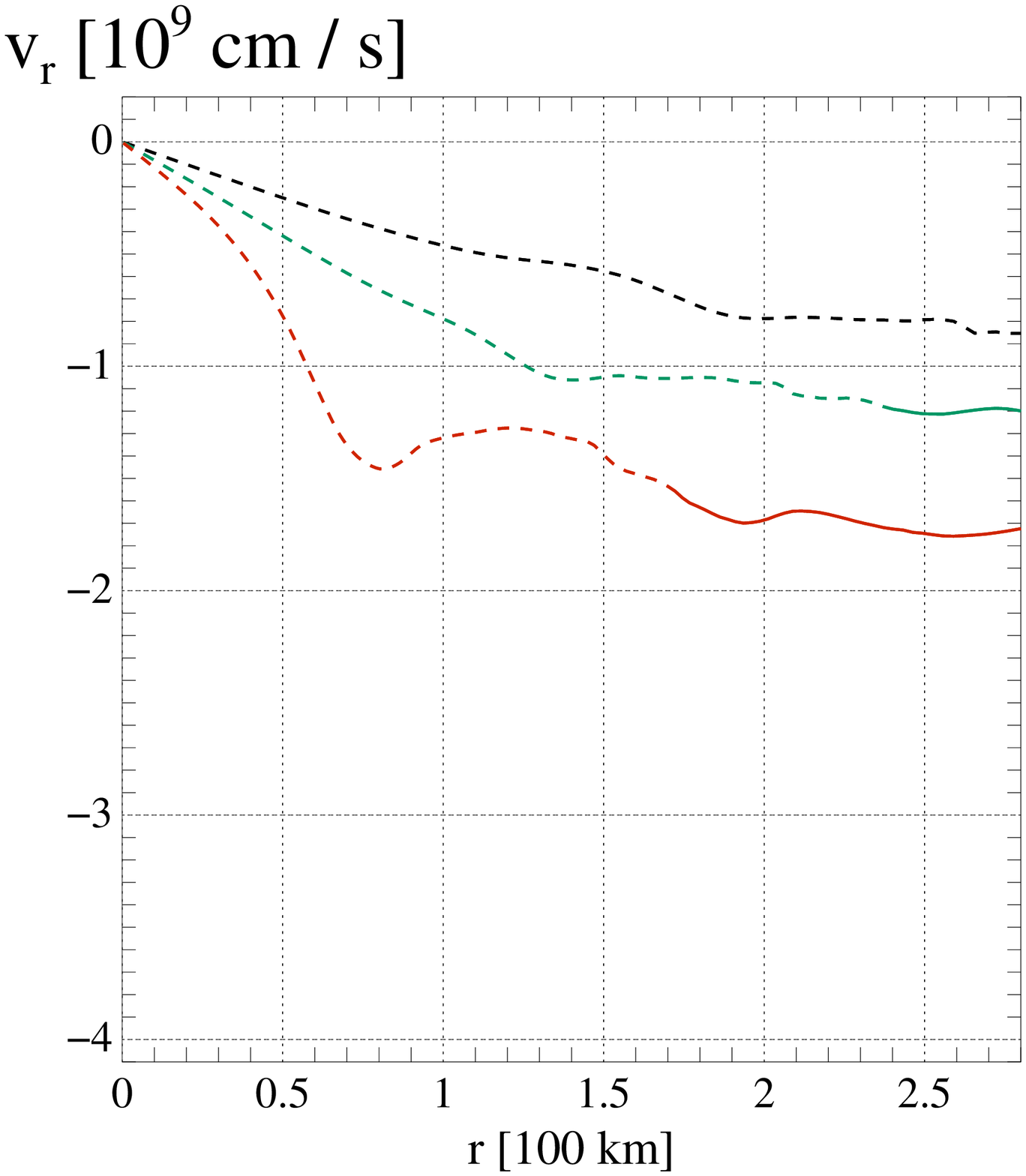}
  \includegraphics[width=5.6cm]{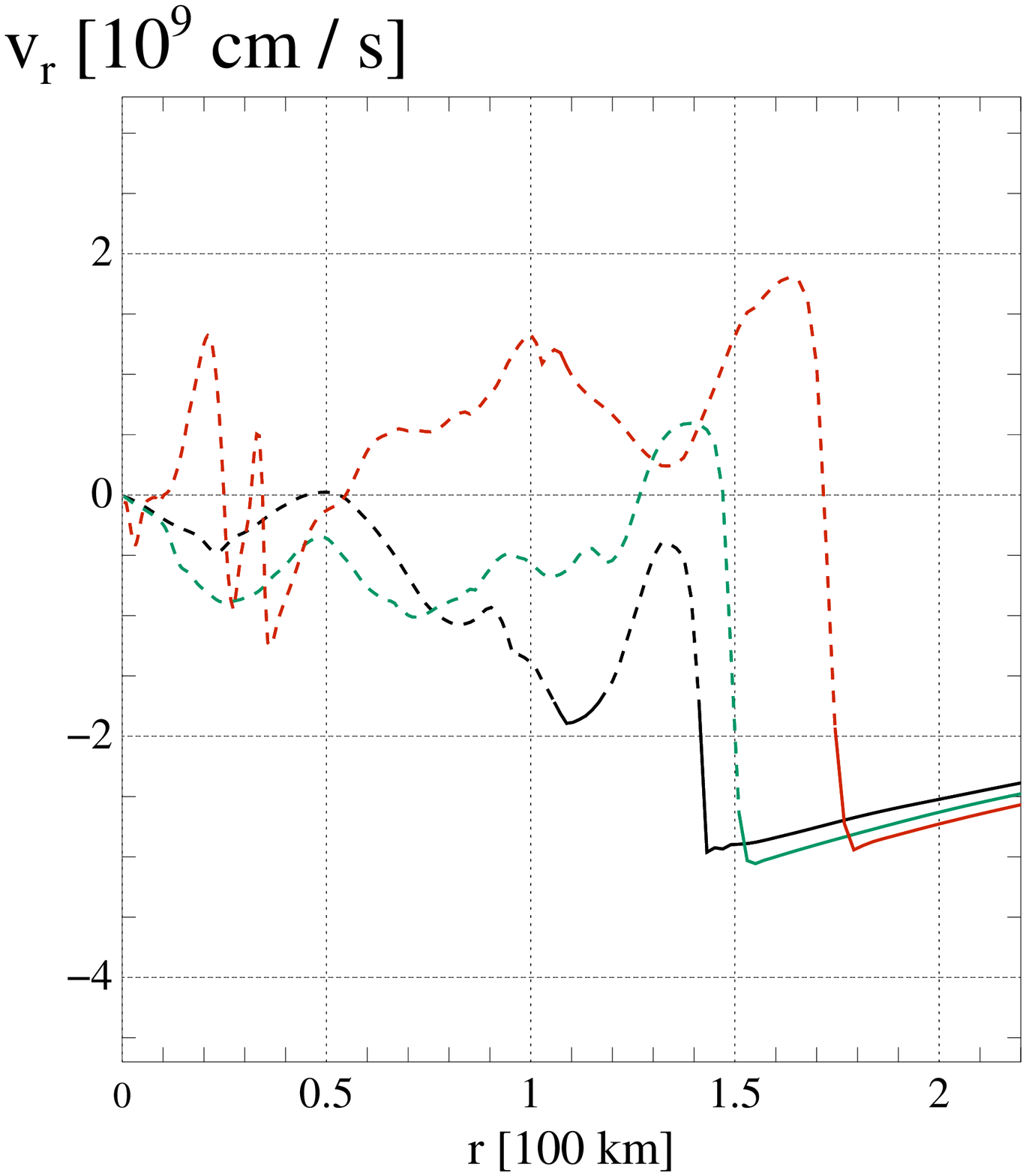}
  \includegraphics[width=5.6cm]{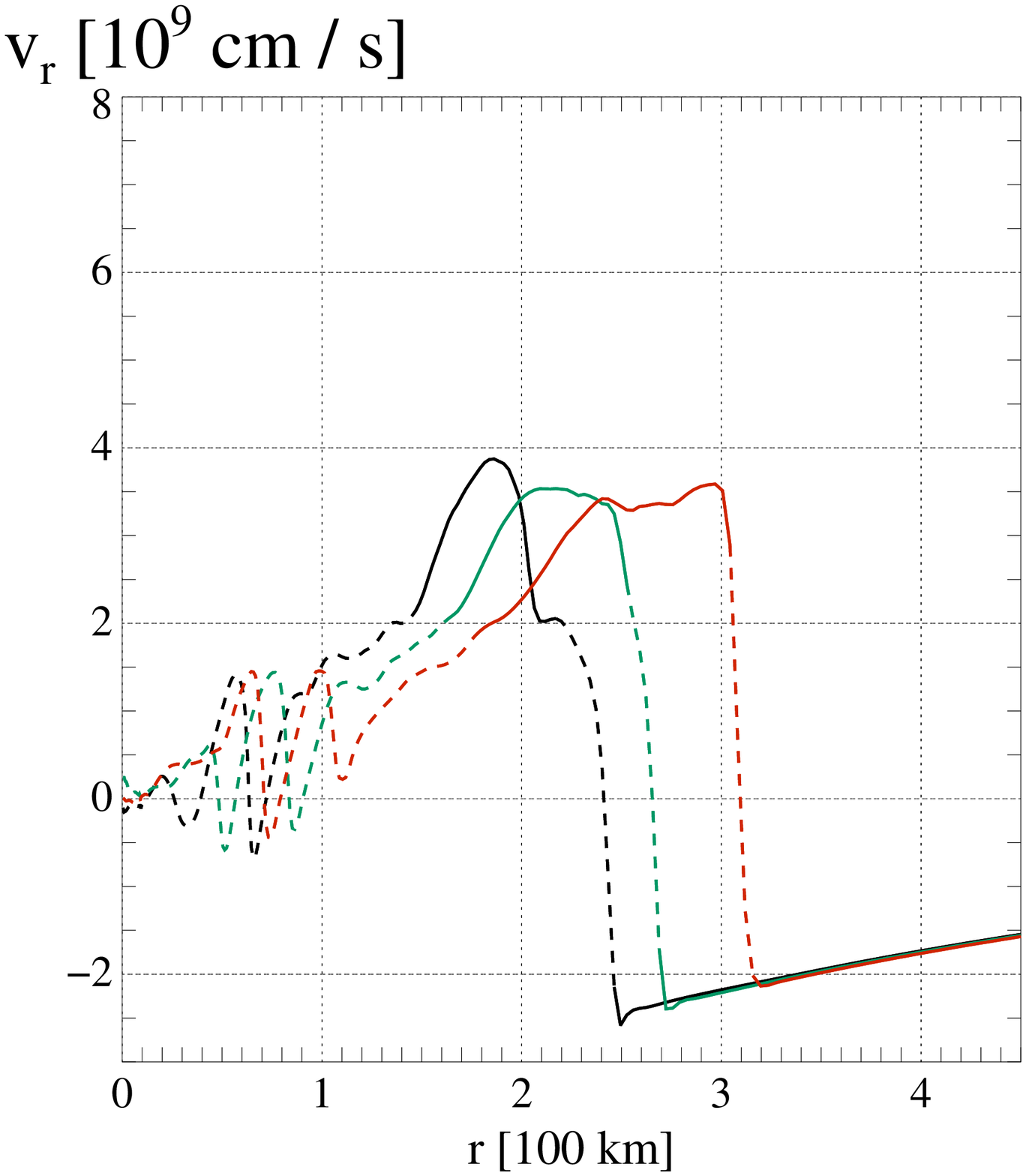}
  \caption{Radial velocity profiles for the strong-field model
    A2B4G1-D3M13-T along the rotation axis (top) and at the equator
    (bottom) at different epochs. Each panel shows three snapshots,
    plotted in black, green, and red. Solid and dashed lines mark
    subsonic and supersonic radial motion, respectively.  The snapshots
    are taken at $t=94.8, 100.4, 105.0\,$ms (left), $t=110.0, 110.7,
    112.3\,$ms (middle), and $t=115.9, 117.0, 117.9\,$ms (right).  A
    color version of the figure can be found in the on-line edition of
    the journal.  }
  \label{Fig:A2B4G1-D3M13_vr}
\end{figure*}

Two shock waves form at high latitudes near the surface of the inner
core (at $r \approx 100\,$km and $r \approx 140\,$km; see
Fig.\,\ref{Fig:A2B4G1-D3M13_vr}, upper left, last snapshot), while
matter near the equator is still falling in. No shock is present there
yet (Fig.\,\ref{Fig:A2B4G1-D3M13_vr}, lower left). About 5\,ms later at
$t\approx 110\,$ms a third polar shock (at $r \approx 30\,$km; see
Fig.\,\ref{Fig:A2B4G1-D3M13_vr}, upper middle, first snapshot) and an
equatorial shock (at $r \approx 130\,$km; see
Fig.\,\ref{Fig:A2B4G1-D3M13_vr}, lower middle, first snapshot) have
formed. The second polar shock, which is stronger and propagates faster
than the first one, is about to merge with it about 6\,ms later at
$t\approx 116\,$ms (Fig.\,\ref{Fig:A2B4G1-D3M13_vr}, upper right, first
snapshot).  Two more polar shock and one additional equatorial one form
up until the end of our simulation at $t\approx 118\,$ms (about 10\,ms
after bounce), when the first and second polar and equatorial shocks
have merged, and are located at a radius of $r_{\mathrm{p}} \approx
410\,$km along the polar direction and $r_{\mathrm{e}} \approx 310\,$km
along the equator, respectively (Fig.\,\ref{Fig:A2B4G1-D3M13_vr}, right,
last snapshots).  The creation and propagation of the various shock
waves gets imprinted on the density distribution, which is strongly
anisotropic and exhibits several distinct discontinuities, their number
and radial location depending on polar angle.

In the corresponding weak-field core, we find an almost isotropic
expansion of the post-shock matter supported by centrifugal forces.  At
$t\approx 109\,$ms the nearly spherically symmetric, leading shock wave
has reached a polar and equatorial radius of $r_{\mathrm{p}} \approx
320\,$km and $r_{\mathrm{e}} \approx 300\,$km, respectively. The density
distribution is almost spherical and shows much less structure than in
the strong-field case.

\subsection{Gravitational wave spectra}
\label{Suk:GW-Spec}

\begin{figure*}[htbp]
  \centering
  \includegraphics[width=6.7cm]{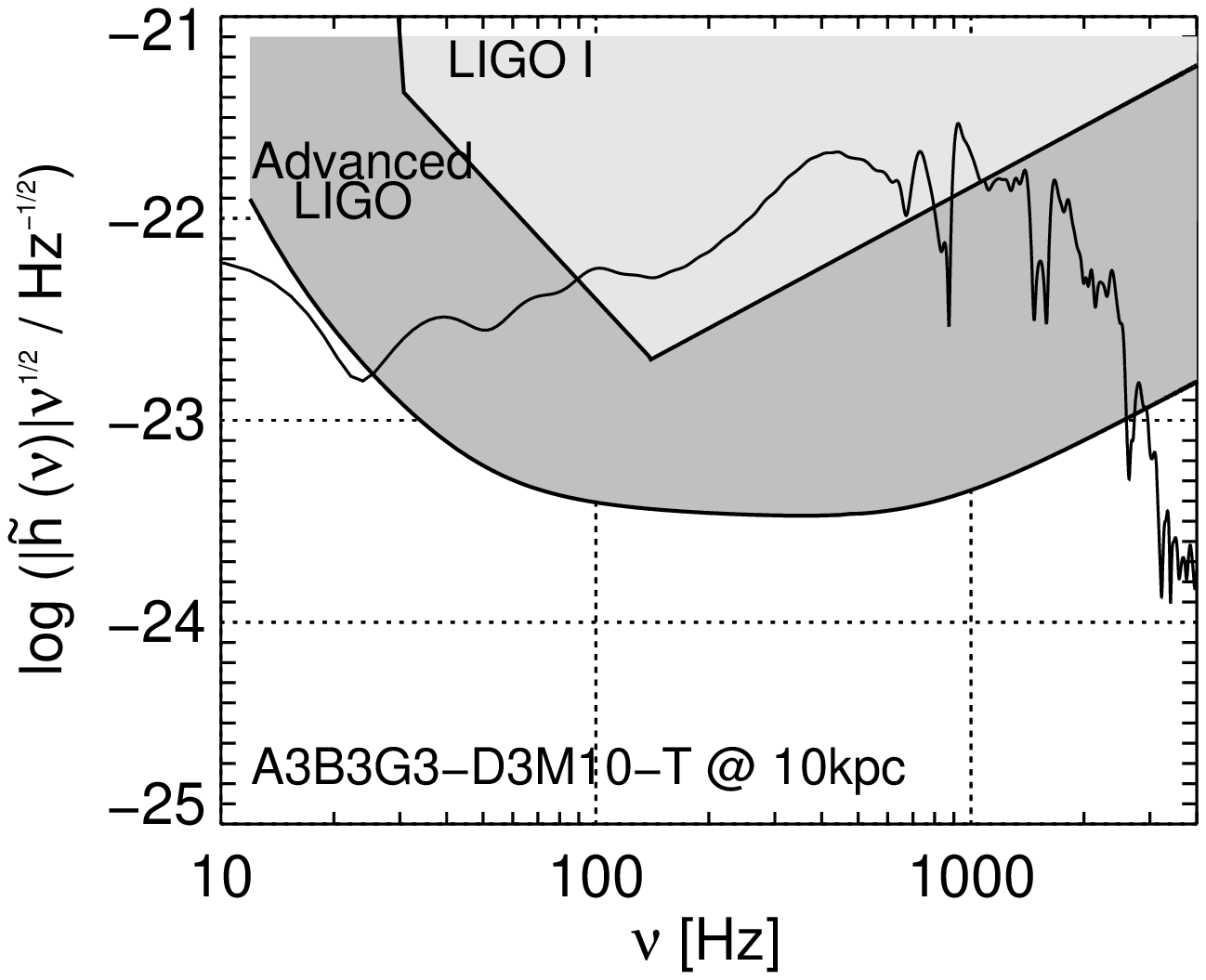}
  \includegraphics[width=6.7cm]{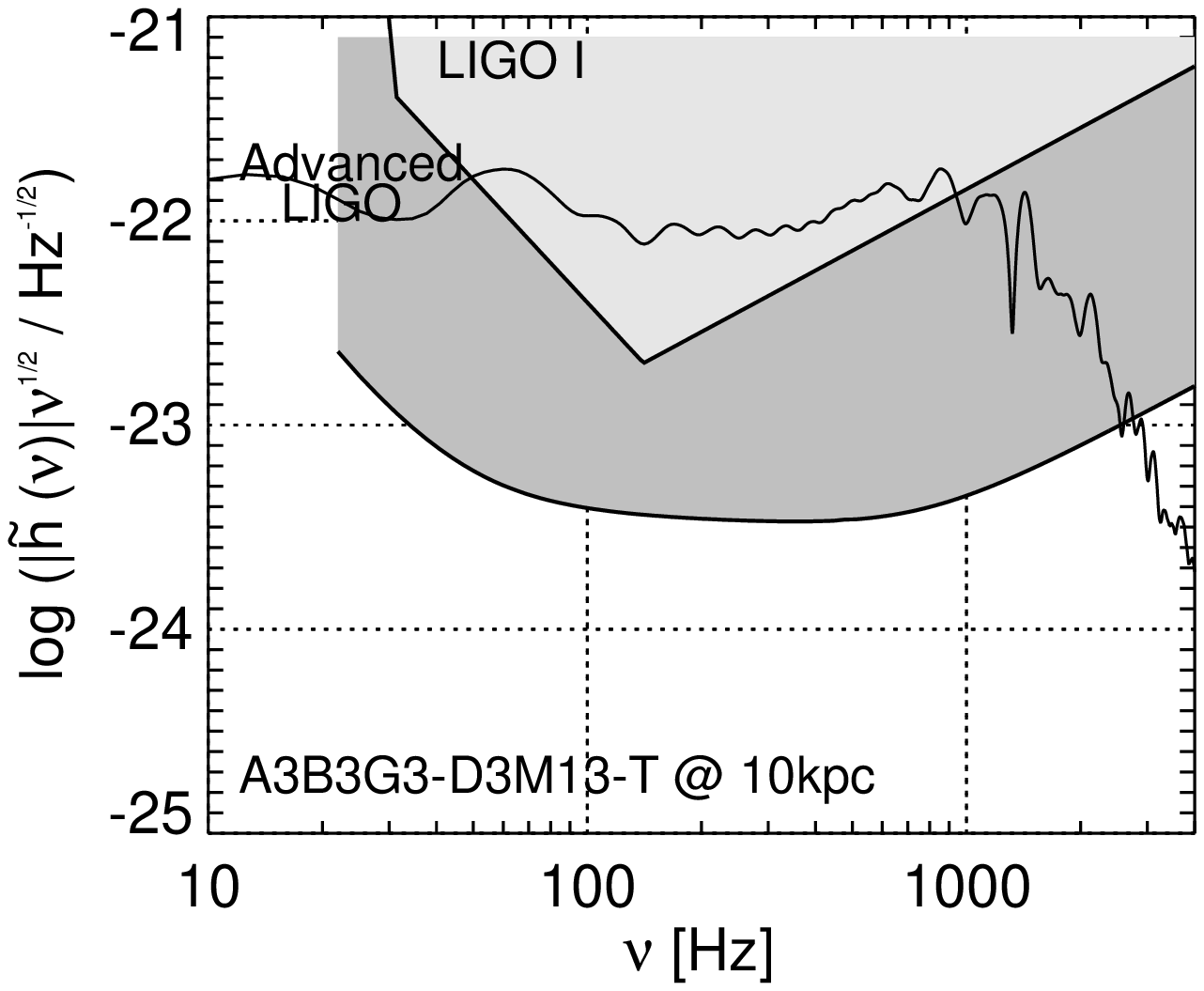}
  \includegraphics[width=6.7cm]{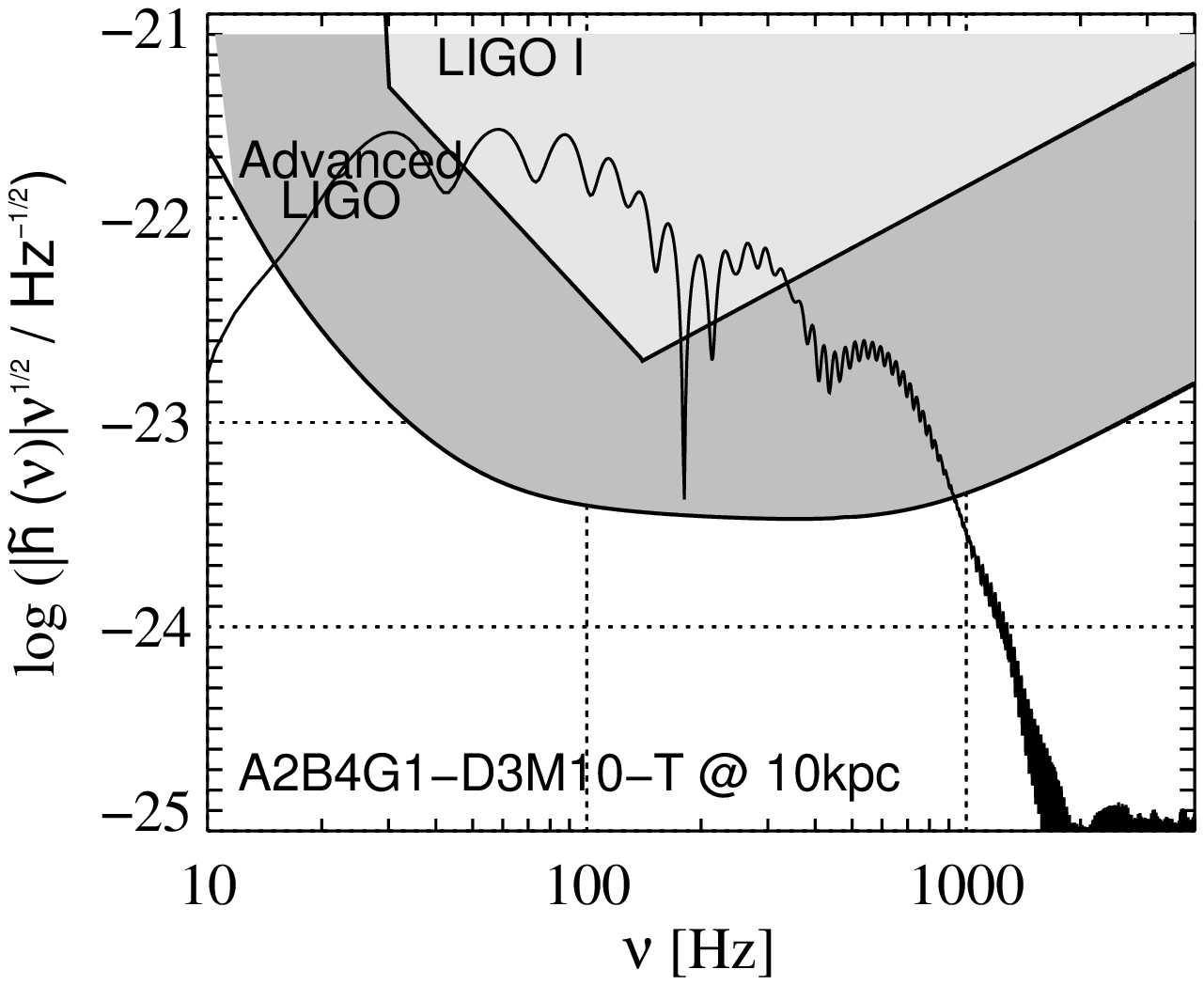}
  \includegraphics[width=6.7cm]{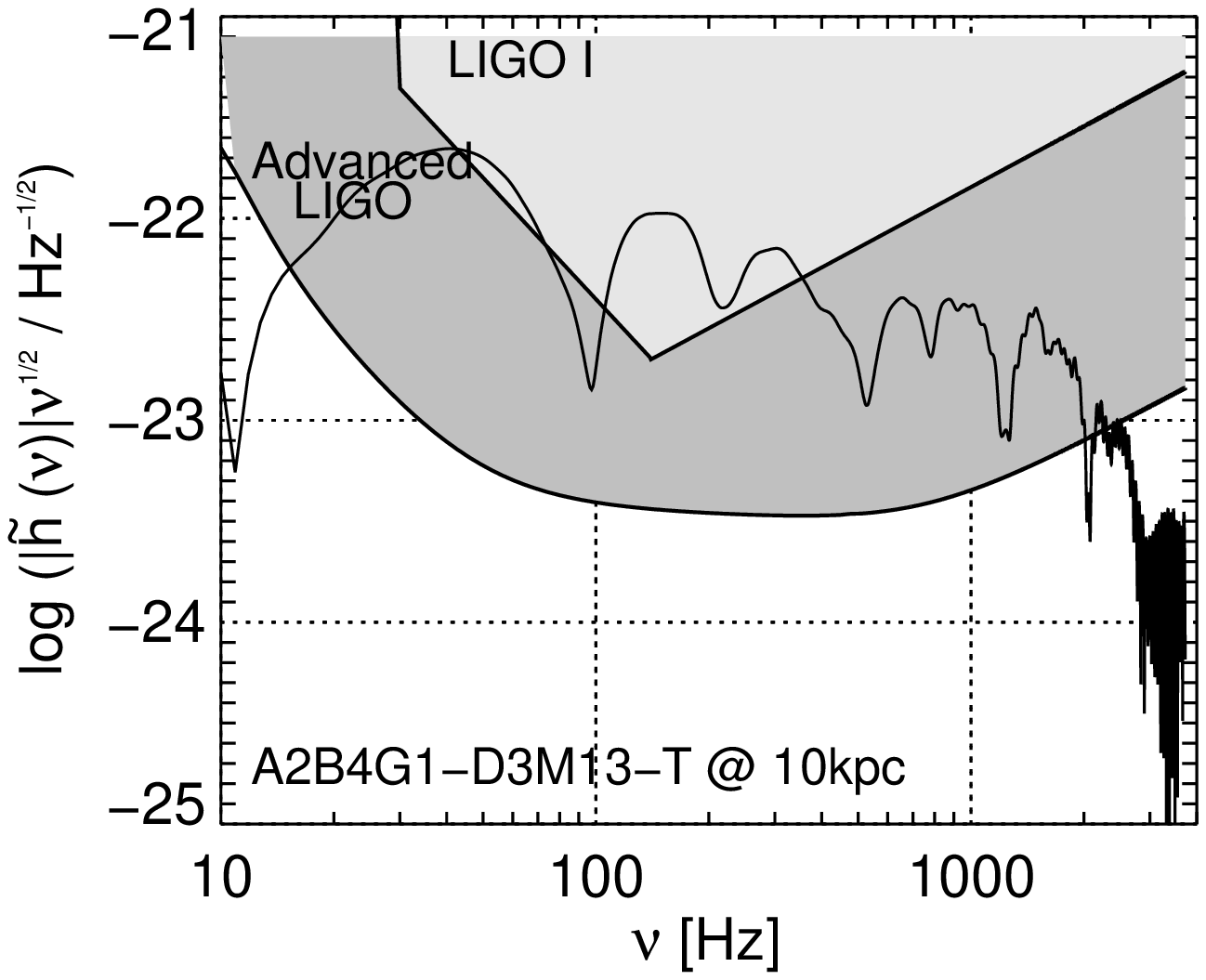}
  \includegraphics[width=6.7cm]{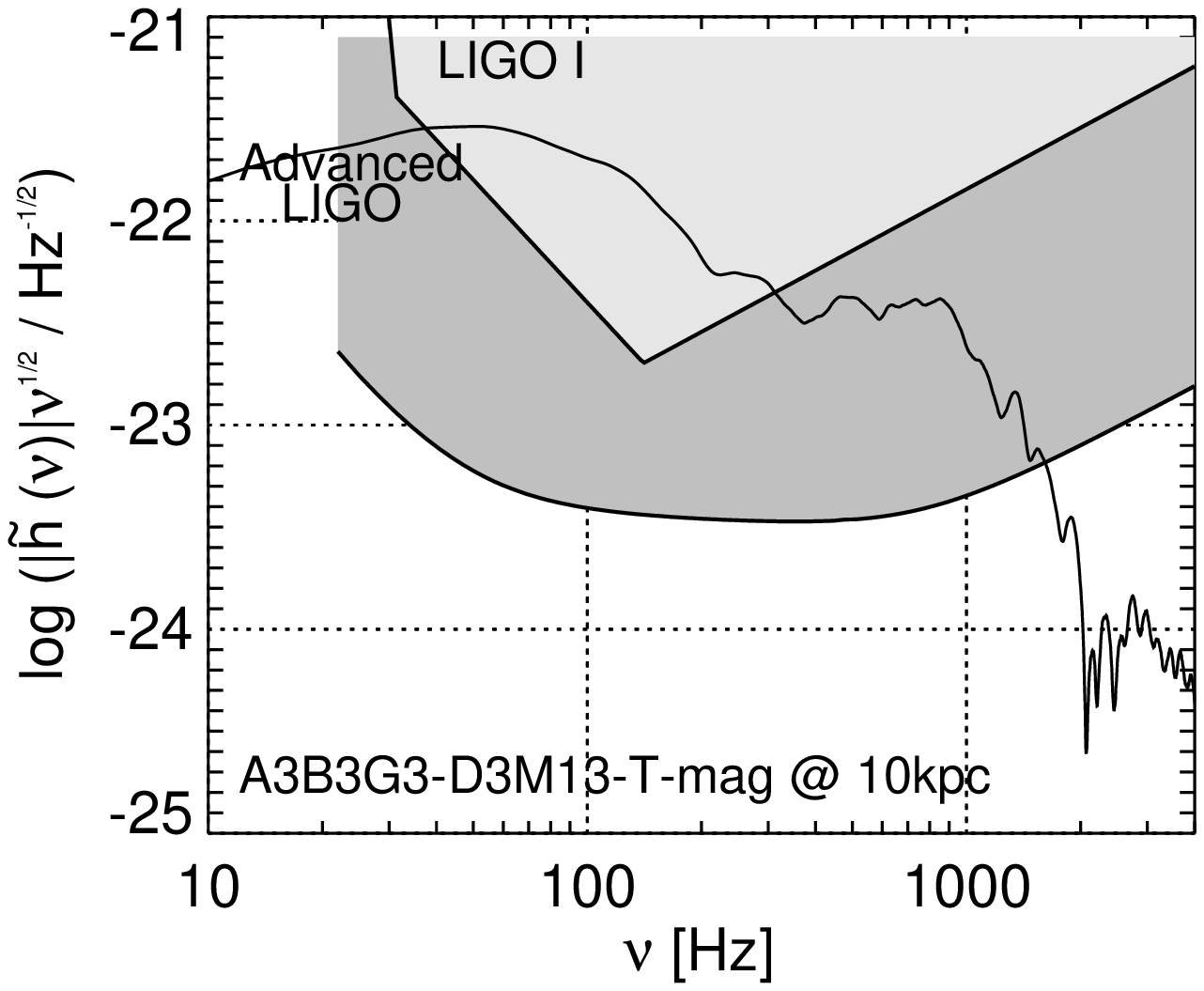}
  \includegraphics[width=6.7cm]{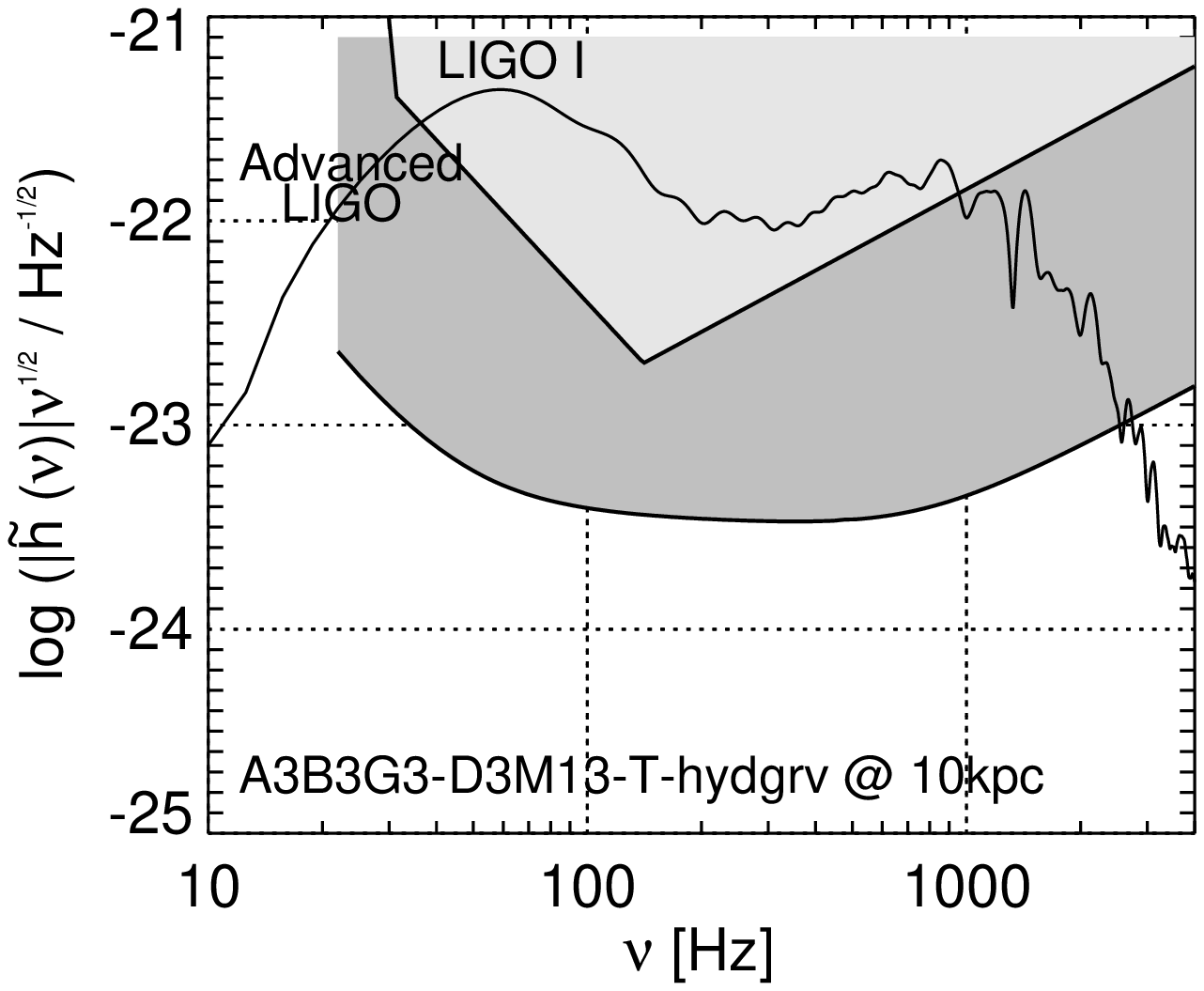}
  \caption{The GW spectral energy distribution $|\tilde{h} (\nu)|
           \nu^{1/2}|$ of a weak-field (A3B3G3-D3M10-T, upper left)
           and strong-field (A3B3G3-D3M13-T, upper right) model
           bouncing due to pressure forces, and of a weak-field
           (A2B4G1-D3M10-T, middle left) and strong-field
           (A2B4G1-D3M13-T, middle right) model bouncing due to
           centrifugal forces. The magnetic and non-magnetic
           contributions to the GW amplitude of model A3B3G3-D3M13-T
           are displayed in the lower left and right panels,
           respectively.  The grey-shaded regions give the detection
           limits of the current (LIGO I) and the Advanced LIGO
           interferometer. The source is assumed to be located at a
           distance of 10\,kpc.}
  \label{Fig:GWspec}
\end{figure*}

We also calculate the spectral energy distribution $|\tilde{h}(\nu)|
\nu^{1/2}$ of the gravitational wave signals emitted by our models.  The
Fourier-transformed amplitudes $\tilde{h}(\nu)$ as a function of the
signal frequency $\nu$ are obtained from the GW signals $h(t)$ using fast
Fourier transforms (for details, see e.g.\,\citealp{Mueller_etal_04}).

Spectra of non-magnetic models were already discussed by \cite{DFM2}.
Typical GW spectra of type\,I (A3B3G3-D3M10-T) and type\,II
(A2B4G1-D3M10-T) models are shown in Fig.\,\ref{Fig:GWspec} for a source
located at a distance of 10\,kpc. The GW spectrum of a type\,I model
peaks at high frequencies ($ \sim 800\,$Hz), whereas the spectrum of a
multiple bounce (type\,II) model possesses a very broad maximum at
frequencies of $\sim 30 - 200\,$Hz, reflecting the different bounce
mechanisms of the inner cores of the two models. In the former case, the
core is very compact, and thus having a short dynamic time scale of a
few milliseconds, leading to rapid ring-down oscillations and a
high-frequency signal. The multiple centrifugal bounces of model
A2B4G1-D3M10-T recurring on comparatively long time scales of $\sim
35\,$ms cause the low-frequency maximum in the spectrum.  This
difference is also present in a comparison of the spectra of the
Newtonian and the TOV version of model A3B3G3-D3M10. The spectrum of the
Newtonian model, being of transition type\,I/II, peaks at considerably
lower frequencies ($\sim 150 - 500\,$Hz) than the one computed with the
effective TOV potential ($\ga 400 \,$Hz).  Most of the weak-field models
are marginally detectable with the LIGO interferometer for a source
located at a distance of 10\,kpc.

At very low-frequencies, the characteristic modification imprinted onto
the GW spectrum by strong magnetic fields is a dominant feature. The
main differences between weak-field and strong-field versions of a given
model are visible in the low frequency range for type\,I models, while
the high-frequency part is affected relatively modestly.  Model
A3B3G3-D3M13-T has a considerable excess of spectral power at
frequencies $\la 100\,$Hz compared to model A3B3G3-D3M10-T. In contrast
to the latter, the spectrum of model A3B3G3-D3M13-T is flat at
frequencies below the peak frequency $\sim 800\,$Hz. This feature can be
attributed to the jet-like outflow, giving rise to a strongly positive
GW amplitude of slow temporal variability.  Both the magnetic and the
non-magnetic contributions to the GW amplitudes have strong
low-frequency contributions reflecting how the outflow is neither purely
magnetic nor purely hydrodynamic, but of a magnetohydrodynamic nature.
Note that the magnetic and non-magnetic contributions to the spectrum
may have different signs. Therefore, the total spectral power is
suppressed by a factor of $\sim 2.5$ with respect to both contributions.
For a source at a distance of 10\,kpc, the flatter spectrum of the
strong-field model improves its detectability with the laser
interferometers such as the LIGO detector.  The high-frequency part of
the spectrum, as well as the frequency and amplitude of the peak in the
spectrum are insensitive to the magnetic field.

For multiple-bounce (type\,II) models there is a substantial difference
between weak-field and strong-field versions both at low frequencies
(like in type\,I models) and high frequencies. A particularly good
example for such a difference are the models of series A2B4G1-D3Mm-T,
ranging from the weak-field type\,II model A2B4G1-D3M10-T to the
type\,IV model A2B4G1-D3M13-T.  Unlike the spectrum of the former model
with its broad low-frequency maximum and a rapid decrease towards high
frequencies, the spectral energy distribution of model A2B4G1-D3M13-T
has a rather strong contribution above $\sim 700\,$Hz. At a frequency of
$\approx 1500\,$Hz the latter signal has about 10 times more power than
the former one. This part of the spectrum is produced by the very rapid
oscillations of the core as it collapses to nuclear matter density, and
the dynamical time scale drastically decreases.


\section{Summary and conclusions}
\label{Sek:Schluss}

We have presented an investigation of the gravitational collapse of
rotating magnetized stellar cores including an approximate relativistic
TOV potential to mimic the effects of general relativity in our
simulations.  The implementation of this potential requires only minor
modifications of an existing Newtonian MHD code and provides a good
approximation to full GR dynamics in the case of the collapse of a
stellar core to a neutron star.  In particular, we are able to reproduce
the change in the bounce mechanism (centrifugal vs.\ pressure) occurring
in some (non-magnetic, rotating) models when changing from Newtonian to
GR gravity.  The maximum densities obtained with the approximate
relativistic TOV potential are, as expected, higher than the
corresponding Newtonian ones. This also holds for the rotational
energies, although some exceptions exist to this rule.  The
gravitational wave signals exhibit the same qualitative behavior as in
full GR, but they differ quantitatively.

Comparing the results obtained with the approximate TOV potential with
those of our previous Newtonian calculations of the magneto-rotational
collapse of stellar cores (described in detail in Paper\,I), we find
that the main difference corresponds to a ``shift'' in the parameter
space. In TOV gravity, one needs faster or more differential rotation to
cause centrifugal effects of similar strength for the same equation of
state.  Given the same initial model and the same equation of state, a
core in TOV gravity will collapse to higher densities.  This deeper
collapse of the TOV model leads in many of our models to a faster
rotation of the inner core than in the corresponding Newtonian case,
causing a more efficient amplification of the magnetic field by
differential rotation ($\Omega$ dynamo).  In principle, an
$\alpha$-$\Omega$-type dynamo could develop due to differential rotation
and possible 3D MHD instabilities, as in the Newtonian case
\citep{Henk02}.  However, in our axisymmetric simulations, the
transformation of toroidal into poloidal fields is suppressed, hence we
are unable to simulate this kind of dynamo.

We find that the magneto-rotational instability can develop in the TOV
models as well as in the Newtonian ones. The growth times and saturation
fields of the magnetic fields resulting from the MRI are within the same
order of magnitude as in the Newtonian case, i.e.\,in the range of
milliseconds and $\ga 10^{15}\,\mathrm{G}$, respectively.  Due to the
grid-resolution problem already discussed for the Newtonian models (see
Paper\,I), we are unable to simulate the evolution of the MRI unless the
seed field is already quite strong.  In models belonging to the latter
class, we find an exponential growth of the poloidal field of the inner
core during the post-bounce evolution by the action of the MRI.

The influence of strong magnetic fields on the dynamics and the GW
signal of the core by braking its rotation and thus triggering a
post-bounce contraction is aided by the deeper TOV potential.  For
type\,II multiple bounce models, we find that magnetic fields affect
the dynamics and GW signal of the core by a similar amount to what is
observed for Newtonian models with a considerably (i.e.\,$\sim10$ times)
stronger initial magnetic field.

For the most extreme type\,II model (A2B4G1-D3M13-T), the rotation rate
$\beta_{\mathrm{rot}}$ already decreases during the final stages of the
initial core collapse.  When the rotation rate reaches its maximum value
of about 12\%, a centrifugally supported shock wave is launched that
fails to explode the core.  The core continues to collapse and
eventually bounces at supra-nuclear densities due to the stiffening of
the equation of state.  The GW signal of this model belongs to the
magnetic-type GW signal (type\,IV) introduced in Paper\,I. At early
epochs it resembles a type\,II signal, but after the launch of the first
shock it shows quite a different nature.  Instead of the long-period
oscillations characteristic of a type\,II signal, the GW amplitude shows
violent oscillations whose frequencies increase as the local
sound-crossing time scale decreases in the core when it collapses to
nuclear matter density.  Such a signal is also produced by a less
magnetized TOV model of the same model series (A2B4G1-D3M12-T) and by
the corresponding strongly magnetized Newtonian model (A2B4G1-D3M13-N).

In some of our models we observe that the GW signal exhibits an almost
constant positive amplitude towards the end of the respective
simulations (see, e.g.\,model A3B3G3-D3M13-T/N), or it oscillates around
a positive value (see, e.g.\,model A3B3G5-D3M13-T/N). This is due to the
presence of the jet-like outflow along the rotation axis in those
models. It is tempting to interpret this behavior of the GW signal as a
``burst with memory'' \citep{BraTho87}, as suggested by T.\,Pradier
(private communication), which results e.g.\,when a blob of matter that
is initially at rest is accelerated during some time interval and moves
at a constant speed afterwards, giving rise to a non-vanishing constant
gravitational wave amplitude (see, e.g.\,\citealp{SegOri01}). However,
we do not think that the GW signals of our models show a ``memory'' effect.
The almost constant GW amplitude in some models is a transient resulting
from the combined action of a decelerating jet-like outflow and the
related magnetic contributions to the total quadrupole moment.

\begin{acknowledgements}
  Some of the simulations were performed using computers of the
  \emph{Rechenzentrum Garching der Max-Planck-Gesellschaft (RZG)}.  MAA
  is a Ram\'on y Cajal Fellow of the Spanish Ministery of Education and
  Science, by which he is also partially supported under the grant
  AYA2004-08067-C03-C01.
\end{acknowledgements}


\appendix

\section{Synopsis of our results}
\label{Sek:Syn}

Tables \ref{Tab:SynopseI} and \ref{Tab:SynopseII} provide an overview of
the dynamic evolution of the flow and the magnetic field and include
information on the resulting gravitational wave signal of all our
models.

\begin{table*}
  \caption[Global parameters of the models I] 
  { Some characteristic model quantities. 
  }
  \label{Tab:SynopseI} 
  \centering
  
  \begin{tabular}{ccclrcccccc}
    \hline\hline
    \rule{0cm}{0.4cm}
    Model & Type & $t_{\mathrm{b}}$ $^{\mathrm{a}}$ & 
    $\rho_{\mathrm{b,14}}$ $^{\mathrm{b}}$ & 
    $A^{\mathrm{E}2}_{20}$ $^{\mathrm{c}}$ & 
    $A^{\mathrm{E}2}_{20; \mathrm{mag}}$ $^{\mathrm{d}}$ & 
    $A^{\mathrm{E}2}_{20; \infty}$ $^{\mathrm{e}}$ &  
    $\beta_{\mathrm{rot}}^{\mathrm{max}}$ $^{\mathrm{f}}$ & 
    $\beta_{\mathrm{mag}}^{\mathrm{max}}$ $^{\mathrm{g}}$ & 
    $t_{\mathrm{m}}$ $^{\mathrm{h}}$ & 
    $\beta_{\mathrm{mag};\phi}^{\mathrm{max}}$ $^{\mathrm{i}}$  \\ 
    & & $[\mathrm{ms}]$ & & $[\mathrm{cm}]$ &    $[\mathrm{cm}]$ &  
        $[\mathrm{cm}]$ & [\%] & [\%] & $[\mathrm{ms}]$ & [\%] \\
    \hline
    A1B3G3-D3M10-T & I & 
    $48.5$ & $4.13$  & $-1110$ & $0.005$ &  $40$ &  $7.3$ & $0.01$!  & $85.1$ &  $0.01$ \\ 
    A1B3G3-D3M12-T & I &
    $48.5$ & $4.14$  & $-1129$ & $22$    &  $80$ &  $7.3$ & $1.3$    & $62.1$ &  $1.1$ \\ 
    A1B3G3-D3M13-T & I &
    $49.6$ & $4.15$! & $-1474$ & $39$    & $400$ &  $6.1$ & $2.8$    & $54.4$ &  $1.5$ \\ 
    \hline
    A2B4G1-D3M10-T & II & 
    $101$  & $0.70$ &  $-673$ & $0.008$ & ---   & $15.9$ & $0.0003$!& $169$ & $0.0003$ \\ 
    A2B4G1-D3M12-T & IV &
    $101$  & $0.72$!&  $-667$ &   $43$  & ---   & $15.8$ & $0.8$!   & $140$ & $0.7$ \\ 
    A2B4G1-D3M13-T & IV &
    $111$  & $0.98$ &  $ 923$ &   $184$ & ---   & $11.7$ & $4.3$    & $113$ & $3.1$ \\ 
    \hline
    A3B3G3-D3M10-T & I & 
    $49.5$ & $3.12$  & $-2044$ & $0.009$ & $100$ & $17.1$ & $0.009$! & $80.6$ & $0.009$ \\ 
    A3B3G3-D3M12-T & I & 
    $49.5$ & $3.13$! & $-2079$ &    $44$ & $100$ & $16.9$ & $1.5$    & $50.6$ & $1.3$ \\ 
    A3B3G3-D3M13-T & I & 
    $51.1$ & $3.31$! & $-1584$ &   $524$ & $700$ & $13.9$ & $4.3$    & $53.6$ & $2.9$ \\ 
    \hline
    \rule{0cm}{0.35cm}
    A3B3G5-D3M10-T & III & 
    $30.4$ & $3.69$  &  $272$ & $10^{-5}$&  $10$ & $10.0$ & $0.002$! & $72.6$ & $0.002$ \\ 
    A3B3G5-D3M12-T & III & 
    $30.4$ & $3.69$  &  $274$ &  $-0.04$ &   $0$ & $10.0$ & $0.6$!   & $38.2$ & $0.5$ \\ 
    A3B3G5-D3M13-T & III & 
    $30.5$ & $3.69$! &  $350$ &    $-35$ & $100$ &  $9.4$ & $5.2$    & $35.1$ & $2.8$ \\ 
    \hline\hline
  \end{tabular}
  \begin{list}{}{}
  \item[$^{\mathrm{a}}$] Time of bounce.
  \item[$^{\mathrm{b}}$] Maximum density at bounce (in units of $10^{14}
    \mathrm{g\,cm}^{-3}$). A density value with an exclamation mark
    indicates that the maximum density of the model exceeds the bounce
    density during the later evolution.
  \item[$^{\mathrm{c}}$] Maximum GW amplitude.
  \item[$^{\mathrm{d}}$] Magnetic contribution to the maximum GW signal
  \item[$^{\mathrm{e}}$] A \emph{rough} mean value of the wave amplitude
    (in cm) at some late epoch; no value is provided when the amplitude
    does not approach a quasi-constant asymptotic value.  A large
    absolute value of this amplitude indicates the presence of an
    aspheric outflow.
  \item[$^{\mathrm{f}}$] Maximum value of the ratio of rotational to
    gravitational energy.
  \item[$^{\mathrm{g}}$] Maximum value of the ratio of magnetic to
    gravitational energy. An exclamation mark indicates that the
    magnetic field is still growing at the end of the simulation.
  \item[$^{\mathrm{h}}$] The time when $\beta_{\mathrm{mag}}$ reaches
    its maximum value.
  \item[$^{\mathrm{i}}$] Maximum value of the ratio of toroidal magnetic
    to gravitational energy.
  \end{list}
\end{table*}

\begin{table*}
  \caption[Global parameters of the models II] 
  {Some characteristic model quantities (name of model given in column
    1) when the core has reached a
    quasi-equilibrium state. 
  }
  \label{Tab:SynopseII}
  \centering
  
  \begin{tabular}{cccccccrr}
    \hline\hline
    \rule{0cm}{0.4cm}
    Model 
    & $t$ $^{\mathrm{a}}$
    & $r_{\mathrm{c}}$  $^{\mathrm{b}}$
    & $M_{\mathrm{c}}$  $^{\mathrm{c}}$
    & ${2\pi}/{\Omega}$  $^{\mathrm{d}}$
    & $|\vec b|$  $^{\mathrm{e}}$
    & $|b_{\phi}|$  $^{\mathrm{f}}$
    & $r_{\mathrm{sh}}^{\mathrm{p}}$  $^{\mathrm{g}}$ 
    & $r_{\mathrm{sh}}^{\mathrm{e}}$  $^{\mathrm{h}}$ \\
    & $[\mathrm{ms}]$ & $[\mathrm{km}]$ & $[{M_{\odot}]}$ 
    & $[\mathrm{ms}]$ &
    $[\mathrm{G}]$ & $[\mathrm{G}]$ & 
    $[\mathrm{km}]$ & $[\mathrm{km}]$ \\
    \hline
    \rule{0cm}{0.35cm}
    A1B3G3-D3M10-T &  $70$ & $19.7$ & $0.58$ & $4.2$ & $7.1\times 10^{13}$ & $7.0\times 10^{13}$ & $528$ & $471$ \\
    A1B3G3-D3M12-T &  $70$ & $19.6$ & $0.56$ & $3.2$ & $1.6\times 10^{15}$ & $1.2\times 10^{15}$ & $549$ & $490$ \\
    A1B3G3-D3M13-T &  $70$ & $18.0$ & $0.60$ & $-1631$ & $2.9\times 10^{15}$ & $2.3\times 10^{14}$ & $1170$ & $563$ \\
    \hline
    \rule{0cm}{0.35cm}
    A2B4G1-D3M10-T & $167$ & $134$  & $1.26$ &  $34$ & $9.8\times 10^{10}$ & $8.5\times 10^{10}$ & --- & ---\\
    A2B4G1-D3M12-T & $141$ & $45.3$ & $0.97$ & $9.0$ & $2.7\times 10^{14}$ & $2.6\times 10^{14}$ & --- & ---\\
    A2B4G1-D3M13-T & $129$ & $23.9$ & $0.91$ & $139$ & $1.9\times 10^{15}$ & $3.4\times 10^{14}$ & --- & ---\\ 
    \hline
    \rule{0cm}{0.35cm}
    A3B3G3-D3M10-T &  $68$ & $22.0$ & $0.64$ & $3.8$ & $9.1\times 10^{13}$ & $9.1\times 10^{13}$ & $532$ & $427$ \\
    A3B3G3-D3M12-T &  $68$ & $25.5$ & $0.55$ & $9.4$ & $6.7\times 10^{14}$ & $5.6\times 10^{14}$ & $545$ & $458$ \\
    A3B3G3-D3M13-T &  $68$ & $23.0$ & $0.54$ & $28$ & $1.3\times 10^{15}$ & $3.9\times 10^{14}$ & $1008$ & $590$ \\
    \hline
    \rule{0cm}{0.35cm}
    A3B3G5-D3M10-T &  $50$ & $12.6$ & $0.19$ & $3.0$ & $4.0\times 10^{13}$ & $4.0\times 10^{13}$ & $241$ & $241$ \\
    A3B3G5-D3M12-T &  $41$ & $10.9$ & $0.17$ & $2.4$ & $4.1\times 10^{15}$ & $3.9\times 10^{15}$ & $144$ & $144$ \\
    A3B3G5-D3M13-T &  $50$ & $13.0$ & $0.24$ & $-31$ & $3.5\times 10^{15}$ & $1.2\times 10^{14}$ & $822$ & $329$ \\
    \hline\hline
  \end{tabular}
  \begin{list}{}{}
  \item[$^{\mathrm{a}}$] Time at which the quantities given were
    determined. For the models of series A2B4G1-D3Mm-T, which do not
    reach a quasi-equilibrium state by the end of the simulation, we
    provide the corresponding quantities for the final model of the
    respective simulation.
  \item[$^{\mathrm{b}}$] The surface radius of the gravitationally bound
    quasi-equilibrium configuration. Since it is still surrounded by an
    (expanding) envelope of high density matter, the definition of its
    surface radius $r_{\mathrm{c}}$ is somewhat uncertain.
  \item[$^{\mathrm{c}}$] The mass of the gravitationally bound
    quasi-equilibrium configuration.
  \item[$^{\mathrm{d}}$] The rotation rate at the surface. The angular
    velocity $\Omega$ is averaged over the angle $\theta$.  Note that
    this quantity varies strongly and on short time scales near the
    surface. Thus, the values provided should be used with care.
    Negative values of the rotation rate signify counter-rotating cores.
  \item[$^{\mathrm{e}}$] The total magnetic field at the surface.  Note
    that this quantity varies strongly and on short time scales near the
    surface. Thus, the values provided should be used with care.
  \item[$^{\mathrm{f}}$] The toroidal magnetic field at the surface.
    Note that this quantity varies strongly and on short time scales
    near the surface. Thus, the values provided should be used with
    care.
  \item[$^{\mathrm{g}}$] The radius of the shock wave at the polar axis.
    No entry here implies that the shock has already left the
    computational grid.
  \item[$^{\mathrm{h}}$] The radius of the shock wave at the equator.
    No entry here implies that the shock has already left the
    computational grid.
  \end{list}
\end{table*}

\bibliographystyle{aa}
\bibliography{4982}

\end{document}